\documentclass[a4paper,11pt]{article}
\usepackage{jheppub} 
\usepackage{orcidlink}
\usepackage{lineno}
\usepackage{slashed}
\usepackage[export]{adjustbox}
\usepackage{physics}
\usepackage{tikz}
\usepackage{ifthen}
\usetikzlibrary{decorations.markings,arrows,arrows.meta}
\usetikzlibrary{decorations.text}
\usepackage[compat=1.1.0]{tikz-feynman}
\usetikzlibrary{bending} 
\usepackage{amssymb}
\usepackage{mathtools}
\usepackage[T1]{fontenc}
\usepackage{FiraSans}
\usetikzlibrary{decorations.markings}
\usepackage{amsmath,amssymb}
\usepackage{float}
\usepackage{graphicx}
\usepackage{menukeys}
\usepackage{slashed}
\usepackage{bbold}
\usepackage{xr}
\usetikzlibrary{arrows,shapes,patterns,calc}
\usepackage[utf8]{inputenc}
\usepackage{simpler-wick}
\usepackage{comment}
\usepackage{amssymb}
\usepackage{amsmath}
\usepackage{amsfonts}
\usepackage{array}
\usepackage{multirow}
\usepackage{psfrag}
\usepackage{xcolor}
\usepackage{caption}
\usepackage{subcaption}
\usepackage{xfrac}
\newcommand{\ccline}[1]{%
    \cline{#1}%
    \noalign{\vskip-\doublerulesep}
    \cline{#1}%
    \noalign{\vskip\doublerulesep}
}
\usepackage{leftindex}
\usepackage{cleveref}
\usepackage{ifthen}
\usepackage{makecell}

\usepackage{tabularx}
\usepackage{colortbl}
\definecolor{babypink}{rgb}{0.96, 0.76, 0.76}

\usepackage{lmodern}

\def\ttlf{\ttfamily\fontseries{l}\selectfont}

\usepackage{listings}
\DeclareCaptionLabelFormat{boldlisting}{\textbf{Listing~#2}}
\newcommand{\uspin}{u(p_1)}
\newcommand{\vbar}{\bar v (p_2)}
\newcommand{\kgluon}[1]{k_{g_{#1}}}
\newcommand{\auxv}{\zeta}
\newcommand{\ii}{\mathrm{i}}
\PassOptionsToPackage{export}{adjustbox}

\def\bare{M}

\def\colorFloops{\delta_{ij} C_F T_F }

\def\unitvector{\hat{\mathbf{k}}}
\def\threshold{t}
\def\blobset{b}
\def\source{\vec{\mathbf{s}}}
\def\loopmomenta{\vec{\mathbf{k}}}
\def\thresholdset{T}

\def\Nf{N_f}

\def\pmtt{\texttt{+\!\!\!\raisebox{-.48ex}-}}
\def\amplFloops{\mathcal{M}^{(2, \tilde{N}_f)}}
\def\singleFeynDiag{\mathcal{G}}
\def\singleFeynDDiag{G}
\def\spinChainFLDiag{\mathcal{D}}
\def\prefactorGIntegrand{\colorFloops \left( \frac{\alpha_s}{4\pi}\right)^2}
\def\fermLoopDiag{\mathcal{V}}

\DeclareMathOperator{\artanh}{artanh}

\newcommand{\planar}{\text{P}}
\newcommand{\nonplanar}{\text{NP}}

\def\twoloopfloopsuv{{R}_{\rm UV} \widehat{F}^{(2,\tilde{N}_f)}}
\def\twoloopfloopsnonplanar{\widehat{F}^{(2,\tilde{N}_f),R}_\nonplanar}
\def\twoloopfloopsplanar{\widehat{F}^{(2,\tilde{N}_f),R}_\planar}

\newcommand{\perm}{\sigma}
\newcommand{\codeTool}[1]{\textsc{#1}}
\newcommand{\codeLanguage}[1]{\textsc{#1}}

\usepackage{xspace}
\newcommand{\vegas}{\codeTool{Vegas}\xspace}
\newcommand{\havana}{\codeTool{Havana}\xspace}

\chardef\MyArticleWithColor=\pdfcolorstackinit page direct{0 g}

\def\centerarc[#1](#2)(#3:#4:#5)
    { \draw[#1] ($(#2)+({#5*cos(#3)},{#5*sin(#3)})$) arc (#3:#4:#5); }

\title{Two-loop QCD corrections for real and off-shell diphoton and triphoton production via quark loops}

\author[a,b]{Dario Kermanschah\orcidlink{0000-0002-9358-744X},}
\author[a]{Matilde Vicini\orcidlink{0009-0003-1914-7328}}
\affiliation[a]{Institute for Theoretical Physics, ETH Zurich, Wolfgang-Pauli-Strasse 27, 8093 Z\"urich, Switzerland}
\emailAdd{d.kermanschah@gmail.com, mvicini@phys.ethz.ch}
\affiliation[b]{Rudolf Peierls Centre for Theoretical Physics, Oxford University, Clarendon Laboratory, Parks Road, Oxford OX1 3PU, UK}
\abstract{
We compute squared matrix elements at next-to-next-to-leading order in perturbative quantum chromodynamics for the production of two or three (on- or off-shell) photons mediated via (light or heavy) quark loops.
Our method handles all cases in a unified framework, using simultaneous subtraction of infrared, ultraviolet, and threshold singularities in loop momentum space to produce a locally finite integrand suitable for numerical integration.
We confirm agreement with available analytic benchmarks at fixed phase-space points and provide new results otherwise.
We also compute the double-virtual corrections to the cross section for on- and off-shell diphoton and on-shell triphoton production by combining the Monte Carlo integration over loop and phase space.
}

\preprint{
\begin{flushright}
OUTP-25-04P 
\end{flushright}
}

\begin{document}
\tikzset{->-/.style={decoration={
    markings,
    mark=at position {#1} with {\arrow{latex}}},postaction={decorate}}
}
\tikzset{
cut/.style={
    opacity=0.4,
    line width=2.5pt,
    shorten <=-15pt,
    shorten >=-15pt}
}
\maketitle
\flushbottom
\section{Introduction}
\label{sec:intro}
The study of electroweak boson production at the Large Hadron Collider (LHC) provides a stringent test of the Standard Model and a window into possible new physics.
Diboson processes have been measured with high precision, reaching experimental uncertainties at the percent level.
Corresponding theoretical predictions are available at next-to-next-to-leading order (NNLO) in quantum chromodynamics (QCD).
At higher final-state multiplicities, triphoton production, measured by ATLAS \cite{ATLAS:2017lpx}, is currently the only triboson process for which NNLO corrections have been computed in massless QCD \cite{Kallweit:2020gcp}.
Triboson processes involving heavy electroweak bosons are rare but have been observed by both ATLAS and CMS \cite{ATLAS:2023avk,ATLAS:2023zkw,ATLAS:2022wmu,CMS:2023rcv,CMS:2025oey,ATLAS:2024nab,CMS:2025hlu,CMS:2020hjs}.
However, the complete NNLO QCD corrections for these processes are not yet known, although desirable.
In fact, as the High-Luminosity LHC delivers increasingly precise measurements, already comparable to theoretical uncertainties in some cases, next-to-leading order (NLO) accuracy will no longer suffice \cite{Huss:2025nlt}.
The last missing ingredient at NNLO is are the double-virtual corrections, for which the required two-loop amplitudes have so far been computed only for $W\gamma\gamma$ production \cite{Badger:2024sqv}.
The double-real and real-virtual contributions, on the other hand, are in principle available through \textsc{MATRIX}~\cite{Grazzini:2017mhc}.
\par
In this work, we compute the gauge-invariant two-loop QCD corrections mediated by fermion loops to the processes 
\begin{align}
    q\bar q \to \gamma \gamma\,, \qquad
    q\bar q \to \gamma^* \gamma^*\,, \qquad
    q\bar q \to \gamma \gamma \gamma\,, \qquad
    q\bar q \to \gamma^* \gamma^* \gamma^*\,,
\end{align}
where the final-state photons may be off-shell, and the quark circulating in the loop can be either light or heavy.
The two-loop matrix elements have previously been computed in massless QCD for the production of $\gamma\gamma$ in ref.~\cite{Anastasiou:2002zn,Caola:2020dfu}, $\gamma^*\gamma^*$ in ref.~\cite{Caola:2014iua,Gehrmann:2015ora}, $\gamma\gamma\gamma$ in ref.~\cite{Chicherin:2020oor,Abreu:2020cwb,Chawdhry:2020for,Abreu:2023bdp}, and the heavy-quark loop contributions to diphoton production have been obtained in refs.~\cite{Becchetti:2023yat,Becchetti:2023wev,Becchetti:2025rrz,Ahmed:2025osb}.
In this paper, we present both evaluations of the two-loop squared matrix elements at fixed phase-space points and the corresponding double-virtual corrections to the NNLO cross section, obtained after convolution with parton distribution functions (PDFs) and phase-space integration.
Where possible, our calculations are validated against existing reference results.
\par
This work not only provides a unified computational framework for two-loop fermionic corrections to multi-boson production but also advances the multi-scale multi-leg frontier by computing the quark-loop contributions to the production of three off-shell photons.
Additionally, we add an extra mass scale by including the heavy-quark loops for $\gamma^* \gamma^*$ and  $\gamma \gamma \gamma$ production.
The $2\to3$ scattering process involves both planar and non-planar penta-box and double-box topologies with three off-shell external legs, the most complex topologies of the full two-loop amplitude, featuring infrared~(IR), ultraviolet~(UV), and threshold singularities.
Obtaining these results in analytic form requires unprecedented effort in analytic techniques, both in integral reduction and in the development of the functional basis.
\par
We extend the Monte Carlo–based framework of ref.~\cite{Kermanschah:2024utt}, initially developed for the $N_f$-contributions to the same class of processes, to the complete set of fermionic two-loop corrections.
Although heavy electroweak  $Z$ and $W^\pm$ bosons are not considered explicitly, their axial couplings can be incorporated as we demonstrated in ref.~\cite{Kermanschah:2024utt}.
The methods presented here are also directly applicable to the other diagrammatic contributions with fermion loops and can be generalised to the production of more than three electroweak vector bosons.
\par
Our method builds on recent advances in constructing locally finite two-loop amplitudes directly in loop momentum space.
The approach introduces local counterterms that cancel ultraviolet and infrared singularities at the integrand level, yielding an expression that is finite everywhere in momentum space.
Initially developed for off-shell photon production in quark-antiquark annihilation at two loops in  quantum electrodynamics (QED)~\cite{Anastasiou:2020sdt}, the method was later extended to QCD~\cite{Anastasiou:2022eym}, with key ingredients for three-loop amplitudes provided in ref.~\cite{Haindl:2025jte} and the extension to gluon-fusion channels presented in ref.~\cite{Anastasiou:2024xvk}.
Most recently, the additional complications when considering on-shell photons in quark-antiquark annihilation were also resolved~\cite{Anastasiou:2025cvy}.
It is important to note that although the fermion-loop contributions considered in this work are finite after integration, the integrand given in terms of Feynman diagrams still contains local UV and IR divergences that must be removed before numerical evaluation, as shown in refs.~\cite{Anastasiou:2020sdt,Anastasiou:2024xvk}. 
\par
Our method further relies on the analytic integration over the energy component of the loop momenta.
This operation results in the loop-tree duality~(LTD)~\cite{Catani:2008xa, Aguilera-Verdugo:2019kbz, Aguilera-Verdugo:2020set,JesusAguilera-Verdugo:2020fsn, Ramirez-Uribe:2022sja, Runkel:2019yrs, Capatti:2019ypt}, or in locally identical but more numerically stable \emph{causal} representations 
from (partially) time-ordered perturbation theory~((P)TOPT)~\cite{Bodwin:1984hc, Collins:1985ue, Sterman:1993hfp, Sterman:1995fz,Sterman:2023xdj,Aguilera-Verdugo:2020set, Aguilera-Verdugo:2020kzc, JesusAguilera-Verdugo:2020fsn,Ramirez-Uribe:2020hes, Sborlini:2021owe, TorresBobadilla:2021ivx, Bobadilla:2021pvr, Benincasa:2021qcb, Kromin:2022txz, Capatti:2020ytd,Capatti:2022mly, Capatti:2023shz}.
Threshold singularities are tackled using local subtraction in momentum space~\cite{Kermanschah:2021wbk,Kilian:2009wy,Kermanschah:2024utt,Vicini:2024ecf},
an alternative approach to contour deformation~\cite{Becker:2012aqa,Becker:2011vg,Soper:1999xk,Capatti:2019edf,Buchta:2015wna,Kromin:2022txz} that offers better control and efficiency.
This method provides a systematic prescription to separately compute dispersive and absorptive parts of any finite $n$-loop integral, where the latter costitutes a finite representation of the optical theorem involving exactly $n+1$ cuts.
Singularities from any additional on-shell cuts cancel locally.
When applied to forward-scattering diagrams, this feature enables the direct numerical integration of fully inclusive cross sections without the need to regularise final-state IR singularities, analogous to refs.~\cite{Capatti:2025khs,AH:2023kor,Capatti:2020xjc,Capatti:2022tit,Soper:1999xk,Soper:1998ye}.
In this work, we apply the threshold subtraction approach to general two-loop thresholds, as previously demonstrated in ref.~\cite{Vicini:2024ecf}, beyond the simpler $N_f$-contributions of ref.~\cite{Kermanschah:2024utt}. 
\par
Finally, the numerical integration is performed using the multi-channel Monte Carlo method~\cite{Hilgart:1992xu,Soper:1999xk} in loop momentum space designed to flatten integrable singularities
and further improve efficiency using the adaptive importance sampling algorithm \vegas \cite{Lepage:2020tgj,Lepage:1977sw,Hahn:2004fe,havana}.
\par
The paper is organized as follows. In section~\ref{sec:framework}, we define the contributions that we compute.
Section~\ref{sec:IR_counterterms} reviews the relevant local IR and UV counterterms.
Threshold singularities and their regularisation via local subtraction are addressed in section~\ref{sec:thresholds}.
In section~\ref{sec:multi-channelling} we present the multi-channel Monte Carlo method and the chosen importance sampling, illustrated with selected examples of multi-loop scalar integrals.
In section~\ref{sec:implementation}, we describe our implementation, and in section~\ref{sec:results}, we provide numerical results for the two-loop squared matrix elements at fixed phase-space points, as well as double-virtual corrections after convolution with the parton distribution functions~(PDFs) and phase-space integration.
In appendices~\ref{sec:individual_me} and~\ref{sec:ps_points}, we show the individual contributions to the numerical results and the values of the phase-space points, respectively.
\section{Framework}
\label{sec:framework}
We study the contributions to two-loop scattering amplitudes for quark–antiquark annihilation into real and virtual photons mediated through closed quark-loops. 
The relevant processes are
\begin{align}
    q(p_1) + \bar q(p_2) &\to \gamma(q_1) +\gamma(q_2)\,, \quad
    & q(p_1) + \bar q(p_2) &\to \gamma(q_1) +\gamma(q_2) +\gamma(q_3)\,,\nonumber \\ \label{eq:processes}
    q(p_1) + \bar q(p_2) &\to \gamma^*(q_1) +\gamma^*(q_2)\,, \quad
   & q(p_1) + \bar q(p_2) &\to \gamma^*(q_1) +\gamma^*(q_2) +\gamma^*(q_3)\,,
\end{align}
where $q$ and $\bar{q}$ denote a massless quark-antiquark pair, and $\gamma$ or $\gamma^*$ an on- or off-shell photon.
The perturbative expansion of the scattering amplitude in the strong coupling $\alpha_s$ reads
\begin{align}
    \bare(\alpha_s) = \bare^{(0)}
    + \left(\frac{\alpha_s}{4\pi}\right)
    \bare^{(1)}
    + \left(\frac{\alpha_s}{4\pi}\right)^2
    \bare^{(2)}
    + \mathcal{O}(\alpha_s^3)\,.
\end{align}
The fermion-loop contributions from $N_f$ quark flavours of mass $m$ to the two-loop amplitude $\bare^{(2)}$, for incoming quarks with colour indices $i$ and $j$, can be decomposed into individually gauge-invariants parts:
\begin{align}
    \delta_{ij}C_F T_F \left( \bare^{(2,N_f)}+\bare^{(2,\tilde{N}_f)}\right)
\end{align}
where $\bare^{(2,N_f)}$ is proportional to $N_f$ and $\bare^{(2,\tilde{N}_f)}$ is proportional to the sum of the quarks' squared charges.
For the latter contribution, we will evaluate the matrix elements
\begin{align}
\label{eq:matrix-element}
    F^{(2,\tilde{N}_f)} &= \sum_h 2\Re\left[\left(M_h^{(0)}\right)^*M_h^{(2,\tilde{N}_f)}\right]
\end{align}
summed over all helicity configurations $h$.
The calculation is performed for both light ($m=0$) and heavy fermions ($m>0$) circulating in the loop.
The hard-scattering contributions to the $N_f$-parts of the same processes were computed in ref.~\cite{Kermanschah:2024utt} with the same techniques.
\par
\begin{figure}[t]
\centering
\begin{subfigure}[b]{0.30\textwidth}
    \centering
\includegraphics[width=\textwidth, page=6]{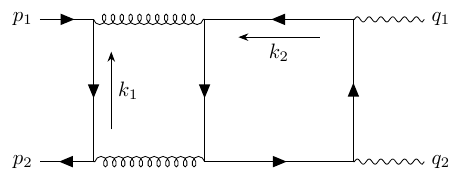}
\caption{}
    \label{fig:subfiga-2to2}
\end{subfigure}
\begin{subfigure}[b]{0.30\textwidth}
    \centering
   \includegraphics[width=\textwidth, page=7]{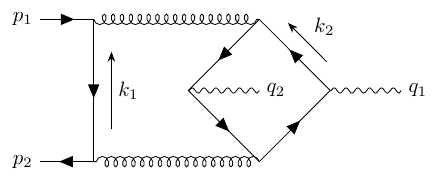}
\caption{}
    \label{fig:subfigb-2to2}
\end{subfigure}
\caption{Examples of planar and non-planar two-loop diagrams contributing to the $2\to 2$ process.}
\label{fig:2to2diags}
\end{figure}

\begin{figure}[b]
\centering
\begin{subfigure}[b]{0.30\textwidth}
    \centering
   \includegraphics[width=\textwidth, page=4]{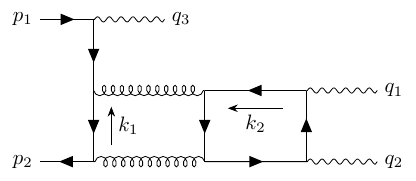}
    \caption{}
    \label{fig:subfig-b}
\end{subfigure}
\begin{subfigure}[b]{0.30\textwidth}
    \centering
   \includegraphics[width=\textwidth, page=5]{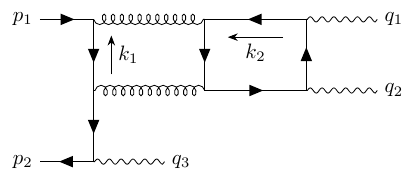}
\caption{}    
    \label{fig:subfig-c}
\end{subfigure}
\begin{subfigure}[b]{0.30\textwidth}
    \centering
\includegraphics[width=\textwidth, page=3]{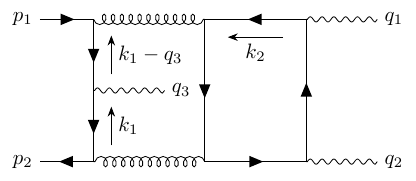}
\caption{}
    \label{fig:subfig-a}
\end{subfigure}
\caption{Examples of planar two-loop diagrams contributing to the $2\to 3$ process.}
\label{fig:box-in-2-to-3}
\end{figure}

For the $2\to2$ processes, the relevant Feynman diagrams are those in figure~\ref{fig:subfiga-2to2} and \ref{fig:subfigb-2to2}, together with permutations of the final-state momenta.
Those contributing to the $2\to3$ process are obtained by attaching the third photon to the incoming quark line, as those in figure~\ref{fig:box-in-2-to-3} for the planar, and analogously for the non-planar diagrams.
\par
After loop integration, the bare amplitude $M^{(2,\tilde{N}_f)}$ is finite and identical to the renormalised amplitude.
However, the loop integrand contains UV divergent and non-factoring IR divergent regions.
These regions need to be locally subtracted with suitable IR and UV counterterms in order to be numerically integrable in $d=4$ spacetime dimensions. 
\section{Construction of IR- and UV- finite amplitudes}\label{sec:IR_counterterms}
\begin{figure}[b]
    \centering
    \includegraphics[width=0.5\textwidth, page=1]{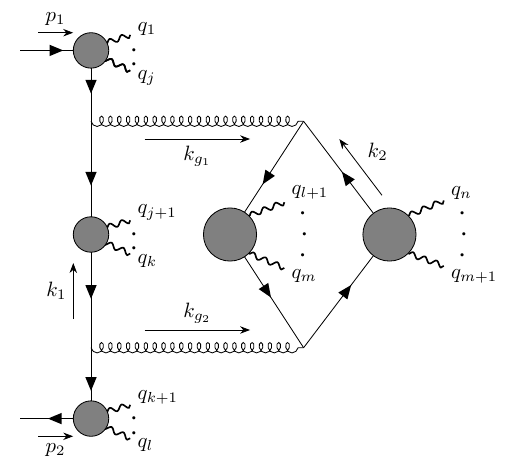}
    \caption{Diagrams that contribute to the amplitude with coefficient $\sum_{f=1}^{\Nf}Q_f^r$, where $r= n - l-1 \geq 1$ is the number of photons attached to the fermion loop in the figure.
    We explicitly assign the loop momenta labels $k_1$ and $k_2$ of the corresponding integrand built with conventional Feynman rules. It is also useful to denote by $\kgluon{1}$ and $\kgluon{2}$ the gluon momenta in the direction specified by the figure. 
    }
    \label{fig:blob}
\end{figure}

We use the diagram in figure~\ref{fig:blob} to represent the class of Feynman diagrams with fermion loops to which at least one photon is attached, and denote it by
\begin{equation}\label{eq:Gdef-blobsets}
   \prefactorGIntegrand \singleFeynDDiag( (q_1,\dots, q_{j}) ,      
   (q_{j+1},\dots, q_{k}),
    (q_{k+1},\dots, q_{l}),
    (q_{l+1},\dots, q_{m}),
    (q_{m+1},\dots, q_{n}))\,.
\end{equation}
Since it will be convenient to work at the loop-integrand level, we will use the convention to denote by calligraphic symbols the integrands of the corresponding (non-calligraphic) quantities, related by
\begin{align}
    \singleFeynDDiag = \int \frac{\dd^d k_1}{\pi^{d/2}} \frac{\dd^d k_2}{\pi^{d/2}} \,\singleFeynDiag\,.
\end{align}
Analogously, the loop integrand of $M^{(2,\tilde{N}_f)}$ will be denoted by $\mathcal{M}^{(2,\tilde{N}_f)}$, which we can then express as the sum
\begin{equation}\label{eq:ampl-blobs-def}
\amplFloops (k_1, k_2)=
\sum_{
    \substack{
    \blobset_1, \dots,\blobset_5 \\
    \sqcup_a \blobset_a = \{q_1,\ldots,q_n\}\,\\
    b_a\,\, {\rm ordered}
    }
    }
\singleFeynDiag\!\left(\blobset_{1},\dots,\blobset_{5};k_1,k_2\right)\,.
\end{equation}
\par
In the following, we will discuss the generic construction of the IR counterterms, for all diagrams of the type in figure~\ref{fig:blob}.
For the processes in~\eqref{eq:processes}, we are only interested in the case $n-l-1=2$, where the two virtual gluons attach the incoming quark line to a fermion loop with four propagators.
By Furry's theorem~\cite{Furry:1937zz,Sterman:1993hfp}, only fermion loops with an even number of photon attachments contribute.
\par
The IR and UV local counterterms for this class of diagrams have been discussed in ref.~\cite{Anastasiou:2020sdt} and~\cite{Anastasiou:2024xvk}. 
We will adopt the notation of ref.~\cite{Anastasiou:2025cvy} and denote the locally IR-finite integrand by
\begin{align}
    \widehat{\mathcal{M}} = \mathcal{M} - \Delta \mathcal{M},
\end{align}
where $\Delta \mathcal{M}$ are local IR counterterms that integrate to zero.
The remaining UV divergences are removed using the $R$-operation~\cite{Bogoliubov:1957gp,Caswell:1981ek,Zimmermann:1969jj,Hepp:1966eg,Chetyrkin:1982nn,Chetyrkin:1984xa,Smirnov:1985yck,Gomes:1974cr,Capatti:2022tit,Herzog:2017bjx,Herzog:2017jgk}, resulting in the locally UV- and IR-finite integrand
\begin{align}
    \widehat{\mathcal{M}}^{(2,\tilde{N}_f),R} 
    =
    \widehat{\mathcal{M}}^{(2,\tilde{N}_f)} - \mathcal{R}_{\rm UV}\widehat{\mathcal{M}}^{(2,\tilde{N}_f)}\,,
\end{align}
where the hat and $R$ superscript may likewise denote this operation for individual diagrams.
\par
Whereas the IR counterterms integrate to zero, the UV counterterms yield a non-vanishing contribution that has to be added back.
As in ref.~\cite{Anastasiou:2020sdt}, we integrate the UV counterterms over $k_2$ analytically in dimensional regularisation.
In the sum of all integrated diagrams, the UV divergences cancel, resulting in a finite integral over $k_1$, which we will perform numerically.
\par

Eventually, we will express the amplitude $M^{(2,\tilde{N}_f)}$ in terms of the building blocks
\begin{align}\label{eq:floops-pieces}
    \bare^{(2,\tilde{N}_f)}
    = \widehat{\bare}^{(2,\tilde{N}_f),R}_\planar
    + \widehat{\bare}^{(2,\tilde{N}_f),R}_\nonplanar
    + {R}_{\rm UV} \widehat{M}^{(2,\tilde{N}_f)}\,, 
\end{align}
where we split the planar and non-planar contributions\footnote{Due to their generally different threshold structure we treat them separately.}, respectively defined as
\begin{align}
\label{eq:planar_and_non_planar}
    \widehat{\bare}^{(2,\tilde{N}_f),R}_\planar
    = 2\sum_{\perm\in S_N} \widehat{G}^R_\planar(\perm)\,, \qquad 
    \widehat{\bare}^{(2,\tilde{N}_f),R}_\nonplanar
    = 2\sum_{\perm\in C_N} \widehat{G}^R_\nonplanar(\perm),
\end{align}
where the factor $2$ stems from the reversed fermion-loop flow.
The sums are over all permutations $S_N$ for the planar and over cyclic permutations $C_N$ for the non-planar diagrams, where $N=2,3$ is the number of final-state photons. 
The planar and non-planar diagrams with a single fermion flow direction and fixed ordering of outgoing momenta are
\begin{align}
\label{eq:planar_and_non_planar_2_to_2}
    \widehat{G}^R_\planar(\perm)
    = \widehat{\singleFeynDDiag}^R(\emptyset,\emptyset,\emptyset,\emptyset,(q_{\perm(2)},q_{\perm(1)}))\,, \qquad
    \widehat{G}^R_\nonplanar(\perm)
    = \widehat{\singleFeynDDiag}^R(\emptyset,\emptyset,\emptyset,(q_{\perm(2)}),(q_{\perm(1)}))
\end{align}
for diphoton production and 
\begin{align}
\label{eq:planar_and_non_planar_2_to_3}
    \widehat{G}^R_\planar(\perm)
    &= \widehat{\singleFeynDDiag}^R((q_{\perm(3)}),\emptyset,\emptyset,\emptyset,(q_{\perm(2)},q_{\perm(1)})) 
    &\widehat{G}^R_\nonplanar(\perm)
    &= \widehat{\singleFeynDDiag}^R((q_{\perm(3)}),\emptyset,\emptyset,(q_{\perm(2)}),(q_{\perm(1)})) \nonumber\\
    &+ \widehat{\singleFeynDDiag}^R(\emptyset,(q_{\perm(3)}),\emptyset,\emptyset,(q_{\perm(2)},q_{\perm(1)}))
    & &+\widehat{\singleFeynDDiag}^R(\emptyset,(q_{\perm(3)}),\emptyset,(q_{\perm(2)}),(q_{\perm(1)})) \\
    &+ \widehat{\singleFeynDDiag}^R(\emptyset,\emptyset,(q_{\perm(3)}),\emptyset,(q_{\perm(2)},q_{\perm(1)}))
    & &+ \widehat{\singleFeynDDiag}^R(\emptyset,\emptyset,(q_{\perm(3)}),(q_{\perm(2)}),(q_{\perm(1)})) \nonumber
\end{align}
for triphoton production.\footnote{Although each summand is individually finite and could be integrated separately, we chose to combine them.}

\subsection{Infrared counterterms}
\label{sec:ir_cts}
We fix the loop momentum routing and denote the gluon momenta by $\kgluon{1}$ and $\kgluon{2}$, as labelled in figure~\ref{fig:blob}.
Explicitly,
\begin{align}
\label{eq:ampl-generic}
\singleFeynDiag(\blobset_1,\cdots, \blobset_5; k_1,k_2) = 
\spinChainFLDiag^{\alpha\beta}
(\blobset_1, \blobset_2,\blobset_3; k_1 )
\fermLoopDiag_{\alpha\beta}(\blobset_4,\blobset_5; k_1,k_2) 
\end{align}
where
\begin{align}\label{eq:ifermion-loop}
\fermLoopDiag^{\alpha\beta}(
\blobset_4, \blobset_5;
k_1,k_2) =
\Tr{ \gamma^{\beta}
B( \blobset_4; \sum_{i=l+1}^{m}q_{i}-k_1-k_2)\gamma^{\alpha}
B( \blobset_5; -k_2)}
\end{align}
is the closed fermion loop subdiagram while
\begin{align}\label{eq:incoming-spin-chain}
\spinChainFLDiag^{\alpha\beta}&(\blobset_1, \blobset_2,\blobset_3; k_1 )
=
\frac{\vbar B(\blobset_3;p_2) \gamma^\beta 
B( \blobset_2; k_1)\gamma^{\alpha}
B( \blobset_1; \sum_{i=1}^jq_{i} -p_1)
\uspin}{
\kgluon{1}^2\kgluon{2}^2
}
\end{align}
is the remaining subgraph and where we defined
\begin{eqnarray}
\includegraphics[width=0.2\textwidth, page=2,valign=c]{figures/blob_figures.pdf} =\includegraphics[width=0.2\textwidth, page=3,valign=c]{figures/blob_figures.pdf}  \equiv  B((q_1, \cdots, q_n); p)\,,
\end{eqnarray}
with the understanding that external quark legs are truncated in $B$.
\par
A set of IR counterterms that render individual diagrams finite and are valid for all possible attachments of external photons was first proposed in ref.~\cite{Anastasiou:2020sdt}.
They exploit the so-called \textit{$\hat \gamma$-prescription}. 
To illustrate the construction of the IR counterterms, we consider a diagram with a collinear singularity when $k_1$ is parallel to $p_2$.
In the limit, the singular part of such diagram is $\spinChainFLDiag$ and, for some arbitrary  $\blobset_1$, $\blobset_2$, it reads 
\begin{equation}
\label{eq:spin-chain-p2-coll}
\spinChainFLDiag^{\alpha\beta} (\blobset_1,\blobset_2,\emptyset;k_1)= 
-
\vbar {\gamma}^\beta
\frac{ (-\slashed{k}_1)  }
{k_1^2\kgluon{2}^2} 
\ldots
\gamma^\alpha
\ldots
 \uspin \,.
\end{equation}
To regulate the collinear-to-$p_2$ singularity, we modify the vertex adjacent to the incoming quark line with momentum $p_2$, replacing $\gamma^\beta$ by
\begin{equation}
\label{eq:gamma-hat}
\hat{\gamma}^\beta\left( {\kgluon{2}}, \auxv \right) 
\equiv 
\gamma^\beta + \kgluon{2}^{\beta} \frac{2 \slashed{\auxv}}{(\kgluon{2}-\auxv)^2 - \auxv^2}\, , 
\end{equation}
where $\auxv$ is an auxiliary vector chosen so that it produces no new pinches from the extra denominator, i.e.~we require that 
\begin{equation}\label{eq:auxv-properties}
p_1\cdot \auxv \,, p_2\cdot \auxv \,, \auxv^2 \neq 0 \,.
\end{equation}
The replacement in eq.~\eqref{eq:gamma-hat} effectively subtracts the unphysical gluon polarisation, while maintaining quadratic denominators. In the collinear limit, $k_1= x p_2\,,\,\,x\in(0,1)$, we have 
\begin{equation}
\vbar \hat \gamma^{\beta}(\kgluon{2},\auxv)
\left(\slashed k_1 \right) \ldots
\xrightarrow{k_1 = x p_2}  0. 
\end{equation}
An analogous modification is made for diagrams which develop a singularity when a gluon is collinear to $p_1$.   
For diagrams with both types of collinear singularities, we apply the corresponding vertex modifications at the same time:
\begin{align}
\label{eq:gamma-hat-ct-incoming-spin-chain}
    \widehat{\spinChainFLDiag}^{\alpha\beta}& (\emptyset,\blobset, \emptyset;k_1)=  
    -
    \vbar \hat{\gamma}^\beta(\kgluon{2},\auxv) 
    \frac{ (-\slashed{k}_1) }
{k_1^2\kgluon{2}^2}
\ldots
\frac{\slashed{p}_1 -\slashed{\kgluon{1}}}{(\kgluon{1}-p_1)^2\kgluon{1}^2}
\hat{\gamma}^\alpha(\kgluon{1}, \auxv)\uspin\,,
\end{align}
such that the IR-finite diagram reads
\begin{align}\label{eq:gamma-hat-ct}
\widehat{\singleFeynDiag}&(\emptyset,b_2,\emptyset,b_4,b_5; k_1,k_2) = 
\widehat{\spinChainFLDiag}^{\alpha\beta} (\emptyset,\blobset_2, \emptyset;k_1) \fermLoopDiag_{\alpha\beta}(b_4,b_5; k_1,k_2)\,,
\end{align}
for some arbitrary ordered lists $\blobset_2,\blobset_4,\blobset_5$ of external photons. 
In this case, the IR local counterterms are inferred as
\begin{equation}
    \Delta \singleFeynDiag  = \singleFeynDiag - \widehat{\singleFeynDiag} \label{eq:uselessDeltaG}
\end{equation}
using $\widehat{\singleFeynDiag}$ from eq.~\eqref{eq:gamma-hat-ct}.
\par
In specific cases, a smaller set of IR counterterms results in a more efficient and compact evaluation.
Whenever there are no bosons attached to the incoming spin chain, we can use the collinear approximations of ref.~\cite{Anastasiou:2024xvk}. 
In this case,
\begin{equation}\label{eq:integrand-pure-ward-ids}
\spinChainFLDiag^{\alpha\beta} (\emptyset,\emptyset,\emptyset;k_1) = 
\ii
\frac{\overline{v}(p_2) \gamma^{\beta} (-\slashed{k}_1)\gamma^{\alpha}u(p_1)}{k_1^2\kgluon{1}^2\kgluon{2}^2}\,.
\end{equation}
In the collinear limits,
\begin{align}\label{eq:ct-p1}
 \singleFeynDiag \xrightarrow[k_1\parallel p_1 ]{} \Delta_1\singleFeynDiag& \equiv 
 -\ii
 \frac{\vbar \gamma^{\beta} \uspin } {k_1^2\kgluon{1}^2}\frac{2p_1\cdot \auxv}{((\kgluon{1}-\auxv)^2-\auxv^2)p_1\cdot p_2} 
\kgluon{1}^{\alpha}\fermLoopDiag_{\alpha\beta}(k_1,k_2)\,,\\
\label{eq:ct-p2}
 \singleFeynDiag \xrightarrow[ k_1\parallel p_2 ]{}  \Delta_2\singleFeynDiag& \equiv 
 -\ii
 \frac{\vbar \gamma^{\alpha} \uspin } {k_1^2\kgluon{2}^2}\frac{-2p_2\cdot \auxv}{((\kgluon{2}-\auxv)^2-\auxv^2)p_1\cdot p_2} 
\kgluon{2}^{\beta}\fermLoopDiag_{\alpha\beta}(k_1,k_2)\,,
\end{align}
where $\auxv$ satisfies the same conditions as eq.~\eqref{eq:auxv-properties}.
Notice that $\Delta_1\singleFeynDiag$ is finite in the $k_1\parallel p_2$ limit and $\Delta_2\singleFeynDiag$ is finite in the $k_1\parallel p_1$ limit, hence they do not re-introduce IR singularities when regularising a single collinear limit and we use directly eqs.~\eqref{eq:ct-p1} and~\eqref{eq:ct-p2} as our local IR counterterms,\footnote{We remark that we slightly modify the counterterm structure proposed in section~5 of ref.~\cite{Anastasiou:2024xvk}, which would have $p_1\cdot (2\auxv-\kgluon{1})$ instead of $2p_1\cdot \auxv$ in the numerator for $\Delta_1\singleFeynDiag$ of eq.~\eqref{eq:ct-p1}, and similarly for $\Delta_2\singleFeynDiag$. Both choices are equivalent in the collinear limits, but our version has the advantages of simplifying the local UV counterterm construction, as well as the numerator evaluation.}
\begin{equation}\label{eq:FD-finite}
 \widehat{\singleFeynDiag} = \singleFeynDiag- \Delta \singleFeynDiag\,,\quad \Delta \singleFeynDiag=\Delta_1\singleFeynDiag+ \Delta_2\singleFeynDiag \,.
\end{equation}
\par
It is also possible to avoid the $\hat \gamma$-prescription for the cases where only a single massless boson attaches to the incoming spin chain between the two gluons.
The diagram in figure~\ref{fig:subfig-a} provides an example of this topology, while for \ref{fig:subfig-b}, \ref{fig:subfig-c}
we need to use the $\hat \gamma$-prescription. 
The integrand for the Feynman diagram in figure~\ref{fig:subfig-a} has
\begin{equation}\label{eq:diag-nmid}
\spinChainFLDiag^{\alpha\beta} (\emptyset, (q_3),\emptyset;k_1) = 
\ii Q_f
\frac{\vbar \gamma^{\beta} (-\slashed{k}_1)\slashed{\epsilon}^*(q_3)(-\slashed{k}_1+\slashed{q}_3)\gamma^{\alpha}\uspin}{k_1^2(k_1-q_3)^2\kgluon{1}^2\kgluon{2}^2}\,,
\end{equation}
where $Q_f$ denotes the electric charge of the fermion labelled by $f$.
The following IR counterterms
\begin{align}
    \label{eq:ct1-internal-photon}
\Delta_1\singleFeynDiag &= 
\ii Q_f
\frac{\vbar \gamma^{\beta} (-\slashed{k}_1)\slashed{\epsilon}^*(q_3)\uspin}{\kgluon{1}^2\kgluon{2}^2(k_1-q_3)^2}    \frac{2p_1\cdot \auxv}{((\kgluon{1}-\auxv)^2-\auxv^2)p_1\cdot q_3}
\kgluon{1}^{\alpha}
\fermLoopDiag_{\alpha\beta}(k_1,k_2)\,, \\
\label{eq:ct2-internal-photon}
\Delta_2 \singleFeynDiag &=
\ii Q_f 
\frac{\vbar \slashed{\epsilon}^*(q_3)(-\slashed{k}_1+\slashed{q}_1)\gamma^{\alpha}\uspin}{k_1^2\kgluon{1}^2\kgluon{2}^2}    \frac{-2p_2\cdot \auxv}{((\kgluon{2}-\auxv)^2-\auxv^2)p_2\cdot q_3}
\kgluon{2}^{\beta}
\fermLoopDiag_{\alpha\beta}(k_1,k_2)\,,
\end{align}
remove the singularities $(k_1-q_3)\parallel p_1$, $k_1\parallel p_2$ respectively.
\par
For all intents and purposes, we set $\auxv = p_1+p_2$ as our auxiliary vector. This ensures that, as we vary the centre-of-mass energy of the scattering, no thresholds that depend on $\auxv$ are introduced. 
\par
We remark that, by using consistent momentum routings in the fermion loop, the IR counterterms of eqs.~\eqref{eq:ct-p1},~\eqref{eq:ct-p2} (and analogously~\eqref{eq:ct1-internal-photon},~\eqref{eq:ct2-internal-photon})
can be combined for the sum of all diagrams (in the same class) in a so-called \textit{shift counterterm}, where the Ward identity cancellations across different diagrams in the collinear limits are made explicit. 
This results in the final expression of the IR counterterms of section 5.3 of ref.~\cite{Anastasiou:2024xvk}. 
We do not combine them here, as we will treat their different threshold singularity structure separately.
\par
The sum of all IR counterterms integrates to zero,
\begin{equation}\label{eq:integrated_IR_cts}
     \Delta \bare^{(2,\tilde{N}_f)}= \sum \Delta G = 0\,,
\end{equation}
due to the QED Ward identity, as shown in refs.~\cite{Anastasiou:2020sdt,Anastasiou:2024xvk}.

\subsection{Ultraviolet counterterms}
\label{sec:uv_cts}
We exploit the construction of section 4.3 of ref.~\cite{Anastasiou:2020sdt}, which we summarise here for completeness.
For the processes under consideration, the only UV divergent subgraphs are the box digrams contributing to the subamplitude
\begin{align}
    A_{\alpha\beta}(q_i,q_j)  = \sum {V}_{\alpha\beta}= \includegraphics[width=0.175\textwidth, page=1,valign=c]{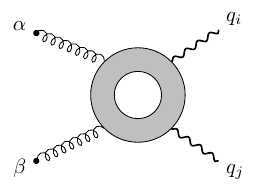}\,.
\end{align}
They are regularised with the $R$-operation, according to
\begin{equation}\label{eq:G-uv-def}
 \widehat{\mathcal{G}}^{R}=\widehat{\mathcal{G}} - \mathcal{R}_{\rm UV} \widehat{\mathcal{G}} = \widehat{\spinChainFLDiag}^{\alpha\beta} (\fermLoopDiag- \mathcal{R}_{\rm UV} \fermLoopDiag)_{\alpha\beta}\,,
\end{equation}
where, for example,
\begin{align}
\mathcal{R}_{\rm UV} \left(\includegraphics[width=0.2\textwidth, page=2,valign=c]{figures/floop.pdf}\right) 
=- 
Q_f^2
\frac{\Tr{\slashed{k}_2\gamma_{\alpha}\slashed{k}_2\slashed{\epsilon}^{*}(q_1)\slashed{k}_2\slashed{\epsilon}^{*}(q_2)\slashed{k}_2\gamma_{\beta}}}{(k_2^2-M^2)^4} \,,
\end{align}
and analogously for the other box diagrams.
After integrating out $k_2$ and summing over all diagrams, the UV divergences cancel and yield
\begin{align}
\label{eq:uv-sum-ferm-loop}
R_{\rm UV}A_{\alpha\beta}(q_i,q_j) =
\ii
\left(\sum_{f=1}^{\Nf} Q_f^2\right)
\frac{4}{3}
\left(g_{\alpha\beta} \epsilon^*(q_i)\cdot\epsilon^*(q_j)+
\epsilon_{\beta}^*(q_i)\epsilon_{\alpha}^*(q_j)
+\epsilon_{\alpha}^*(q_i)\epsilon_{\beta}^*(q_j)
\right)\,.
\end{align} 
The full UV counterterm is obtained by contracting with the incoming fermion chains $\widehat{\mathcal{D}}$, yielding
\begin{align}
\label{eq:integrated_uv_2_to_2}
    R_\text{UV}M^{(2,\tilde{N}_f)}
    = \int \frac{\dd^d k_1}{\pi^{d/2}} \widehat{D}^{\alpha\beta}(\emptyset,\emptyset,\emptyset)R_\text{UV}A_{\alpha\beta}(q_1,q_2)
\end{align}
for diphoton production and
\begin{align}
\label{eq:integrated_uv_2_to_3}
    R_\text{UV}M^{(2,\tilde{N}_f)}
    = 
    \sum_{\perm\in C_3}
    \int \frac{\dd^d k_1}{\pi^{d/2}}
    &\left(\widehat{D}^{\alpha\beta}((q_{\perm(3)}),\emptyset,\emptyset)
    + \widehat{D}^{\alpha\beta}(\emptyset,(q_{\perm(3)}),\emptyset)
    + \widehat{D}^{\alpha\beta}(\emptyset,\emptyset,(q_{\perm(3)})
    \right) \nonumber\\
    &\times R_\text{UV}A_{\alpha\beta}(q_{\perm(1)},q_{\perm(2)})
\end{align}
for triphoton production.
The remaining integral over $k_1$ is locally finite in $d=4$ dimensions and we perform it numerically.
The full contribution $\bare^{(2,\tilde{N}_f)}$ is then combined using the locally finite pieces according to eq.~\eqref{eq:floops-pieces}.

\newcommand{\tba}{{\color{orange}\threshold_1^{s}}}
\newcommand{\tcf}{{\color{gray}\threshold_2^{s}}}
\newcommand{\tadf}{{\color{magenta}\threshold_3^{s}}}
\newcommand{\tbcd}{{\color{teal}\threshold_4^{s}}}
\newcommand{\taeg}{{\color{purple}\threshold_5^{s}}}

\newcommand{\tef}{{\color{purple}\threshold_1^{s_2}}}
\newcommand{\tbde}{{\color{cyan}\threshold_2^{s_2}}}
\newcommand{\tdg}{{\color{olive}\threshold_3^{s_2}}}
\newcommand{\tacg}{{\color{brown}\threshold_4^{s_2}}}

\newcommand{\tce}{{\color{green}\threshold_1^{s_1}}}
\newcommand{\tade}{{\color{blue}\threshold_2^{s_1}}}
\newcommand{\tbcg}{{\color{pink}\threshold_3^{s_1}}}

\newcommand{\thi}{{\color{red}\threshold^{s_3}}}

\section{Treatment of thresholds}\label{sec:thresholds}
\begin{figure}[t]
\centering
\begin{subfigure}[b]{\textwidth}
\centering
\hfill
\begin{minipage}{0.65\textwidth}
\begin{tikzpicture}[scale=3]
\coordinate (a) at (1.5,1);
\coordinate (b) at (2.5,1);
\coordinate (c) at (3,0.5);
\coordinate (d) at (2.5,0);
\coordinate (e) at (1.5,0);
\draw[cut,orange] (a) -- node[opacity=1,left] {$\tba$} (e);
\draw[cut,gray] (b) -- node[opacity=1,left] {$\tcf$} (d);
\draw[cut,blue] (a) -- node[opacity=1,pos=0.44,above] {$\tade$} (c);
\draw[cut,green] (b) -- node[opacity=1,above right] {$\!\tce$} (c);
\draw[cut,violet] (d) -- node[opacity=1,below right] {$\!\tef$} (c);
\draw[cut,cyan] (e) -- node[opacity=1,pos=0.44,below] {$\tbde$} (c);
\draw[cut,magenta] (a) -- node[opacity=1,pos=0.33,left] {$\tadf~$} (d);
\draw[cut,teal] (e) -- node[opacity=1,pos=0.33,left] {$\tbcd~$} (b);
\begin{feynman}
\vertex (A) at (0.5,1);
\vertex (B) at (1,1);
\vertex (C) at (2,1);
\vertex (D) at (3,1);
\vertex (E) at (3.5,1);
\vertex (F) at (0.5,0);
\vertex (G) at (1,0);
\vertex (H) at (2,0);
\vertex (I) at (3,0);
\vertex (J) at (3.5,0);
\diagram*{
(A) -- [fermion, momentum=$p_1$] (B),
(I) -- [boson, momentum'=$q_2$] (J),
(G) -- [fermion, reversed momentum=$p_2$] (F),
(D) -- [boson, momentum=$q_1$] (E),
(B) -- [gluon] (C),
(G) -- [gluon] (H),
(B) -- [fermion, reversed momentum'=$k_1$] (G),
(D) -- [fermion, ultra thick, momentum'=$k_2$] (C),
(H) -- [fermion, ultra thick] (I),
(C) -- [fermion, ultra thick] (H),
(I) -- [fermion, ultra thick] (D)
};
\end{feynman}
\end{tikzpicture}
\end{minipage}%
\hfill
\begin{minipage}{0.3\textwidth}
\caption{Planar two-loop box diagram with fixed ordering of photon momenta $G_\text{P}$.}
\end{minipage}
\hfill
\end{subfigure}
\begin{subfigure}[b]{\textwidth}
\centering
\hfill
\begin{minipage}{0.65\textwidth}
\begin{tikzpicture}[scale=3]
\draw[cut,orange] (1.25,1) -- node[opacity=1,left] {$\tba$} (1.25,0);
\draw[cut,blue]  (2,1) node[opacity=1,above right] {$\!\tade$} -- (2.75,0.25);
\draw[cut,green] (2.75,0.75) -- node[opacity=1,right] {$\tce$} (2.75,0.25);
\draw[cut,cyan] (1.5,0) .. controls (1.5,1) .. node[opacity=1,pos=0.08,left] {$\tbde\!$} (2.75,0.25);
\draw[cut,teal] (1.75,0) .. controls (1.75,0.75) .. node[opacity=1,pos=0.05,right] {$\tbcd$} (2.75,0.75);
\draw[cut,brown] (1.5,1) .. controls (1.5,0) .. node[opacity=1,pos=0.08,left] {$\tacg\!$} (2.75,0.75);
\draw[cut,purple] (1.75,1) .. controls (1.75,0.25) .. node[opacity=1,pos=0.05,right] {$\taeg$} (2.75,0.25);
\draw[cut,pink] (2,0) node[opacity=1,below right] {$\!\tbcg$} -- (2.75,0.75);
\draw[cut,olive] (2.25,0.75) -- node[opacity=1,left] {$\tdg\!$} (2.25,0.25);

\begin{feynman}
\vertex (A) at (0.5,1);
\vertex (B) at (1,1);
\vertex (C) at (2.5,1);
\vertex (D) at (3,0.5);
\vertex (E) at (3.5,0.5);
\vertex (F) at (0.5,0);
\vertex (G) at (1,0);
\vertex (H) at (2,0.5);
\vertex (I) at (2.5,0);
\vertex (J) at (1.5,0.5);
\diagram*{
(A) -- [fermion, momentum=$p_1$] (B),
(H) -- [boson, momentum=$q_2$] (J),
(G) -- [fermion, reversed momentum=$p_2$] (F),
(D) -- [boson, momentum'=$q_1$] (E),
(B) -- [gluon] (C),
(I) -- [gluon] (G),
(B) -- [fermion, reversed momentum'=$k_1$] (G),
(D) -- [fermion, ultra thick, momentum'=$k_2$] (C),
(H) -- [fermion, ultra thick] (I),
(C) -- [fermion, ultra thick] (H),
(I) -- [fermion, ultra thick] (D)
};
\end{feynman}
\end{tikzpicture}
\end{minipage}%
\hfill
\begin{minipage}{0.3\textwidth}
\caption{Non-planar two-loop box diagram with fixed ordering of photon momenta $G_\text{NP}$.}
\end{minipage}
\hfill
\end{subfigure}
\caption{Possible Cutkosky cuts identifying the threshold singularities.}
\label{fig:2to2thresholds}
\end{figure}
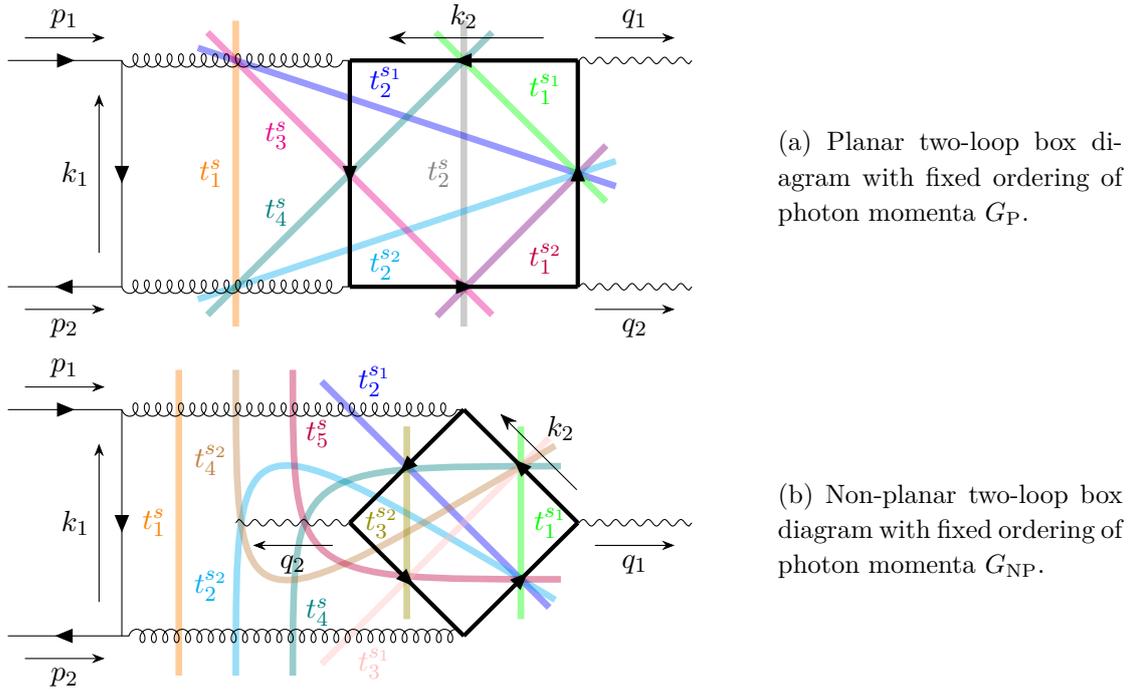

The possible thresholds appearing in the relevant contributions to the $2\to2$ process are given by the Cutkosky cuts in figure~\ref{fig:2to2thresholds}, from which their defining equations can be inferred.
With the chosen momentum routing we have for example:
\begin{align}
    \tadf(\vec{k}_1,\vec{k}_2):\
    \sqrt{\left|\vec{k}_1-\vec{p}_2\right|^2} + \sqrt{\left|\vec{k}_1+\vec{k}_2+\vec{p}_1\right|^2+m^2} +
    \sqrt{\left|\vec{k}_2\right|^2+m^2} - p_1^0 - p_2^0= 0\,.
\end{align}
Whether the singular surfaces have real solutions for the spatial loop momenta depends on the external momenta and internal masses (cf.~\cite{Capatti:2019edf}) and is summarised in table~\ref{tab:thresholds2}.
\par
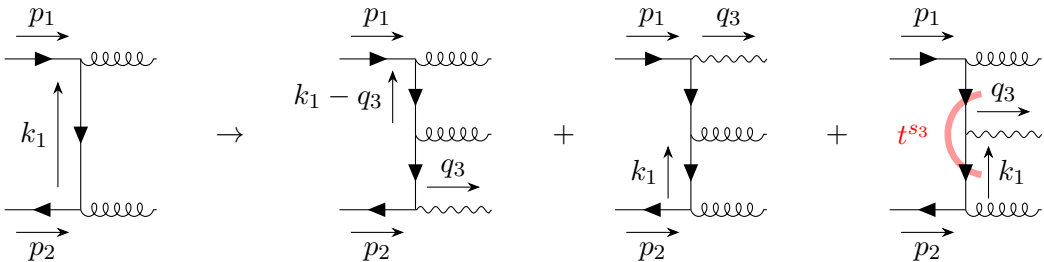
\begin{figure}[b]
\centering
\begin{minipage}{0.15\textwidth}
\begin{tikzpicture}[scale=2]
\begin{feynman}
\vertex (A) at (0.5,1);
\vertex (B) at (1,1);
\vertex (C) at (1.5,1);
\vertex (F) at (0.5,0);;
\vertex (G) at (1,0);
\vertex (H) at (1.5,0);
\diagram*{
(A) -- [fermion, momentum=$p_1$] (B),
(G) -- [fermion, reversed momentum=$p_2$] (F),
(B) -- [fermion, reversed momentum'=$k_1$] (G),
(B) -- [gluon] (C),
(G) -- [gluon] (H)
};
\end{feynman}
\end{tikzpicture}
\end{minipage}
$\quad\to\quad$
\begin{minipage}{0.2\textwidth}
\begin{tikzpicture}[scale=2]
\begin{feynman}
\vertex (A) at (0.5,1);
\vertex (B) at (1,1);
\vertex (C) at (1.5,1);
\vertex (D) at (1,0.5);
\vertex (E) at (1.5,0.5);
\vertex (F) at (0.5,0);;
\vertex (G) at (1,0);
\vertex (H) at (1.5,0);
\diagram*{
(A) -- [fermion, momentum=$p_1$] (B),
(G) -- [fermion, reversed momentum=$p_2$] (F),
(B) -- [fermion, reversed momentum'=$k_1-q_3$] (D),
(D) -- [fermion] (G),
(B) -- [gluon] (C),
(G) -- [boson, momentum=$q_3$] (H),
(D) -- [gluon] (E),
};
\end{feynman}
\end{tikzpicture}
\end{minipage}
$\quad+\quad$
\begin{minipage}{0.15\textwidth}
\begin{tikzpicture}[scale=2]
\begin{feynman}
\vertex (A) at (0.5,1);
\vertex (B) at (1,1);
\vertex (C) at (1.5,1);
\vertex (D) at (1,0.5);
\vertex (E) at (1.5,0.5);
\vertex (F) at (0.5,0);;
\vertex (G) at (1,0);
\vertex (H) at (1.5,0);
\diagram*{
(A) -- [fermion, momentum=$p_1$] (B),
(G) -- [fermion, reversed momentum=$p_2$] (F),
(B) -- [fermion] (D),
(D) -- [fermion, reversed momentum'=$k_1$] (G),
(B) -- [boson, momentum=$q_3$] (C),
(G) -- [gluon] (H),
(D) -- [gluon] (E),
};
\end{feynman}
\end{tikzpicture}
\end{minipage}
$\quad+\quad$
\begin{minipage}{0.15\textwidth}
\begin{tikzpicture}[scale=2]
\begin{feynman}
\tikzset{
cut2/.style={
    opacity=0.4,
    line width=2.5pt,
    shorten <=-6pt,
    shorten >=-6pt}
}
\draw[cut2,red] (1,0.75) to[out=-170, in=170] node[opacity=1,left] {$\thi$} (1,0.25);

\vertex (A) at (0.5,1);
\vertex (B) at (1,1);
\vertex (C) at (1.5,1);
\vertex (D) at (1,0.5);
\vertex (E) at (1.5,0.5);
\vertex (F) at (0.5,0);;
\vertex (G) at (1,0);
\vertex (H) at (1.5,0);
\diagram*{
(A) -- [fermion, momentum=$p_1$] (B),
(G) -- [fermion, reversed momentum=$p_2$] (F),
(B) -- [fermion] (D),
(D) -- [fermion, reversed momentum=$k_1$] (G),
(B) -- [gluon] (C),
(G) -- [gluon] (H),
(D) -- [boson, momentum=$q_3$] (E),
};
\end{feynman}
\end{tikzpicture}
\end{minipage}
\caption{Replacement that turns the planar and non-planar diagrams, $G_\text{P}$ and $G_\text{NP}$, from the $2\to2$ in fig.~\ref{fig:2to2thresholds} into those of the $2\to3$ process, with thresholds.}
\label{fig:2to3thresholds}
\end{figure}

The relevant topologies for the $2\to3$ are similar to the $2\to2$ process. They are obtained by attaching the third photon to the incoming spin line,
as shown in figure~\ref{fig:2to3thresholds}.
The thresholds remain unchanged unless an off-shell photon attaches inside the loop, in which case a new threshold singularity $\thi$ appears.

\subsection{Threshold subtraction}

\begin{table}[b]
    \centering
    \def\arraystretch{1.3}%
    \begin{tabular}{l|l|l|l|l|l|l|}
    \multirow{2}{*}{Part} & \multicolumn{4}{c|}{Thresholds $\thresholdset$} & \multicolumn{2}{c|}{Channels in rest frame of $q_1 + q_2$} \\\cline{2-7}
    & $s_{12}>0$ & $s_{12}>4m^2$ & $s_1>4m^2$ & $s_2>4m^2$ & Groups $\thresholdset_c$ & Source $\source_c=\left(\vec{s}_1,\vec{s}_2\right)$ \\\hline\hline
    $\widehat{G}^R_\text{P}$ & $\tba$ & $\tcf,\tadf,\tbcd$ & $\tce,\tade$ & $\tef,\tbde$ & $\thresholdset$ & $\vec{s}_1=\vec{p}_2, \vec{s}_2=\vec{0}$\\\hline
    \multirow{2}{*}{$\widehat{G}^R_\text{NP}$} & \multirow{2}{*}{$\tba$} & \multirow{2}{*}{$\tbcd, \taeg$} & \multirow{2}{*}{$\tce,\tade,\tbcg$} & \multirow{2}{*}{$\tbde,\tdg,\tacg$} & $\thresholdset\setminus \{\taeg\}$ & $\vec{s}_1=\vec{p}_2, \vec{s}_2=\vec{0}$ \\\cline{6-7}
    & & & & & $\thresholdset\setminus \{\tbcd\}$ & $\vec{s}_1=\vec{p}_2, \vec{s}_2=\vec{q}_2$ \\\hline
    \end{tabular}
    \caption{Thresholds and channels for the finite contributions corresponding to the planar and non-planar $2\to2$ diagrams in figure~\ref{fig:2to2thresholds}.}
    \label{tab:thresholds2}
\end{table}

Each diagram, together with the subtractions in section~\ref{sec:IR_counterterms}, produces an IR- and UV-finite integrand.
We exploit this by treating the threshold singularities of each such integrand separately, making their independent Monte Carlo integration possible.
\par
We follow the threshold subtraction method of 
refs.~\cite{Kermanschah:2021wbk,Kermanschah:2024utt} to construct an integrand in which the $\ii \epsilon$ prescription is explicitly removed, allowing for direct Monte Carlo integration in the physical region.
The dispersive part is obtained by introducing a counterterm for each Cutkosky cut that locally cancels the corresponding singularity.
By design, only the absorptive part of these counterterms, namely the residue at the pole, survives upon integration, reproducing the sum over all cuts.
With the parametrisations discussed below, any residual singularities in individual residues arising from additional on-shell cuts cancel in the sum.
For the processes considered here, the absorptive part does not contribute to the matrix element, but it must generally be included for parity-violating processes as shown in ref.~\cite{Kermanschah:2024utt}.
\par
The method relies on parametrising the threshold surface
\begin{equation}
   \threshold(\vec{k}_1,\vec{k}_2) = 0\,,
\end{equation}
in terms of the radial coordinate $r$ of six-dimensional spherical coordinates, 
$(\vec{k}_1,\vec{k}_2)= r \unitvector+\source$, 
shifted by a vector $\source=(\vec{s}_1,\vec{s}_2)$, referred to as the \emph{source}.
We require that the source lies inside the surface, $\threshold(\source)<0$, in order to guarantee a unique solution $r=r^*(\unitvector)$ for each direction $\unitvector$, satisfying $\threshold(r^*(\unitvector)\unitvector+\source)=0$.
For cuts involving only two on-shell energies, we obtain the root analytically (cf.~\cite{Kermanschah:2021wbk}), while in all other cases we compute it numerically using Brent's method~\cite{Brent:1973:AMD,vorot_roots_2022_crate}.
\par
In order for higher-order poles arising from intersections of multiple surfaces\footnote{
That is, triple cuts at one loop, quadruple cuts at two loops, and so on.
} to cancel automatically without introducing additional singularities, not only in the sum of residues but also in the sum of counterterms, the source $\source$ must lie inside all intersecting surfaces simultaneously.
In practice, however, it is not always possible to find a single source that lies inside all thresholds of a given loop integrand. 
In such cases, we partition the set of thresholds $\thresholdset$ into subsets $\thresholdset_c$, which can generally overlap, with $\bigcup_c \thresholdset_c=\thresholdset$, so that a suitable source $\source_c$ can be chosen for each subset. 
If the intersection of two threshold surfaces produces a double pole of the integrand, they must belong to the same subset. 
Double poles arise only when a common loop-momentum basis exists among the cut propagators involved in the thresholds. 
Surfaces that do not intersect, or whose intersections do not produce double poles, can be assigned to different subsets.
\par
Following ref.~\cite{Kermanschah:2024utt}, we define an integration channel $m_c \mathcal{I}$ for each subset $\thresholdset_c$. 
To ensure that only these thresholds contribute, we assign the weight
\begin{align}
\label{eq:multi_channel_factor}
    m_c = 
    \frac{
        \left(
            \prod_{\threshold \in \thresholdset \setminus \thresholdset_c} \frac{\threshold}{E_\text{CM}}
        \right)^2
    }{
        \sum_c \left(
            \prod_{\threshold \in \thresholdset \setminus \thresholdset_c} \frac{\threshold}{E_\text{CM}}
        \right)^2
    } \,,
\end{align}
normalized such that $\sum_c m_c = 1$, with $E_\text{CM}$ denoting the center-of-mass energy of the scattering. 
For each channel, the threshold counterterms are then computed separately with respect to the source $\source_c$.
\par
This multi-channelling approach disentangles conflicting threshold parametrisations and thereby extends the threshold subtraction method to generic configurations.
Importantly, it also ensures the local cancellation of any remaining poles within individual Cutkosky cuts in their sum.\footnote{
Ref.~\cite{Capatti:2025khs} applied this approach to achieve the local cancellation of threshold and final-state IR singularities in subgraphs of $2 \to 2$ forward-scattering graphs, enabling the direct numerical integration of inclusive cross sections.
}
\par
Below, we summarise our choice of channels and sources for the various contributions we have computed. 
We emphasise that the choice of sources is not unique, and that the subsets can, in some cases, be further split or merged depending on the Lorentz reference frame or the chosen momentum routing. 
However, the existence of threshold singularities and higher-order poles in the integrand is a Lorentz invariant property.

\subsubsection*{Diphoton}
In table~\ref{tab:thresholds2}, we list the thresholds and their most general grouping along with the corresponding sources valid in the centre-of-mass frame of $p_1+p_2=q_1+q_2$, for the topologies shown in figure~\ref{fig:2to2thresholds} with a planar and non-planar quark loop.
\par
For the planar topology, threshold subtraction was previously demonstrated in ref.~\cite{Vicini:2024ecf}, requiring only a single channel with source $\source=(\vec{p}_2,\vec{0})$ in all cases, whether for on- or off-shell photons and light- or heavy-quark loops. 
\par
For the non-planar topology, two channels are generally required if $\taeg$ and $\tbcd$ exist: one including all thresholds except $\taeg$, and one including all except $\tbcd$, with sources $(\vec{p}_2,\vec{0})$ and $(\vec{p}_2, \vec{q}_2)$, respectively.
These two channels can be merged if the surfaces $\taeg$ and $\tbcd$ either do not exist, or if they overlap, which occurs when $m$ is zero or sufficiently small. 
In this case, a single channel including all thresholds can be used, with source at $(\vec{p}_2,\vec{q}_2/2)$.

\subsubsection*{Triphoton}
For the $2\to3$ diagrams, the treatment of thresholds builds on the $2\to2$ case discussed above.
If $s_3=0$, no additional thresholds arise in the $2\to2$ subtopologies compared to the $2\to2$ scattering kinematics. 
In the rest frame of the two photons attached to the fermion loop, namely of $q_1+q_2$, we can therefore use the same grouping and parametrisations as in the $2\to2$ case.
Although the threshold singularities can be regularised in any frame, we observed that the integration performs significantly better in the rest frame of $q_1+q_2$ than in the centre-of-mass frame of the incoming quark-antiquark pair.\footnote{
This improvement is likely due to the simplified local singularity structure, with overlapping (integrable) collinear singularities and coinciding threshold singularities, which may enable the \vegas grid to adapt more efficiently.
}
\par
When $s_3>0$, an additional threshold $\thi$ appears, for which new channels have to be introduced.
For the planar topology, we keep the original channel from the $2\to2$ case but remove $\thi$ from the set of all thresholds.
We then introduce one additional channel for $\thi$ and, if present, the thresholds $\tce$, $\tcf$ and $\tef$, with source $(\vec{q}_3, \vec{0})$.
For the non-planar topology, we also keep existing channels from the $2\to2$ case but exclude $\thi$, and add two additional channels: $\{\thi, \tce\}$ with source $(\vec{q}_3, \vec{0})$, and $\{\thi, \tdg\}$ with source $(\vec{q}_3, -\vec{p}_1)$.
If either $\tce$ or $\tdg$, or both, do not exist, these channels can again be merged.

\newcommand{\hypercube}{\mathbf{x}}

\section{Multi-channel Monte Carlo}
\label{sec:multi-channelling}
After subtracting IR, UV, and threshold counterterms, the integrand still exhibits integrable singularities, such as those discussed in sections~\ref{sec:soft_channels} and~\ref{sec:coll_channels} below.
Although these singularities do not make the integral divergent, they increase the variance and slow down the convergence in Monte Carlo integration.
The \vegas algorithm~\cite{Lepage:1977sw,Lepage:2020tgj,Hahn:2004fe} can adaptively reduce the variance when the integration variables align with the enhancements, but it struggles with non-factorisable structures~\cite{Ohl:1998jn,Jadach:1999sf}, which must first be disentangled.
\par
Integrable singularities can be locally removed via a suitable change of variables, but no single global transformation can flatten all singularities in our integrands.
To address this, we employ multi-channel Monte Carlo integration, where each channel $c$ corresponds to a specific change of variables $\phi^c$ that flattens a subset of singularities.
By combining multiple channels, we can reduce the variance across the entire integration domain.
\par
This approach is commonly used in phase-space integration \cite{Berends:1984gf,Hilgart:1992xu,Roth:1999kk,Ilyin:1996gy,Kleiss:1985gy,Kleiss:1991rn,Kleiss:1994qy,Papadopoulos:2000tt,Denner:2002cg,Ohl:1998jn,Jadach:1991ty} and has also been applied in loop momentum space~\cite{Soper:1999xk,Becker:2012aqa}.
For completeness, we summarise the approach here.
Given a set of parametrisations $\phi^c$ from the unit hypercube to the integration domain, the integral can be expressed as a weighted sum over channels
\begin{align}
    I &= \int \mathrm{d}\loopmomenta \, \mathcal{I}(\loopmomenta)
      = \int \mathrm{d}\hypercube \sum_c \frac{\mathcal{I}(\phi^c(\hypercube))}{\rho(\phi^c(\hypercube))}\,,
\end{align}
where the total sampling density is given in terms of the Jacobian determinants of the maps $\phi^c$, as
\begin{align}
    \rho(\loopmomenta) = \sum_c \rho^c(\loopmomenta)\,,
    \qquad
    \frac{1}{\rho^c(\phi^c(\hypercube))} = \left| \det \frac{\partial \phi^c(\hypercube)}{\partial \hypercube} \right|\,.
\end{align}
The Monte Carlo estimate and error using $N$ samples is
\begin{align}
    I \approx I_N = \frac{1}{N}\sum_i f_i\,,
    \qquad
    \Delta I_N = \sqrt{\frac{1}{N-1}\left[ \left(\frac{1}{N}\sum_i f_i^2\right) - I_N^2 \right]}\,,
\end{align}
with
\begin{align}
    f_i = \frac{\mathcal{I}(\phi^{c_i}(\hypercube_i))}{\rho(\phi^{c_i}(\hypercube_i))}.
\end{align}
Here, $\hypercube_i$ are uniformly sampled points in the hypercube, and $c_i$ are the sampled channel indices.
The closer the sampling density $\rho$ approximates the integrand $\mathcal{I}$, the better the convergence of the Monte Carlo integration.
\par
Channel indices can either be chosen as integers directly or derived from a continuous variable by introducing an extra integration dimension to the \vegas hypercube.
Since the channels are independent, it is advantageous to use a separate adaptive grid for each, as implemented in the library \havana~\cite{havana,Capatti:2020xjc}, to further improve convergence.
\par
To compute the dispersive part of an $n$-loop integral using the multi-channel approach, we introduce $n$ maps, where each parametrises an independent propagator momentum $\vec{q}_i$.
The list $\vec{\mathbf{q}}=(\vec{q}_1,\dots,\vec{q}_n)$ defines a loop momentum basis and relates to the original loop momenta via the basis transformation (cf.~\cite{Capatti:2019ypt}):
\begin{align}
\label{eq:basis_trsf}
    \loopmomenta = \mathbf{S}\cdot\vec{\mathbf{q}}-\vec{\mathbf{b}}\,,
\end{align}
where $\mathbf{S}$ is an invertible matrix and $\vec{\mathbf{b}}$ is a shift given in terms of external momenta.
\par
Each momentum $\vec{q}_i$ is parametrised by a map $\phi_i$ with weight $\rho_i$.
The full parametrisation for a single channel then reads
\begin{align}
    \vec{\mathbf{q}} = \phi^c(\hypercube) = (\phi_1(\vec{x}_1),\dots,\phi_n(\vec{x}_n))\,,
\end{align}
with weight
\begin{align}
\label{eq:channel_jac}
    \rho^c(\loopmomenta) = \prod_{i=1}^n \rho_i(\vec{q}_i)\,.
\end{align}
\par
Below, we discuss two such maps $\phi_i$, each designed for a specific class of integrable singularities.
Note that, although they tackle two main sources of large integrand variance, some singularities may remain.\footnote{
For example, threshold counterterms can project existing collinear singularities along the radial direction, spreading the integrable singularity.
}

\subsection{Soft channels}
\label{sec:soft_channels}
The integrand contains denominators of the form
\begin{align}
    \mathcal{I} \sim \frac{1}{\sqrt{|\vec{q}_i|^2 + m_i^2}}\,,
\end{align}
which produce an integrable singularity at $\vec{q}_i = 0$ if $m_i = 0$.
This singularity can be removed by parametrising $\vec{q}_i$ in spherical coordinates,
\begin{align}
    \vec{q}_i = r \, \big(\sin\theta\cos\varphi, \, \sin\theta\sin\varphi, \, \cos\theta\big)\,,
\end{align}
with the radius and angles mapped from hypercube variables $\vec{x}_i = (x_1, x_2, x_3)$ as
\begin{align}
\label{eq:rad_map}
    r = E_\text{CM} \frac{x_1}{1 - x_1}\,, \qquad
    \varphi = 2 \pi x_2\,, \qquad
    \cos\theta = 2 x_3 - 1\,.
\end{align}
This defines the map $\vec{q}_i = \phi_i(\vec{x}_i)$, whose Jacobian determinant reads
\begin{align}
    \frac{1}{\rho_i(\vec{q}_i)}
    = \frac{4\pi}{E_\text{CM}} |\vec{q}_i|^2 \left(E_\text{CM} + |\vec{q}_i|\right)^2.
\end{align}
The Jacobian not only cancels the integrable singularity at $\vec{q}_i = 0$ but also flattens the integrand in the UV, if its superficial degree of divergence is $-1$.\footnote{
The example integrals in table~\ref{tab:soft_mc} drop off faster in the UV, which could in principle be accounted for by adjusting the radial map, but we did not attempt this.
}
Alternative radial maps that performed similarly in our tests include $r = \tan(\frac{\pi}{2}x)$ (cf.~\cite{Becker:2012aqa}).
\par
For all loop momentum bases that can be formed from the propagator momenta $\vec{q}_i$, we define a corresponding sampling channel for each, following eqs.~\eqref{eq:basis_trsf} to \eqref{eq:channel_jac}.
\par
A related multi-channel method addressing the same integrable singularities is described in ref.~\cite{Capatti:2019edf}.  
As in the approach we present here, channels are split according to spanning trees (or loop momentum bases).  
However, that method flattens singularities by introducing a tailored unity factor and summing over all channels for each sample.  
In contrast, we use the actual hypercube Jacobian and select the integration channel probabilistically, keeping the evaluation time independent of the number of channels.  
This improves Monte Carlo convergence, especially when separate sampling grids are employed.
A related approach, based on tropical sampling of loop momentum configurations, was introduced in ref.~\cite{Borinsky:2025asc}.

\par
\begin{table}[t]
\centering
\resizebox{\textwidth}{!}{%
\begin{tabular}{|lll|ll|ll|}
\hline
\multirow{2}{*}{Diag.} & \multirow{2}{*}{Exp.} & \multirow{2}{*}{Reference} & \multicolumn{2}{c|}{Adaptive Monte Carlo} &  \multicolumn{2}{c|}{Naive Monte Carlo} \\\cline{4-7}
& & & Multi-channel  & Single-channel & Multi-channel & Single-channel \\
\hline\hline
\ttlf\texttt{1L3P} & \ttlf$\texttt{10}^{\texttt{-5}}$ & \ttlf\texttt{\phantom{+}9.76546} & \ttlf\texttt{\phantom{+}9.76546(12)} &  \ttlf\texttt{\phantom{+}9.76527(16)} & \ttlf\texttt{\phantom{+}9.76546(28)} &  \ttlf\texttt{\phantom{+}9.76546(98)} \\
\ttlf\texttt{2L6P} & \ttlf$\texttt{10}^{\texttt{-4}}$ & \ttlf\texttt{\phantom{+}1.1339(5)} & \ttlf\texttt{\phantom{+}1.133755(48)} & \ttlf\texttt{\phantom{+}1.13329(26)} & \ttlf\texttt{\phantom{+}1.13370(33)} & \ttlf\texttt{\phantom{+}1.089(18)} \\
\ttlf\texttt{3L2P} & \ttlf$\texttt{10}^{\texttt{-6}}$ & \ttlf\texttt{\phantom{+}5.26647} & \ttlf\texttt{\phantom{+}5.26628(22)} & \ttlf\texttt{\phantom{+}5.2106(69)} & \ttlf\texttt{\phantom{+}5.26652(44)} & \ttlf\texttt{\phantom{+}5.297(90)} \\
\ttlf\texttt{4L2P} & \ttlf$\texttt{10}^{\texttt{-8}}$ & \ttlf\texttt{\phantom{+}8.36515} & \ttlf\texttt{\phantom{+}8.3681(10)} & \ttlf\texttt{\phantom{+}7.382(50)} & \ttlf\texttt{\phantom{+}8.3648(22)} & \ttlf\texttt{\phantom{+}7.89(13)} \\
\ttlf\texttt{4L4P} & \ttlf$\texttt{10}^{\texttt{-14}}$ & \ttlf\texttt{\phantom{+}2.6919} & \ttlf\texttt{\phantom{+}2.69220(20)} & \ttlf\texttt{\phantom{+}2.6770(24)} & \ttlf\texttt{\phantom{+}2.69202(77)} & \ttlf\texttt{\phantom{+}2.684(15)} \\\hline
\ttlf\texttt{2L4P} & \ttlf$\texttt{10}^{\texttt{-6}}$ & \ttlf\texttt{-1.0841} & \ttlf\texttt{-1.0812(19)} &  \ttlf\texttt{-1.0819(18)} & \ttlf\texttt{-1.080(34)} &  \ttlf\texttt{-1.046(98)} \\
\ttlf\texttt{3L4P} & \ttlf$\texttt{10}^{\texttt{-9}}$ & \ttlf\texttt{-3.4003} & \ttlf\texttt{-3.330(97)} & 
\ttlf\texttt{-3.428(74)} &  \ttlf\texttt{-3.15(57)} & \ttlf\texttt{\phantom{+}8.4(9.6)} \\
\hline
\end{tabular}%
}
\caption{
Comparison of integration strategies for scalar Feynman diagrams at fixed phase-space points. 
The first five configurations correspond to Euclidean kinematics and are taken from ref.~\cite{Borinsky:2025asc}, where a similar study with a specialised sampling algorithm was presented.
The last two correspond to the dispersive parts of the double- and triple-box diagrams at physical phase-space points from ref.~\cite{Capatti:2019edf}, evaluated with threshold subtraction.
The precise phase space points and the $n$-loop $m$-point diagrams $n$L$m$P are defined in the cited references. 
All results are based on $10^9$ Monte Carlo samples. 
The adaptive strategy updates the grid iteratively with the \vegas algorithm, using {\ttlf\texttt{nstart}}=$10^7$ and {\ttlf\texttt{nincrease}}=$10^6$, while the naive approach does not use grid adaptation. 
The adaptive multi-channel method assigns a separate grid to each channel, as implemented in \havana.
The multi-channel variant combines all parametrisations described in section~\ref{sec:soft_channels} according to section~\ref{sec:multi-channelling}, whereas the single-channel variant uses just one such parametrisation globally. 
}
\label{tab:soft_mc}
\end{table}
In table~\ref{tab:soft_mc}, we compare single-channel with multi-channel Monte Carlo integration using the soft parametrisations described above, both with and without grid adaptation.
We test Euclidean scalar loop integrals from ref.~\cite{Borinsky:2025asc}, for which we observe improvements of the same order of magnitude, and two integrals with physical kinematics from ref.~\cite{Vicini:2024ecf}.
In naive Monte Carlo, multi-channel integration reduces the error by up to two orders of magnitude.
With grid adaptation, the gain is smaller, suggesting that the grid adapts effectively.
However, as seen in the single-channel naive Monte Carlo of the 3L4P integral, large weights from remaining integrable singularities can spoil the integration, and in the 4L2P integral, the error can be significantly underestimated, leaving a bias even with adaptive integration.
Multi-channel integration is therefore generally preferred and necessary, as it provides more reliable estimates from the first iteration, reduces the risk of underestimating the error, and allows for more accurate grid adaptation.

\subsection{Collinear channels}\label{sec:coll_channels}
If the integrand contains two massless adjacent propagators with momenta $\vec{q}_i$ and $\vec{q}_i - \vec{p}$, with $p^2 = 0$, it behaves as
\begin{align}
    \mathcal{I} \sim \frac{1}{|\vec{q}_i|} \frac{1}{|\vec{q}_i - \vec{p}|} \frac{1}{(|\vec{q}_i| + |\vec{q}_i - \vec{p}| - |\vec{p}|)^\alpha}\,,
\end{align}
close to the collinear configuration $\vec{q}_i = x \vec{p}$ with $0 \le x \le 1$ that lies on the singular surface $|\vec{q}_i| + |\vec{q}_i - \vec{p}| - |\vec{p}| = 0$.
It is integrable if $\alpha \leq 1$.  
\par
Integrable singularities of this type can be removed by parametrising $\vec{q}_i$ in elliptic coordinates,
\begin{align}
    \vec{q}_i = R\cdot\frac{|\vec{p}|}{2} \big(\sinh\mu \sin\nu \cos\varphi, \, \sinh\mu \sin\nu \sin\varphi, \, \cosh\mu \cos\nu + 1\big)\,,
\end{align}
where $R$ rotates $\hat e_z$ into $\hat p$, with hypercube variables $\vec{x} = (x_1, x_2, x_3)$ mapped as
\begin{align}
    \mu = \artanh x_1\,, \qquad
    \varphi = 2 \pi x_2\,, \qquad
    \cos\nu = 2 x_3 - 1\,.
\end{align}
This defines the map $\vec{q}_i = \phi_{i,\vec{p}}(\vec{x})$, whose Jacobian determinant is
\begin{align}
    \frac{1}{\rho_{i,\vec{p}}(\vec{q}_i)}
    = \frac{2\pi}{|\vec{p}|^2} |\vec{q}_i| \, |\vec{q}_i - \vec{p}| \, (|\vec{q}_i| + |\vec{q}_i - \vec{p}|)^2 
    \sqrt{(|\vec{q}_i| + |\vec{q}_i - \vec{p}|)^2 - |\vec{p}|^2}\,.
\end{align}
This Jacobian cancels both the soft endpoints and the collinear singularity exactly if $\alpha=\frac{1}{2}$.
\par
In an $n$-loop integrand, there may exist multiple singular (pinch) surfaces of the form
\begin{align}
    \Big|\sum_{i=1}^r \vec{q}_i - \vec{p}\,\Big| - \Big|\vec{p}\,\Big| + \sum_{i=1}^r \Big|\vec{q}_i\Big| = 0,
\end{align}
where the $\vec{q}_i$ are $r$ independent loop momenta with $r \leq n$. 
This condition is satisfied when all $\vec{q}_i$ are collinear to $\vec{p}$, i.e.~$\vec{q}_i = x_i \vec{p}$ with $x_i > 0$ and $\sum_i x_i \le 1$.  
We identify all these pinch surfaces, each specifying $r$ independent propagator momenta and a collinear momentum $\vec{p}$.  
We complete these $r$ momenta to a full loop momentum basis by adding $n-r$ independent propagator momenta.  
Each pinch surface, together with its basis completion, then defines a sampling channel according to eqs.~\eqref{eq:basis_trsf} to \eqref{eq:channel_jac}, where the collinear map $\phi_{i,\vec{p}}$ is applied to the $r$ momenta and the soft map from the previous section to the remaining $n-r$ momenta.
\par
\begin{table}[t]
\centering
\resizebox{\textwidth}{!}{%
\begin{tabular}{|lll|ll|l|ll|l|}
\hline
\multirow{3}{*}{Diag.} & \multirow{3}{*}{Exp.} & \multirow{3}{*}{Reference} & \multicolumn{3}{c|}{Adaptive Monte Carlo} &  \multicolumn{3}{c|}{Naive Monte Carlo} \\\cline{4-9}
& & & \multicolumn{2}{c|}{Multi-channel}  & \multirow{2}{*}{Single-channel} & \multicolumn{2}{c|}{Multi-channel} & \multirow{2}{*}{Single-channel} \\\cline{4-5}\cline{7-8}
& & & Coll. + soft & Soft only & & Coll. + soft & Soft only & \\
\hline\hline
\ttlf$\overline{\texttt{1L4P}}$ & \ttlf$\texttt{10}^{\texttt{-2}}$ & \ttlf\texttt{-2.29293} & \ttlf\texttt{-2.2930(10)} & \ttlf\texttt{-2.2937(10)} &  \ttlf\texttt{-2.2943(14)} & 
\ttlf\texttt{-2.2962(21)} & \ttlf\texttt{-2.2942(27)} &  \ttlf\texttt{-2.240(41)} \\
\ttlf$\overline{\texttt{2L4P}}$ & \ttlf$\texttt{10}^{\texttt{-6}}$ & \ttlf\texttt{-5.48045} & \ttlf\texttt{-5.480(16)} & \ttlf\texttt{-5.490(29)} & \ttlf\texttt{-5.489(21)} & \ttlf\texttt{-5.26(20)} & \ttlf\texttt{-4.68(62)} & \ttlf\texttt{-5.7(2.0)} \\
\hline
\end{tabular}%
}
\caption{
Comparison of integration strategies for the dispersive part of two scalar integrals with integrable collinear singularities: the IR-subtracted box diagram of ref.~\cite{Anastasiou:2018rib}, and the double-box diagram with massless propagators
$k_1^2,(k_1-p_2)^2,(k_1-p_1-p_2)^2,(k_1-k_2)^2,k_2^2,(k_2-q_1)^2,(k_2-p_1-p_2)^2$
and a numerator
$k_1^x k_1^y k_2^x k_2^y$ that ensures IR finiteness, which we also computed analytically for the benchmark reference.
The Monte Carlo variants are as in table~\ref{tab:soft_mc}, with multi-channelling results shown for both collinear+soft channels and for soft channels only.
Both are evaluated with threshold subtraction at the physical phase space point
$p_1=(\tfrac{1}{2},0,0,\tfrac{1}{2}),\; p_2=(\tfrac{1}{2},0,0,-\tfrac{1}{2}),\; q_1=(\tfrac{1}{2},\tfrac{\sqrt{2}}{3},0,-\tfrac{1}{6}),\; q_2=p_1+p_2-q_1$.
}
\label{tab:collinear_mc}
\end{table}
In table~\ref{tab:collinear_mc}, we show the effect of the collinear (and soft) channels on the Monte Carlo integration for two finite scalar integrals with physical kinematics, both with and without grid adaptation.
The example integrands contain IR singularities regularised by local counterterms (in $\overline{\mathrm{1L4P}}$) or numerator suppression (in $\overline{\mathrm{2L4P}}$), much like our fermion-loop integrands.
The differences between single-channel and multi-channel integration are consistent with the observations discussed for table~\ref{tab:soft_mc} in the previous section.
Adding the collinear channels on top of the soft channels results in similar performance for $\overline{\mathrm{1L4P}}$, while for $\overline{\mathrm{2L4P}}$, it reduces the error by roughly factors of 2 and 3 for adaptive and naive Monte Carlo, respectively.
We also note a bias in the naive multi-channel integration with only soft channels for $\overline{\mathrm{2L4P}}$, likely caused by a large weight near a collinear singularity.

\section{Implementation}\label{sec:implementation}
Our pipeline extends the implementation developed in ref.~\cite{Kermanschah:2024utt} and automates the generation of the integrand, including threshold counterterms in numerically stable representations, using \codeLanguage{Python} and \codeTool{FORM}~\cite{Vermaseren:2000nd,Ruijl:2017dtg,Ueda:2020wqk}.  
The resulting expressions are compiled into a \codeLanguage{Rust} library, which is called from the integrator, also written in \codeLanguage{Rust}.  
The integrator performs helicity sums over selected left and right amplitudes, as well as dispersive and absorptive integrals over loop momentum and phase space.  
\par
The required input is a finite integrand expressed in terms of propagators with numerators, which can be Feynman diagrams supplemented with appropriate local IR and UV counterterms.  
We generated the diagrams using \codeTool{Qgraf}~\cite{Nogueira:1991ex}, while the IR and UV counterterms are produced with \codeTool{FORM} and \codeLanguage{Python}, linking to \codeTool{Symbolica}~\cite{symbolica-16} for some symbolic manipulations.  
For threshold subtraction, the relevant channels and their sources must be provided.  
\par
The Monte Carlo integration is parallelised across cores, allowing the program to run on a personal computer or a single cluster node, with multi-node parallelisation planned for the future.  
The source code is currently under active development and is not yet publicly available.
\par
In the following paragraphs, we discuss the main additions compared to the implementation of ref.~\cite{Kermanschah:2024utt}: new integrand representations, optimisation and compilation of large expressions, and multi-channelling importance sampling.
\par
For each set of propagators with a numerator, we generate the LTD representation of ref.~\cite{Capatti:2019ypt}, as well as a locally equivalent but more numerically stable representation.  
The latter is constructed algebraically using partial fractioning without reference to an underlying graph, and can therefore differ from the graphical cross-free family~(CFF) representation algorithm~\cite{Capatti:2022mly,Capatti:2023shz}. 
Although our algorithm can in principle generate spurious denominator factors similar to those in TOPT, we verified that no such terms arose for the diagrams considered in this work.
The LTD representation is used only for generating factorised threshold counterterms, which at two loops correspond to thresholds that cut both loops simultaneously.  
Additionally, for threshold configurations involving a cut in only one of the two loops, we construct a \textit{mixed} representation.  
We apply the LTD algorithm solely to the loop momenta involved in the threshold, while the remaining uncut loop momenta are treated using our alternative CFF algorithm.  
This is achieved by a change of basis in loop momentum space to align with the cut, which is then inverted in the final expression.  
In this way, the residue at the threshold remains compact, while the remaining loops are represented in the numerically more stable form.\footnote{This procedure is automated and implemented for arbitrary threshold residues with remaining loops at any loop order.}
Although threshold counterterms derived from the full CFF representation offer comparable numerical stability, they would take longer to evaluate.
\par
An important objective was the reduction of integrand evaluation time.
Due to the presence of contracted spin chains with traces, the numerators are much longer than in the $\Nf$-part.
To accelerate the trace evaluation, we explicitly expanded the numerators of each diagram into components using the library \codeTool{spenso}~\cite{symbolica-community,spenso} and optimised the resulting expressions.
Combined with optimisation of the denominators, performed with either \codeTool{Symbolica} or \codeTool{FORM}~\cite{Kuipers:2013pba,Ruijl:2013epa}, this resulted in a notable speed-up.
An even larger improvement was achieved by symbolically substituting the loop energies in the numerator by the on-shell conditions imposed by the energy flows, followed by a simultaneous optimisation of the multiple expressions with \codeTool{Symbolica}.
However, this simultaneous optimisation drastically increased the integrand generation time.
For the $\gamma^*\gamma^*\gamma^*$ fermion-loop contributions it did not go through in our current implementation\footnote{The IR counterterms for $q\bar q \to \gamma^*\gamma^*\gamma^*$ have up to 10 propagators, whose expression is significantly larger than for the simpler cases after integration of the loop energies with CFF.\label{foot:opt_2to3massive}
}, and we resorted back to the optimised numerators with generic loop-energy components (see evaluation times in table~\ref{tab:timings}).
\par
If the expressions are too large, optimisation in \codeLanguage{Rust} can become a bottleneck.
We found that compiling the optimised expressions into a static \codeLanguage{C} library is often orders of magnitude faster, while achieving comparable evaluation times.
The \codeLanguage{Rust} integrand then calls this static library.
\par
The multi-channel importance sampling described in section~\ref{sec:multi-channelling} required refactoring our \codeLanguage{Rust} integrator.
The parametrisations for the importance sampling are produced automatically from the Feynman diagrams at integrand generation time, contrary to the code used for ref.~\cite{Kermanschah:2024utt}, where the hypercube variables were mapped to the loop momenta globally at the start of the evaluation.

\section{Numerical results}
\label{sec:results}

We will show numerical results for the fermion-loop mediated contributions for four different processes, both with light and heavy quark flavours propagating in the closed fermion loop computed with the presented method.
In the next section~\ref{sec:matrix_elements}, we will show matrix elements evaluated at fixed phase space points.
In section~\ref{sec:cross-sect-level}, we will show virtual corrections after phase-space integration and convolution with PDFs for on- and off-shell diphoton and for triphoton production via light-quark loops.
In section~\ref{sec:timings}, we will discuss the integrand evaluation times for the contributions considered.
All results reported here have been obtained using double precision arithmetic. 

\subsection{Squared matrix elements}
\label{sec:matrix_elements}
In table~\ref{tab:me}, we show the two-loop contributions to the squared matrix elements for four processes with light quark-loops and for three processes with heavy quark-loops.
In these results the colour factors are extracted, the quark charges, QED couplings and $N_f$ are set to one.
For each process, we show evaluations at three phase-space points, given in appendix~\ref{sec:ps_points}.
All phase-space points were generated with \codeTool{RAMBO}~\cite{Kleiss:1985gy}, except those used for the $2\to2$ matrix elements via heavy-quark loops, which were manually generated from the invariants used in refs.~\cite{Becchetti:2025rrz,Ahmed:2025osb}.
The off-shell photon masses were taken to be either the $Z$ boson mass, $m_{\gamma^*} = m_Z$, or $m_{\gamma_2^*} = 50\,\text{GeV}$.
\par
In appendix~\ref{sec:individual_me}, we report the individual planar and non-planar contributions, together with the integrated UV counterterms, each computed with separate Monte~Carlo integrations and combined according to eqs.~\eqref{eq:floops-pieces} and \eqref{eq:matrix-element}.
We show these contributions separately to make the cancellations among them visible.
Such cancellations can be expected when computing gauge-dependent finite pieces of amplitudes separately, as also pointed out in related numerical computations of refs.~\cite{AH:2023kor,Capatti:2025khs}, and their severity varies significantly for different kinematic configurations.
In the matrix-element evaluations, we observed global cancellations between planar and non-planar contributions of up to two digits, as discussed in more detail in appendix~\ref{sec:individual_me}.
This can possibly be mitigated, when locally combining all diagrammatic contributions with suitable momentum routings, as local cancellations have been demonstrated in the collinear limit in ref.~\cite{Anastasiou:2024xvk}.
\par
Although the matrix elements do not depend on a UV regularisation scale at integrated level, they locally  depend on a UV mass $M$, which we set to $M=\sqrt{s}=E_\text{CM}$.
For all results, we targeted a relative precision $\Delta\,[\%]$ below one percent and found agreement with the reference results, where available, within two standard deviations $\Delta\,[\sigma]$ of the Monte~Carlo error.
We used the \vegas settings {\ttlf\texttt{n\_start}}=$10^7$ and {\ttlf\texttt{n\_increase}}=$10^6$, and imposed a small cutoff of $10^{-7}$ on the radial hypercube variables to avoid numerical instabilities near soft points.
This cutoff introduces a negligible error, as the integrand remains finite in this region thanks to our multi-channel importance sampling.
The number of Monte Carlo samples ranges from $10^8$ to $10^{10}$ for the computation of the IR- and UV-finite planar and non-planar contributions, with more details given in section~\ref{sec:individual_me}.
\par
The benchmark results for the light-fermion-loop contributions were obtained using the expressions of ref.~\cite{Anastasiou:2002zn} for $q\bar{q} \to \gamma\gamma$ and ref.~\cite{Gehrmann:2015ora} for $q\bar{q} \to \gamma^* \gamma^*$\footnote{We used ref.~\cite{Frellesvig:2016ske} for the numerical evaluation of $\mathrm{Li}_{2,2}(x,y)$.}, and the \codeTool{FivePointAmplitudes} library~\cite{Chicherin:2020oor,Abreu:2020cwb, Abreu:2023bdp} for $q\bar{q} \to \gamma\gamma\gamma$.
The heavy-quark-loop contribution to $q\bar{q} \to \gamma\gamma$ was computed in refs.~\cite{Becchetti:2023wev,Becchetti:2023yat,Becchetti:2025rrz,Ahmed:2025osb}, and we compared to ref.~\cite{Becchetti:2025rrz}.
\begin{table}[t]
\centering
\begin{tabular}{llrclrr}
\hline\hline
Process & PSP & Exp. & Reference & \ \ \ \ \  \ \ \ \ Result & $\mathtt{\Delta \ [\sigma]}$ & $\Delta \ [\%]$ \\
\hline\hline\multirow{3}{*}{\makecell[l]{$q\bar{q}\to\gamma\gamma$\\$m=0$}}
& \multirow{1}{*}{1}
& \ttlf$\texttt{10}^{\texttt{+2}}$
& {\ttlf\texttt{-7.2127}}
& {\ttlf\texttt{-7.2574~$\pmtt$~0.0458}}
& {\ttlf\texttt{0.976}}
& {\ttlf\texttt{0.631}}
\\
[\doublerulesep]\cline{2-2}\cline{3-3}\cline{4-4}\cline{5-5}\cline{6-6}\cline{7-7}& \multirow{1}{*}{2}
& \ttlf$\texttt{10}^{\texttt{+2}}$
& {\ttlf\texttt{-6.7813}}
& {\ttlf\texttt{-6.7649~$\pmtt$~0.0421}}
& {\ttlf\texttt{0.391}}
& {\ttlf\texttt{0.622}}
\\
[\doublerulesep]\cline{2-2}\cline{3-3}\cline{4-4}\cline{5-5}\cline{6-6}\cline{7-7}& \multirow{1}{*}{3}
& \ttlf$\texttt{10}^{\texttt{+2}}$
& {\ttlf\texttt{-8.4533}}
& {\ttlf\texttt{-8.5154~$\pmtt$~0.0558}}
& {\ttlf\texttt{{1.113}}}
& {\ttlf\texttt{0.655}}
\\
[\doublerulesep]\ccline{1-1}\ccline{2-2}\ccline{3-3}\ccline{4-4}\ccline{5-5}\ccline{6-6}\ccline{7-7}\multirow{3}{*}{\makecell[l]{$q\bar{q}\to\gamma^*\gamma^*$\\$m=0$}}
& \multirow{1}{*}{1}
& \ttlf$\texttt{10}^{\texttt{+2}}$
& {\ttlf\texttt{-4.6281}}
& {\ttlf\texttt{-4.5791~$\pmtt$~0.0347}}
& {\ttlf\texttt{{1.411}}}
& {\ttlf\texttt{0.758}}
\\
[\doublerulesep]\cline{2-2}\cline{3-3}\cline{4-4}\cline{5-5}\cline{6-6}\cline{7-7}& \multirow{1}{*}{2}
& \ttlf$\texttt{10}^{\texttt{+2}}$
& {\ttlf\texttt{-4.5407}}
& {\ttlf\texttt{-4.5137~$\pmtt$~0.0352}}
& {\ttlf\texttt{0.768}}
& {\ttlf\texttt{0.780}}
\\
[\doublerulesep]\cline{2-2}\cline{3-3}\cline{4-4}\cline{5-5}\cline{6-6}\cline{7-7}& \multirow{1}{*}{3}
& \ttlf$\texttt{10}^{\texttt{+2}}$
& {\ttlf\texttt{-4.3357}}
& {\ttlf\texttt{-4.2968~$\pmtt$~0.0339}}
& {\ttlf\texttt{{1.149}}}
& {\ttlf\texttt{0.788}}
\\
[\doublerulesep]\ccline{1-1}\ccline{2-2}\ccline{3-3}\ccline{4-4}\ccline{5-5}\ccline{6-6}\ccline{7-7}\multirow{3}{*}{\makecell[l]{$q\bar{q}\to\gamma^*\gamma_2^*$\\$m=0$}}
& \multirow{1}{*}{1}
& \ttlf$\texttt{10}^{\texttt{+2}}$
& {\ttlf\texttt{-5.2774}}
& {\ttlf\texttt{-5.2742~$\pmtt$~0.0087}}
& {\ttlf\texttt{0.363}}
& {\ttlf\texttt{0.164}}
\\
[\doublerulesep]\cline{2-2}\cline{3-3}\cline{4-4}\cline{5-5}\cline{6-6}\cline{7-7}& \multirow{1}{*}{2}
& \ttlf$\texttt{10}^{\texttt{+2}}$
& {\ttlf\texttt{-5.1052}}
& {\ttlf\texttt{-5.0972~$\pmtt$~0.0087}}
& {\ttlf\texttt{0.913}}
& {\ttlf\texttt{0.170}}
\\
[\doublerulesep]\cline{2-2}\cline{3-3}\cline{4-4}\cline{5-5}\cline{6-6}\cline{7-7}& \multirow{1}{*}{3}
& \ttlf$\texttt{10}^{\texttt{+2}}$
& {\ttlf\texttt{-5.3523}}
& {\ttlf\texttt{-5.3433~$\pmtt$~0.0110}}
& {\ttlf\texttt{0.821}}
& {\ttlf\texttt{0.206}}
\\
[\doublerulesep]\ccline{1-1}\ccline{2-2}\ccline{3-3}\ccline{4-4}\ccline{5-5}\ccline{6-6}\ccline{7-7}\multirow{3}{*}{\makecell[l]{$q\bar{q}\to\gamma\gamma\gamma$\\$m=0$}}
& \multirow{1}{*}{1}
& \ttlf$\texttt{10}^{\texttt{-1}}$
& {\ttlf\texttt{-3.3918}}
& {\ttlf\texttt{-3.4066~$\pmtt$~0.0159}}
& {\ttlf\texttt{0.929}}
& {\ttlf\texttt{0.467}}
\\
[\doublerulesep]\cline{2-2}\cline{3-3}\cline{4-4}\cline{5-5}\cline{6-6}\cline{7-7}& \multirow{1}{*}{2}
& \ttlf$\texttt{10}^{\texttt{+0}}$
& {\ttlf\texttt{-1.4313}}
& {\ttlf\texttt{-1.4335~$\pmtt$~0.0101}}
& {\ttlf\texttt{0.217}}
& {\ttlf\texttt{0.707}}
\\
[\doublerulesep]\cline{2-2}\cline{3-3}\cline{4-4}\cline{5-5}\cline{6-6}\cline{7-7}& \multirow{1}{*}{3}
& \ttlf$\texttt{10}^{\texttt{-1}}$
& {\ttlf\texttt{-3.3987}}
& {\ttlf\texttt{-3.3930~$\pmtt$~0.0167}}
& {\ttlf\texttt{0.341}}
& {\ttlf\texttt{0.493}}
\\
[\doublerulesep]\ccline{1-1}\ccline{2-2}\ccline{3-3}\ccline{4-4}\ccline{5-5}\ccline{6-6}\ccline{7-7}\multirow{3}{*}{\makecell[l]{$q\bar{q}\to\gamma^*\gamma^*\gamma^*$\\$m=0$}}
& \multirow{1}{*}{1}
& \ttlf$\texttt{10}^{\texttt{-1}}$
& 
& {\ttlf\texttt{-2.1869~$\pmtt$~0.0209}}
& 
& {\ttlf\texttt{0.956}}
\\
[\doublerulesep]\cline{2-2}\cline{3-3}\cline{4-4}\cline{5-5}\cline{6-6}\cline{7-7}& \multirow{1}{*}{2}
& \ttlf$\texttt{10}^{\texttt{-1}}$
& 
& {\ttlf\texttt{-4.8302~$\pmtt$~0.0481}}
& 
& {\ttlf\texttt{0.996}}
\\
[\doublerulesep]\cline{2-2}\cline{3-3}\cline{4-4}\cline{5-5}\cline{6-6}\cline{7-7}& \multirow{1}{*}{3}
& \ttlf$\texttt{10}^{\texttt{-1}}$
& 
& {\ttlf\texttt{-2.1826~$\pmtt$~0.0208}}
& 
& {\ttlf\texttt{0.951}}
\\
[\doublerulesep]\ccline{1-1}\ccline{2-2}\ccline{3-3}\ccline{4-4}\ccline{5-5}\ccline{6-6}\ccline{7-7}\multirow{3}{*}{\makecell[l]{$q\bar{q}\to\gamma\gamma$\\$m>0$}}
& \multirow{1}{*}{1}
& \ttlf$\texttt{10}^{\texttt{+0}}$
& {\ttlf\texttt{ 4.0175}}
& {\ttlf\texttt{~4.0709~$\pmtt$~0.0358}}
& {\ttlf\texttt{{1.491}}}
& {\ttlf\texttt{0.879}}
\\
[\doublerulesep]\cline{2-2}\cline{3-3}\cline{4-4}\cline{5-5}\cline{6-6}\cline{7-7}& \multirow{1}{*}{2}
& \ttlf$\texttt{10}^{\texttt{+1}}$
& {\ttlf\texttt{-4.9939}}
& {\ttlf\texttt{-4.9853~$\pmtt$~0.0098}}
& {\ttlf\texttt{0.883}}
& {\ttlf\texttt{0.196}}
\\
[\doublerulesep]\cline{2-2}\cline{3-3}\cline{4-4}\cline{5-5}\cline{6-6}\cline{7-7}& \multirow{1}{*}{3}
& \ttlf$\texttt{10}^{\texttt{+2}}$
& {\ttlf\texttt{-2.8525}}
& {\ttlf\texttt{-2.8432~$\pmtt$~0.0075}}
& {\ttlf\texttt{{1.236}}}
& {\ttlf\texttt{0.264}}
\\
[\doublerulesep]\ccline{1-1}\ccline{2-2}\ccline{3-3}\ccline{4-4}\ccline{5-5}\ccline{6-6}\ccline{7-7}\multirow{3}{*}{\makecell[l]{$q\bar{q}\to\gamma^*\gamma^*$\\$m>0$}}
& \multirow{1}{*}{1}
& \ttlf$\texttt{10}^{\texttt{+0}}$
& 
& {\ttlf\texttt{~4.0862~$\pmtt$~0.0392}}
& 
& {\ttlf\texttt{0.959}}
\\
[\doublerulesep]\cline{2-2}\cline{3-3}\cline{4-4}\cline{5-5}\cline{6-6}\cline{7-7}& \multirow{1}{*}{2}
& \ttlf$\texttt{10}^{\texttt{+1}}$
& 
& {\ttlf\texttt{-5.3944~$\pmtt$~0.0185}}
& 
& {\ttlf\texttt{0.343}}
\\
[\doublerulesep]\cline{2-2}\cline{3-3}\cline{4-4}\cline{5-5}\cline{6-6}\cline{7-7}& \multirow{1}{*}{3}
& \ttlf$\texttt{10}^{\texttt{+2}}$
& 
& {\ttlf\texttt{-2.9477~$\pmtt$~0.0118}}
& 
& {\ttlf\texttt{0.401}}
\\
[\doublerulesep]\ccline{1-1}\ccline{2-2}\ccline{3-3}\ccline{4-4}\ccline{5-5}\ccline{6-6}\ccline{7-7}\multirow{3}{*}{\makecell[l]{$q\bar{q}\to\gamma\gamma\gamma$\\$m>0$}}
& \multirow{1}{*}{1}
& \ttlf$\texttt{10}^{\texttt{-2}}$
& 
& {\ttlf\texttt{-5.5712~$\pmtt$~0.0538}}
& 
& {\ttlf\texttt{0.966}}
\\
[\doublerulesep]\cline{2-2}\cline{3-3}\cline{4-4}\cline{5-5}\cline{6-6}\cline{7-7}& \multirow{1}{*}{2}
& \ttlf$\texttt{10}^{\texttt{-1}}$
& 
& {\ttlf\texttt{-3.6549~$\pmtt$~0.0312}}
& 
& {\ttlf\texttt{0.854}}
\\
[\doublerulesep]\cline{2-2}\cline{3-3}\cline{4-4}\cline{5-5}\cline{6-6}\cline{7-7}& \multirow{1}{*}{3}
& \ttlf$\texttt{10}^{\texttt{-2}}$
& 
& {\ttlf\texttt{-3.5927~$\pmtt$~0.0334}}
& 
& {\ttlf\texttt{0.931}}
\\
\cline{3-3}\cline{4-4}\cline{5-5}\cline{6-6}\cline{7-7}\hline\hline
\end{tabular}%
\caption{\label{tab:me}Fermion-loop contributions $F^{(2,\tilde{N}_f)}$ mediated by light-quark ($m=0$) and heavy-quark ($m>0$) loops to the two-loop squared matrix elements, as defined in eq.~\eqref{eq:matrix-element}, evaluated at the phase-space points listed in section~\ref{sec:ps_points}.}%

\end{table}

\def\csvirtanalytic{\rm benchmark}
\def\csvirt{{\sigma^{(2,\tilde{N}_f)}}}
\def\csvirtuv{{R}_{\rm UV} \widehat{\sigma}^{(2,\tilde{N}_f)}}
\def\csvirtnonplanar{\widehat{\sigma}^{(2,\tilde{N}_f),R}_\nonplanar}
\def\csvirtplanar{\widehat{\sigma}^{(2,\tilde{N}_f),R}_\planar}

\subsection{Double-virtual corrections}\label{sec:cross-sect-level}
We demonstrate for on- and off-shell diboson production and triphoton production via light fermion loops that the presented integrand representations can readily be used for the computation of double-virtual corrections to cross sections.
In table~\ref{tab:cross_section0}, we show the phase-space integrated contribution in the quark-antiquark channel
\begin{align}
\label{eq:cs_virt}
    \csvirt(s,\mu_F)
    = \sum_{a,b}\int \dd x_1 \dd x_2 \dd \Pi_2
    f_a(x_1,\mu_F) f_b(x_2,\mu_F)
    \frac{F^{(2,\tilde{N}_f)}(x_1 x_2 s)}{2x_1x_2s}\,\mathcal{O},
\end{align}
where $f_{a,b}(x_{1,2},\mu_F)$ are the PDFs of partons $a,b$ with momentum fraction $x_{1,2}$, respectively.
By $\dd \Pi_2$ we denote the $2$-body phase space measure and $\mathcal{O}$ is an observable function.
We chose $\mathcal{O}$ to impose a phase-space cut requiring minimal transverse momenta of $p_T^\text{min} = 50$\,GeV for on-shell photons and $25$\,GeV for off-shell photons.
The results include the electric charge quantum numbers of the incoming quarks, while colour factors, QED couplings, and the sum of squared charges of the quarks circulating in the loop are factored out.
The reference results were obtained by integrating the analytic expressions for $\gamma\gamma$ \cite{Anastasiou:2002zn}, $\gamma^*\gamma^*$ \cite{Gehrmann:2015ora} and $\gamma\gamma\gamma$ \cite{Abreu:2023bdp} using our own phase-space generator~\cite{Kermanschah:2024utt}.
\par

\begin{table}[t]
\centering
\renewcommand{\arraystretch}{1.35}
\begin{tabular}{llrrlr}
\hline\hline
Process & Part & $N_p \ [10^8]$ & Exp. & \ \ \ \ \  \ \ \ \ Result & $\Delta \ [\%]$ \\
\hline\hline\multirow{5}{*}{\makecell[l]{$q\bar{q}\to\gamma\gamma$\\$m=0$}}
& \multirow{1}{*}{${\csvirtplanar}$}
& \ttlf{$\texttt{1000.8}$}
& {\ttlf\ttlf$\texttt{10}^{\texttt{-4}}$}
& {\ttlf\texttt{-5.4142~$\pmtt$~0.0062}}
& {\ttlf\texttt{0.115}}
\\
\cline{2-2}\cline{3-3}\cline{4-4}\cline{5-5}\cline{6-6}& \multirow{1}{*}{${\csvirtnonplanar}$}
& \ttlf{$\texttt{1000.8}$}
& {\ttlf\ttlf$\texttt{10}^{\texttt{-4}}$}
& {\ttlf\texttt{~3.9034~$\pmtt$~0.0030}}
& {\ttlf\texttt{0.076}}
\\
\cline{2-2}\cline{3-3}\cline{4-4}\cline{5-5}\cline{6-6}& \multirow{1}{*}{$\csvirtuv$}
& \ttlf{$\texttt{1.4}$}
& {\ttlf\ttlf$\texttt{10}^{\texttt{-4}}$}
& {\ttlf\texttt{-1.0357~$\pmtt$~0.0005}}
& {\ttlf\texttt{0.048}}
\\
\ccline{2-2}\ccline{3-3}\ccline{4-4}\ccline{5-5}\ccline{6-6}& \multirow{1}{*}{$\boldsymbol\csvirt$}
& 
& {$\texttt{10}^{\texttt{-3}}$}
& {\texttt{-1.4886~$\pmtt$~0.0026}}
& {\texttt{0.173}}
\\
\cline{2-2}\cline{3-3}\cline{4-4}\cline{5-5}\cline{6-6}& \multirow{1}{*}{$\boldsymbol\csvirtanalytic$}
& {$\texttt{0.1}$}
& {$\texttt{10}^{\texttt{-3}}$}
& {\texttt{-1.4905~$\pmtt$~0.0004}}
& {\texttt{0.025}}
\\
\ccline{1-1}\ccline{2-2}\ccline{3-3}\ccline{4-4}\ccline{5-5}\ccline{6-6}\multirow{5}{*}{\makecell[l]{$q\bar{q}\to\gamma^*\gamma^*$\\$m=0$}}
& \multirow{1}{*}{${\csvirtplanar}$}
& \ttlf{$\texttt{352.0}$}
& {\ttlf\ttlf$\texttt{10}^{\texttt{-4}}$}
& {\ttlf\texttt{-1.0802~$\pmtt$~0.0004}}
& {\ttlf\texttt{0.035}}
\\
\cline{2-2}\cline{3-3}\cline{4-4}\cline{5-5}\cline{6-6}& \multirow{1}{*}{${\csvirtnonplanar}$}
& \ttlf{$\texttt{529.3}$}
& {\ttlf\ttlf$\texttt{10}^{\texttt{-4}}$}
& {\ttlf\texttt{~2.8959~$\pmtt$~0.0014}}
& {\ttlf\texttt{0.048}}
\\
\cline{2-2}\cline{3-3}\cline{4-4}\cline{5-5}\cline{6-6}& \multirow{1}{*}{$\csvirtuv$}
& \ttlf{$\texttt{1.1}$}
& {\ttlf\ttlf$\texttt{10}^{\texttt{-5}}$}
& {\ttlf\texttt{-3.5849~$\pmtt$~0.0017}}
& {\ttlf\texttt{0.047}}
\\
\ccline{2-2}\ccline{3-3}\ccline{4-4}\ccline{5-5}\ccline{6-6}& \multirow{1}{*}{$\boldsymbol\csvirt$}
& 
& {$\texttt{10}^{\texttt{-4}}$}
& {\texttt{~1.1126~$\pmtt$~0.0032}}
& {\texttt{0.287}}
\\
\cline{2-2}\cline{3-3}\cline{4-4}\cline{5-5}\cline{6-6}& \multirow{1}{*}{$\boldsymbol\csvirtanalytic$}
& {$\texttt{0.1}$}
& {$\texttt{10}^{\texttt{-4}}$}
& {\texttt{~1.1187~$\pmtt$~0.0005}}
& {\texttt{0.043}}
\\
\ccline{1-1}\ccline{2-2}\ccline{3-3}\ccline{4-4}\ccline{5-5}\ccline{6-6}\multirow{5}{*}{\makecell[l]{$q\bar{q}\to\gamma\gamma\gamma$\\$m=0$}}
& \multirow{1}{*}{${\csvirtplanar}$}
& \ttlf{$\texttt{1808.6}$}
& {\ttlf\ttlf$\texttt{10}^{\texttt{-6}}$}
& {\ttlf\texttt{-4.0711~$\pmtt$~0.1084}}
& {\ttlf\texttt{{2.662}}}
\\
\cline{2-2}\cline{3-3}\cline{4-4}\cline{5-5}\cline{6-6}& \multirow{1}{*}{${\csvirtnonplanar}$}
& \ttlf{$\texttt{1570.3}$}
& {\ttlf\ttlf$\texttt{10}^{\texttt{-6}}$}
& {\ttlf\texttt{~0.7717~$\pmtt$~0.3021}}
& {\ttlf\texttt{{39.142}}}
\\
\cline{2-2}\cline{3-3}\cline{4-4}\cline{5-5}\cline{6-6}& \multirow{1}{*}{$\csvirtuv$}
& \ttlf{$\texttt{46.1}$}
& {\ttlf\ttlf$\texttt{10}^{\texttt{-7}}$}
& {\ttlf\texttt{-6.9927~$\pmtt$~0.0035}}
& {\ttlf\texttt{0.049}}
\\
\ccline{2-2}\ccline{3-3}\ccline{4-4}\ccline{5-5}\ccline{6-6}& \multirow{1}{*}{$\boldsymbol\csvirt$}
& 
& {$\texttt{10}^{\texttt{-5}}$}
& {\texttt{-4.6321~$\pmtt$~0.2231}}
& {\texttt{{4.816}}}
\\
\cline{2-2}\cline{3-3}\cline{4-4}\cline{5-5}\cline{6-6}& \multirow{1}{*}{$\boldsymbol\csvirtanalytic$}
& {$\texttt{0.02}$}
& {$\texttt{10}^{\texttt{-5}}$}
& {\texttt{-4.2678~$\pmtt$~0.0084}}
& {\texttt{0.197}}
\\
\cline{3-3}\cline{4-4}\cline{5-5}\cline{6-6}\hline\hline
\end{tabular}%
\caption{\label{tab:cross_section0}Double-virtual corrections mediated by a light-quark loop, as defined in eq.~\eqref{eq:cs_virt}, including contributions from planar (P) and non-planar (NP) diagrams and integrated UV counterterms.}%

\end{table}

We exploit the fact that the phase-space integrated planar and non-planar contributions, $\csvirtplanar$ and $\csvirtnonplanar$, as well as the integrated UV counterterms $\csvirtuv$, are identical by symmetry for any fixed final-state momentum ordering.
Thus, it suffices to evaluate a single ordering and assign the corresponding symmetry factor.
As a result, the virtual corrections require fewer contributions than the matrix element at a fixed phase-space point.
\par
The table shows the individual Monte Carlo integrations obtained using $N_p$ samples and the combined contribution, targeting sub-percent precision $\Delta\,[\%]$.
The collision energy is fixed at $E_\text{CM}\equiv\sqrt{s}=13$\,TeV.
We chose the \texttt{CT10} NLO PDF set with \texttt{id 11000}~\cite{Lai:2010vv} of \textsc{LHAPDF} \cite{Buckley:2014ana} using five light flavours.
The factorisation scales, UV masses and off-shell photon mass are set to $\mu_F=M=M_{\gamma^*}=M_Z=91.1876\,\text{GeV}$.
Our results, with Monte Carlo errors below $0.3\%$ for diphoton and $5\%$ for triphoton production are consistent with the reference results within two standard deviations.
\par

The phase-space integration for fermion-loop mediated contribution to the $2\to3$ process is significantly more complex than for the $2\to2$ processes.
This is not only because the integration space is 13-dimensional, but also because the threshold parametrisation described in section~\ref{sec:thresholds} requires boosting to the rest frame of the two photons attached to the quark loop at each phase-space point.
Possible avenues to improve integration efficiency, which we have not explored here, include implementing multi-channel phase-space integration and training the adaptive grid first on the tree-level contribution.
For practical reasons, we have not computed the phase-space integration over the heavy-quark loop contributions, since the threshold singularity structure varies across the phase space and the relevant threshold counterterms must be selected at runtime, a feature we leave for future work.

\subsection{Timing and performance}
\label{sec:timings}

\begin{table}[b]
\centering
\renewcommand{\arraystretch}{1.35}
\begin{tabular}{lccc}
\hline\hline
Process & $\twoloopfloopsplanar$ & $\twoloopfloopsnonplanar$ & $\twoloopfloopsuv$ \\
\hline\hline
\makecell[l]{$q\bar{q}\to\gamma\gamma$, $m=0$}
& \ttlf\texttt{0.118} & \ttlf\texttt{0.117} & \ttlf\texttt{0.006} \\
\hline
\makecell[l]{$q\bar{q}\to\gamma^*\gamma^*$, $m=0$}
& \ttlf\texttt{0.395} & \ttlf\texttt{0.590} & \ttlf\texttt{0.023} \\
\hline
\makecell[l]{$q\bar{q}\to\gamma\gamma\gamma$, $m=0$}
& \ttlf\texttt{1.289} & \ttlf\texttt{1.396} & \ttlf\texttt{0.082} \\
\hline
\makecell[l]{$q\bar{q}\to\gamma^*\gamma^*\gamma^*$, $m=0$}
& \ttlf\texttt{28.38} & \ttlf\texttt{34.14} & \ttlf\texttt{0.625} \\
\hline
\makecell[l]{$q\bar{q}\to\gamma\gamma$, $m>0$}
& \ttlf\texttt{0.148} & \ttlf\texttt{0.156} & \ttlf\texttt{0.006} \\
\hline
\makecell[l]{$q\bar{q}\to\gamma^*\gamma^*$, $m>0$}
& \ttlf\texttt{0.641} & \ttlf\texttt{0.683} & \ttlf\texttt{0.024} \\
\hline
\makecell[l]{$q\bar{q}\to\gamma\gamma\gamma$, $m>0$}
& \ttlf\texttt{1.594} & \ttlf\texttt{2.912} & \ttlf\texttt{0.061} \\
\hline\hline
\end{tabular}%
\caption{\label{tab:timings}Average single-core integrand evaluation times in milliseconds using double-precision arithmetic.
}
\end{table}

In table~\ref{tab:timings}, we show the integrand evaluation time per sample per threshold channel in milliseconds, averaged over $10^{4}$ samples on a single core of an \texttt{AMD EPYC~7742} 64-core processor (2 sockets, base clock 2.25~GHz, boost clock up to 3.4~GHz).
The integrands shown correspond to the IR- and UV-subtracted planar and non-planar fermion-loop contributions to the matrix element including threshold counterterms, as well as the semi-integrated UV counterterms, for a single attachment of final-state photons.
These results incorporate expression simplifications, compilation optimisations, and employ multi-channel Monte Carlo parametrisations.
Off-shell final-state photons increase the evaluation time, since more helicity configurations must be summed over and additional threshold counterterms are required.
Introducing a heavy quark in the loop has a smaller but noticeable effect on the evaluation time.
In general, evaluation times grow gradually with increasing complexity, except in the special case of $\gamma^* \gamma^* \gamma^*$.
There, the evaluation was significantly slower because the larger expressions containing diagrams with up to 10 propagators prevented us from applying the same code optimisations, as mentioned in section~\ref{sec:implementation} and footnote~\ref{foot:opt_2to3massive}.
\par

Our integrations were carried out on the Euler high-performance computing cluster at ETH Zurich, with each Monte Carlo run executed on a single node and parallelised across 128 cores.
The wall-times to achieve below one-percent uncertainty on the squared  matrix elements ranged from a few hours to a few days, necessitating approximately $10^8$ to $10^{10}$ Monte Carlo samples, depending on process complexity (see tables in section~\ref{sec:individual_me}).
Notably, the integrations over loop and phase space converged on comparable timescales to those of the matrix elements evaluated at fixed phase-space points for the $2\to 2$ processes, whereas for the higher-dimensional $2\to 3$ case, the difference in convergence times was found to be larger.
In turn, the phase-space–integrated matrix elements receive fewer diagrammatic contributions thanks to symmetry.
\section{Conclusions}
In this paper, we numerically computed the fermion-loop mediated double-virtual corrections for the production of two and three on- and off-shell photons involving both heavy and light quarks in the loop.
These gauge-invariant contributions involve the topologically most complicated integrals that enter the two-loop amplitudes.
Together with ref.~\cite{Kermanschah:2024utt}, we demonstrated that two-loop electroweak boson production amplitudes with quark loops can be treated within a unified computational framework, which can accommodate both massive final states and heavy internal quarks.
\par
We validated our computation of fermionic two-loop matrix elements against known analytic results for $q\bar{q}\to\gamma\gamma$, $\gamma^*\gamma^*$, and $\gamma\gamma\gamma$ in massless QCD, as well as for $q\bar{q}\to\gamma\gamma$ via a heavy-quark loop.
New results are presented for $q\bar{q}\to\gamma^*\gamma^*$ and $q\bar{q}\to\gamma\gamma\gamma$ with heavy-quark loops and for $q\bar{q}\to\gamma^*\gamma^*\gamma^*$ with light-quark loops, advancing the known two-loop contributions to five-point processes.
\par
Our method relies on direct numerical integration over loop momenta using Monte Carlo methods.
Although the considered amplitude contributions are finite after integration, the loop integrands exhibit divergent regions when naively constructed from Feynman diagrams.
Consequently, IR and UV singularities must be subtracted using counterterms, as those proposed in~\cite{Anastasiou:2020sdt,Anastasiou:2024xvk}, to obtain a finite integrand in $d=4$ dimensions.
Compared to the approach of ref.~\cite{Kermanschah:2024utt}, this necessitated new types of IR and UV counterterms and threshold parametrisations involving two-loop momenta~\cite{Kermanschah:2021wbk,Vicini:2024ecf} for planar and non-planar two-loop Feynman diagrams with four-point fermion loops.
Among these diagrams, the penta-box topologies are the most challenging, involving up to eight kinematic scales, and their analytic expressions are beyond reach with existing techniques.
To handle integrable singularities from soft and collinear configurations, we further applied multi-channel importance-sampling.  
Finally, we improved the codebase to be more flexible and efficient, including new constructions of compact, more numerically stable expressions along with their optimisation and compilation.
\par
Axial couplings from $W$ and $Z$ bosons in the final state can be included with relatively little additional effort, as demonstrated in ref.~\cite{Kermanschah:2024utt} for the $N_f$-part, and additional fermion-loop Feynman diagrams can be treated analogously.
Similarly, when generalising the presented method to fermion-loop contributions for the production of more than three bosons, we merely expect increased evaluation times due to larger integrands, more helicity configurations, and the need for additional threshold counterterms.
\par
Future improvements could include importance sampling in both loop momentum and phase space, along with a more refined decomposition into, possibly adaptive, integration channels.
As a next milestone, we intend to compute the remaining contributions to full two-loop electroweak production amplitudes using these numerical techniques.
\acknowledgments
We thank C.~Anastasiou for his encouragement and for providing computational resources and J.~Karlen for valuable discussions.
We acknowledge the use of the Euler cluster at ETH Zurich for much of the numerical computations.
This work was supported by the Swiss National Science Foundation through its Postdoc.Mobility funding scheme (grant numbers 211062 and 230593) and project funding scheme (grant number 10001706).
\newpage
\appendix
\section{Tables of planar and non-planar contributions}
\label{sec:individual_me}
Tables~\ref{tab:me_single0},~\ref{tab:me_single1},~\ref{tab:me_single2},~\ref{tab:me_single3} and~\ref{tab:me_single4} show the planar and non-planar contributions, $\twoloopfloopsplanar$ and $\twoloopfloopsnonplanar$, together with the integrated UV counterterms, $\twoloopfloopsuv$, each obtained from independent Monte Carlo integrations.
These contributions are evaluated for each permutation of final-state momenta in di- and triphoton production at three phase-space points provided in appendix~\ref{sec:ps_points}, and are defined in eqs.~\eqref{eq:planar_and_non_planar_2_to_2} and \eqref{eq:integrated_uv_2_to_2} for diphoton and eqs.~\eqref{eq:planar_and_non_planar_2_to_3} and \eqref{eq:integrated_uv_2_to_3} for triphoton production.
They are then combined according to eqs.~\eqref{eq:floops-pieces}, \eqref{eq:planar_and_non_planar} and \eqref{eq:matrix-element} to obtain the gauge-invariant fermion-loop contributions $F^{(2,\tilde{N}_f)}$ to the two-loop matrix elements in table~\ref{tab:me}, with the Monte Carlo errors propagated in quadrature.
\par
The table headers are, in this order, the process including whether the quark in the closed loop is heavy ($m>0$) or light ($m=0$), the phase-space point, the part under consideration, the permutation of final-state momenta, the number of Monte Carlo samples $N_p$, the factored power of ten, the Monte Carlo estimate, and the relative error $\Delta\,[\%]$.
\par
Because of cancellations, achieving percent-level precision for the full fermion-loop contribution required sub-percent precision for the individual terms.
For light-quark loops ($m=0$), $0.5\%$ was sufficient for on-shell photon production, but up to $0.2\%$ was needed for off-shell production.
Cancellations were more severe for heavy-quark loops ($m>0$), particularly in on- and off-shell diphoton production.
Phase-space points 1 and 2 required precision up to $0.01\%$ or better.
For these points the matrix elements are around one order of magnitude smaller than for point 3, and they only have a single $s$-channel threshold corresponding to cutting both gluons, whereas point 3 includes all $s$-channel cuts.

\raggedbottom

\begin{table}[H]
\centering
\renewcommand{\arraystretch}{1.13}
\resizebox{\columnwidth}{!}{%
\begin{tabular}{llllrrlr}
\hline\hline
Process & PSP & Part & Perm. & $N_p \ [10^8]$ & Exp. & \ \ \ \ \  \ \ \ \ Result & $\Delta \ [\%]$ \\
\hline\hline\multirow{12}{*}{\makecell[l]{$q\bar{q}\to\gamma\gamma$\\$m=0$}}
& \multirow{4}{*}{1}
& \multirow{2}{*}{${\twoloopfloopsplanar}$}
& {\ttlf1}
& {\ttlf\ttlf{$\texttt{5.8}$}}
& {\ttlf\ttlf$\texttt{10}^{\texttt{+2}}$}
& {\ttlf\texttt{-2.7439~$\pmtt$~0.0132}}
& {\ttlf\texttt{0.480}}
\\
\cline{4-4}\cline{5-5}\cline{6-6}\cline{7-7}\cline{8-8}& & & {\ttlf2}
& {\ttlf\ttlf{$\texttt{3.1}$}}
& {\ttlf\ttlf$\texttt{10}^{\texttt{+2}}$}
& {\ttlf\texttt{-3.0673~$\pmtt$~0.0148}}
& {\ttlf\texttt{0.483}}
\\
\cline{3-3}\cline{4-4}\cline{5-5}\cline{6-6}\cline{7-7}\cline{8-8}& & \multirow{1}{*}{${\twoloopfloopsnonplanar}$}
& {\ttlf1}
& {\ttlf\ttlf{$\texttt{1.6}$}}
& {\ttlf\ttlf$\texttt{10}^{\texttt{+2}}$}
& {\ttlf\texttt{~2.4299~$\pmtt$~0.0114}}
& {\ttlf\texttt{0.469}}
\\
\cline{3-3}\cline{4-4}\cline{5-5}\cline{6-6}\cline{7-7}\cline{8-8}& & \multirow{1}{*}{$\twoloopfloopsuv$}
& {\ttlf1}
& {\ttlf\ttlf{$\texttt{0.1}$}}
& {\ttlf\ttlf$\texttt{10}^{\texttt{+1}}$}
& {\ttlf\texttt{-4.9488~$\pmtt$~0.0064}}
& {\ttlf\texttt{0.129}}
\\
[\doublerulesep]\ccline{2-2}\ccline{3-3}\ccline{4-4}\ccline{5-5}\ccline{6-6}\ccline{7-7}\ccline{8-8}& \multirow{4}{*}{2}
& \multirow{2}{*}{${\twoloopfloopsplanar}$}
& {\ttlf1}
& {\ttlf\ttlf{$\texttt{2.8}$}}
& {\ttlf\ttlf$\texttt{10}^{\texttt{+2}}$}
& {\ttlf\texttt{-2.8053~$\pmtt$~0.0130}}
& {\ttlf\texttt{0.463}}
\\
\cline{4-4}\cline{5-5}\cline{6-6}\cline{7-7}\cline{8-8}& & & {\ttlf2}
& {\ttlf\ttlf{$\texttt{4.5}$}}
& {\ttlf\ttlf$\texttt{10}^{\texttt{+2}}$}
& {\ttlf\texttt{-2.6029~$\pmtt$~0.0126}}
& {\ttlf\texttt{0.483}}
\\
\cline{3-3}\cline{4-4}\cline{5-5}\cline{6-6}\cline{7-7}\cline{8-8}& & \multirow{1}{*}{${\twoloopfloopsnonplanar}$}
& {\ttlf1}
& {\ttlf\ttlf{$\texttt{1.4}$}}
& {\ttlf\ttlf$\texttt{10}^{\texttt{+2}}$}
& {\ttlf\texttt{~2.2544~$\pmtt$~0.0108}}
& {\ttlf\texttt{0.477}}
\\
\cline{3-3}\cline{4-4}\cline{5-5}\cline{6-6}\cline{7-7}\cline{8-8}& & \multirow{1}{*}{$\twoloopfloopsuv$}
& {\ttlf1}
& {\ttlf\ttlf{$\texttt{0.1}$}}
& {\ttlf\ttlf$\texttt{10}^{\texttt{+1}}$}
& {\ttlf\texttt{-4.5720~$\pmtt$~0.0047}}
& {\ttlf\texttt{0.102}}
\\
[\doublerulesep]\ccline{2-2}\ccline{3-3}\ccline{4-4}\ccline{5-5}\ccline{6-6}\ccline{7-7}\ccline{8-8}& \multirow{4}{*}{3}
& \multirow{2}{*}{${\twoloopfloopsplanar}$}
& {\ttlf1}
& {\ttlf\ttlf{$\texttt{13.3}$}}
& {\ttlf\ttlf$\texttt{10}^{\texttt{+2}}$}
& {\ttlf\texttt{-3.0095~$\pmtt$~0.0148}}
& {\ttlf\texttt{0.491}}
\\
\cline{4-4}\cline{5-5}\cline{6-6}\cline{7-7}\cline{8-8}& & & {\ttlf2}
& {\ttlf\ttlf{$\texttt{2.3}$}}
& {\ttlf\ttlf$\texttt{10}^{\texttt{+2}}$}
& {\ttlf\texttt{-3.8462~$\pmtt$~0.0192}}
& {\ttlf\texttt{0.499}}
\\
\cline{3-3}\cline{4-4}\cline{5-5}\cline{6-6}\cline{7-7}\cline{8-8}& & \multirow{1}{*}{${\twoloopfloopsnonplanar}$}
& {\ttlf1}
& {\ttlf\ttlf{$\texttt{2.5}$}}
& {\ttlf\ttlf$\texttt{10}^{\texttt{+2}}$}
& {\ttlf\texttt{~2.9083~$\pmtt$~0.0138}}
& {\ttlf\texttt{0.474}}
\\
\cline{3-3}\cline{4-4}\cline{5-5}\cline{6-6}\cline{7-7}\cline{8-8}& & \multirow{1}{*}{$\twoloopfloopsuv$}
& {\ttlf1}
& {\ttlf\ttlf{$\texttt{0.1}$}}
& {\ttlf\ttlf$\texttt{10}^{\texttt{+1}}$}
& {\ttlf\texttt{-6.2058~$\pmtt$~0.0138}}
& {\ttlf\texttt{0.222}}
\\
[\doublerulesep]\ccline{1-1}\ccline{2-2}\ccline{3-3}\ccline{4-4}\ccline{5-5}\ccline{6-6}\ccline{7-7}\ccline{8-8}\multirow{12}{*}{\makecell[l]{$q\bar{q}\to\gamma^*\gamma^*$\\$m=0$}}
& \multirow{4}{*}{1}
& \multirow{2}{*}{${\twoloopfloopsplanar}$}
& {\ttlf1}
& {\ttlf\ttlf{$\texttt{6.6}$}}
& {\ttlf\ttlf$\texttt{10}^{\texttt{+2}}$}
& {\ttlf\texttt{-4.0349~$\pmtt$~0.0079}}
& {\ttlf\texttt{0.195}}
\\
\cline{4-4}\cline{5-5}\cline{6-6}\cline{7-7}\cline{8-8}& & & {\ttlf2}
& {\ttlf\ttlf{$\texttt{2.8}$}}
& {\ttlf\ttlf$\texttt{10}^{\texttt{+2}}$}
& {\ttlf\texttt{-4.3676~$\pmtt$~0.0084}}
& {\ttlf\texttt{0.192}}
\\
\cline{3-3}\cline{4-4}\cline{5-5}\cline{6-6}\cline{7-7}\cline{8-8}& & \multirow{1}{*}{${\twoloopfloopsnonplanar}$}
& {\ttlf1}
& {\ttlf\ttlf{$\texttt{3.6}$}}
& {\ttlf\ttlf$\texttt{10}^{\texttt{+2}}$}
& {\ttlf\texttt{~6.5489~$\pmtt$~0.0130}}
& {\ttlf\texttt{0.199}}
\\
\cline{3-3}\cline{4-4}\cline{5-5}\cline{6-6}\cline{7-7}\cline{8-8}& & \multirow{1}{*}{$\twoloopfloopsuv$}
& {\ttlf1}
& {\ttlf\ttlf{$\texttt{0.3}$}}
& {\ttlf\ttlf$\texttt{10}^{\texttt{+1}}$}
& {\ttlf\texttt{-8.7177~$\pmtt$~0.0063}}
& {\ttlf\texttt{0.072}}
\\
[\doublerulesep]\ccline{2-2}\ccline{3-3}\ccline{4-4}\ccline{5-5}\ccline{6-6}\ccline{7-7}\ccline{8-8}& \multirow{4}{*}{2}
& \multirow{2}{*}{${\twoloopfloopsplanar}$}
& {\ttlf1}
& {\ttlf\ttlf{$\texttt{2.8}$}}
& {\ttlf\ttlf$\texttt{10}^{\texttt{+2}}$}
& {\ttlf\texttt{-4.3028~$\pmtt$~0.0082}}
& {\ttlf\texttt{0.191}}
\\
\cline{4-4}\cline{5-5}\cline{6-6}\cline{7-7}\cline{8-8}& & & {\ttlf2}
& {\ttlf\ttlf{$\texttt{4.5}$}}
& {\ttlf\ttlf$\texttt{10}^{\texttt{+2}}$}
& {\ttlf\texttt{-4.1280~$\pmtt$~0.0082}}
& {\ttlf\texttt{0.199}}
\\
\cline{3-3}\cline{4-4}\cline{5-5}\cline{6-6}\cline{7-7}\cline{8-8}& & \multirow{1}{*}{${\twoloopfloopsnonplanar}$}
& {\ttlf1}
& {\ttlf\ttlf{$\texttt{3.1}$}}
& {\ttlf\ttlf$\texttt{10}^{\texttt{+2}}$}
& {\ttlf\texttt{~6.6090~$\pmtt$~0.0132}}
& {\ttlf\texttt{0.200}}
\\
\cline{3-3}\cline{4-4}\cline{5-5}\cline{6-6}\cline{7-7}\cline{8-8}& & \multirow{1}{*}{$\twoloopfloopsuv$}
& {\ttlf1}
& {\ttlf\ttlf{$\texttt{0.3}$}}
& {\ttlf\ttlf$\texttt{10}^{\texttt{+1}}$}
& {\ttlf\texttt{-8.7002~$\pmtt$~0.0051}}
& {\ttlf\texttt{0.059}}
\\
[\doublerulesep]\ccline{2-2}\ccline{3-3}\ccline{4-4}\ccline{5-5}\ccline{6-6}\ccline{7-7}\ccline{8-8}& \multirow{4}{*}{3}
& \multirow{2}{*}{${\twoloopfloopsplanar}$}
& {\ttlf1}
& {\ttlf\ttlf{$\texttt{21.0}$}}
& {\ttlf\ttlf$\texttt{10}^{\texttt{+2}}$}
& {\ttlf\texttt{-3.5016~$\pmtt$~0.0069}}
& {\ttlf\texttt{0.196}}
\\
\cline{4-4}\cline{5-5}\cline{6-6}\cline{7-7}\cline{8-8}& & & {\ttlf2}
& {\ttlf\ttlf{$\texttt{3.6}$}}
& {\ttlf\ttlf$\texttt{10}^{\texttt{+2}}$}
& {\ttlf\texttt{-4.6141~$\pmtt$~0.0091}}
& {\ttlf\texttt{0.196}}
\\
\cline{3-3}\cline{4-4}\cline{5-5}\cline{6-6}\cline{7-7}\cline{8-8}& & \multirow{1}{*}{${\twoloopfloopsnonplanar}$}
& {\ttlf1}
& {\ttlf\ttlf{$\texttt{8.6}$}}
& {\ttlf\ttlf$\texttt{10}^{\texttt{+2}}$}
& {\ttlf\texttt{~6.4078~$\pmtt$~0.0125}}
& {\ttlf\texttt{0.196}}
\\
\cline{3-3}\cline{4-4}\cline{5-5}\cline{6-6}\cline{7-7}\cline{8-8}& & \multirow{1}{*}{$\twoloopfloopsuv$}
& {\ttlf1}
& {\ttlf\ttlf{$\texttt{0.3}$}}
& {\ttlf\ttlf$\texttt{10}^{\texttt{+1}}$}
& {\ttlf\texttt{-8.8084~$\pmtt$~0.0109}}
& {\ttlf\texttt{0.123}}
\\
[\doublerulesep]\ccline{1-1}\ccline{2-2}\ccline{3-3}\ccline{4-4}\ccline{5-5}\ccline{6-6}\ccline{7-7}\ccline{8-8}\multirow{12}{*}{\makecell[l]{$q\bar{q}\to\gamma^*\gamma_2^*$\\$m=0$}}
& \multirow{4}{*}{1}
& \multirow{2}{*}{${\twoloopfloopsplanar}$}
& {\ttlf1}
& {\ttlf\ttlf{$\texttt{125.2}$}}
& {\ttlf\ttlf$\texttt{10}^{\texttt{+2}}$}
& {\ttlf\texttt{-4.1127~$\pmtt$~0.0020}}
& {\ttlf\texttt{0.050}}
\\
\cline{4-4}\cline{5-5}\cline{6-6}\cline{7-7}\cline{8-8}& & & {\ttlf2}
& {\ttlf\ttlf{$\texttt{59.5}$}}
& {\ttlf\ttlf$\texttt{10}^{\texttt{+2}}$}
& {\ttlf\texttt{-4.3755~$\pmtt$~0.0022}}
& {\ttlf\texttt{0.050}}
\\
\cline{3-3}\cline{4-4}\cline{5-5}\cline{6-6}\cline{7-7}\cline{8-8}& & \multirow{1}{*}{${\twoloopfloopsnonplanar}$}
& {\ttlf1}
& {\ttlf\ttlf{$\texttt{87.3}$}}
& {\ttlf\ttlf$\texttt{10}^{\texttt{+2}}$}
& {\ttlf\texttt{~6.2833~$\pmtt$~0.0031}}
& {\ttlf\texttt{0.050}}
\\
\cline{3-3}\cline{4-4}\cline{5-5}\cline{6-6}\cline{7-7}\cline{8-8}& & \multirow{1}{*}{$\twoloopfloopsuv$}
& {\ttlf1}
& {\ttlf\ttlf{$\texttt{1.3}$}}
& {\ttlf\ttlf$\texttt{10}^{\texttt{+1}}$}
& {\ttlf\texttt{-8.6443~$\pmtt$~0.0027}}
& {\ttlf\texttt{0.031}}
\\
[\doublerulesep]\ccline{2-2}\ccline{3-3}\ccline{4-4}\ccline{5-5}\ccline{6-6}\ccline{7-7}\ccline{8-8}& \multirow{4}{*}{2}
& \multirow{2}{*}{${\twoloopfloopsplanar}$}
& {\ttlf1}
& {\ttlf\ttlf{$\texttt{53.1}$}}
& {\ttlf\ttlf$\texttt{10}^{\texttt{+2}}$}
& {\ttlf\texttt{-4.3082~$\pmtt$~0.0021}}
& {\ttlf\texttt{0.049}}
\\
\cline{4-4}\cline{5-5}\cline{6-6}\cline{7-7}\cline{8-8}& & & {\ttlf2}
& {\ttlf\ttlf{$\texttt{88.7}$}}
& {\ttlf\ttlf$\texttt{10}^{\texttt{+2}}$}
& {\ttlf\texttt{-4.1702~$\pmtt$~0.0021}}
& {\ttlf\texttt{0.050}}
\\
\cline{3-3}\cline{4-4}\cline{5-5}\cline{6-6}\cline{7-7}\cline{8-8}& & \multirow{1}{*}{${\twoloopfloopsnonplanar}$}
& {\ttlf1}
& {\ttlf\ttlf{$\texttt{62.8}$}}
& {\ttlf\ttlf$\texttt{10}^{\texttt{+2}}$}
& {\ttlf\texttt{~6.3614~$\pmtt$~0.0031}}
& {\ttlf\texttt{0.049}}
\\
\cline{3-3}\cline{4-4}\cline{5-5}\cline{6-6}\cline{7-7}\cline{8-8}& & \multirow{1}{*}{$\twoloopfloopsuv$}
& {\ttlf1}
& {\ttlf\ttlf{$\texttt{1.3}$}}
& {\ttlf\ttlf$\texttt{10}^{\texttt{+1}}$}
& {\ttlf\texttt{-8.6339~$\pmtt$~0.0022}}
& {\ttlf\texttt{0.025}}
\\
[\doublerulesep]\ccline{2-2}\ccline{3-3}\ccline{4-4}\ccline{5-5}\ccline{6-6}\ccline{7-7}\ccline{8-8}& \multirow{4}{*}{3}
& \multirow{2}{*}{${\twoloopfloopsplanar}$}
& {\ttlf1}
& {\ttlf\ttlf{$\texttt{293.6}$}}
& {\ttlf\ttlf$\texttt{10}^{\texttt{+2}}$}
& {\ttlf\texttt{-3.6445~$\pmtt$~0.0039}}
& {\ttlf\texttt{0.108}}
\\
\cline{4-4}\cline{5-5}\cline{6-6}\cline{7-7}\cline{8-8}& & & {\ttlf2}
& {\ttlf\ttlf{$\texttt{88.7}$}}
& {\ttlf\ttlf$\texttt{10}^{\texttt{+2}}$}
& {\ttlf\texttt{-4.6022~$\pmtt$~0.0023}}
& {\ttlf\texttt{0.050}}
\\
\cline{3-3}\cline{4-4}\cline{5-5}\cline{6-6}\cline{7-7}\cline{8-8}& & \multirow{1}{*}{${\twoloopfloopsnonplanar}$}
& {\ttlf1}
& {\ttlf\ttlf{$\texttt{216.9}$}}
& {\ttlf\ttlf$\texttt{10}^{\texttt{+2}}$}
& {\ttlf\texttt{~6.0103~$\pmtt$~0.0031}}
& {\ttlf\texttt{0.051}}
\\
\cline{3-3}\cline{4-4}\cline{5-5}\cline{6-6}\cline{7-7}\cline{8-8}& & \multirow{1}{*}{$\twoloopfloopsuv$}
& {\ttlf1}
& {\ttlf\ttlf{$\texttt{1.3}$}}
& {\ttlf\ttlf$\texttt{10}^{\texttt{+1}}$}
& {\ttlf\texttt{-8.7043~$\pmtt$~0.0043}}
& {\ttlf\texttt{0.049}}
\\
\cline{4-4}\cline{5-5}\cline{6-6}\cline{7-7}\cline{8-8}\hline\hline
\end{tabular}%
}%
\caption{\label{tab:me_single0}Contributions to the two-loop squared matrix elements of $q\bar{q}\to \gamma\gamma$, $q\bar{q}\to \gamma^*\gamma^*$ and $q\bar{q}\to \gamma^*\gamma_2^*$ via a light-quark $(m=0)$ loop, as defined in section~\ref{sec:individual_me}.}%

\end{table}

\begin{table}[H]
\centering
\renewcommand{\arraystretch}{1.1}
\resizebox{\columnwidth}
{!}{%
\begin{tabular}{llllrrlr}
\hline\hline
Process & PSP & Part & Perm. & $N_p \ [10^8]$ & Exp. & \ \ \ \ \  \ \ \ \ Result & $\Delta \ [\%]$ \\
\hline\hline\multirow{36}{*}{\makecell[l]{$q\bar{q}\to\gamma\gamma\gamma$\\$m=0$}}
& \multirow{12}{*}{1}
& \multirow{6}{*}{${\twoloopfloopsplanar}$}
& {\ttlf1}
& {\ttlf\ttlf{$\texttt{9.9}$}}
& {\ttlf\ttlf$\texttt{10}^{\texttt{-2}}$}
& {\ttlf\texttt{-4.0972~$\pmtt$~0.0204}}
& {\ttlf\texttt{0.498}}
\\
\cline{4-4}\cline{5-5}\cline{6-6}\cline{7-7}\cline{8-8}& & & {\ttlf2}
& {\ttlf\ttlf{$\texttt{8.2}$}}
& {\ttlf\ttlf$\texttt{10}^{\texttt{-2}}$}
& {\ttlf\texttt{-4.1937~$\pmtt$~0.0209}}
& {\ttlf\texttt{0.497}}
\\
\cline{4-4}\cline{5-5}\cline{6-6}\cline{7-7}\cline{8-8}& & & {\ttlf3}
& {\ttlf\ttlf{$\texttt{4.5}$}}
& {\ttlf\ttlf$\texttt{10}^{\texttt{-2}}$}
& {\ttlf\texttt{-6.3926~$\pmtt$~0.0305}}
& {\ttlf\texttt{0.477}}
\\
\cline{4-4}\cline{5-5}\cline{6-6}\cline{7-7}\cline{8-8}& & & {\ttlf4}
& {\ttlf\ttlf{$\texttt{8.6}$}}
& {\ttlf\ttlf$\texttt{10}^{\texttt{-2}}$}
& {\ttlf\texttt{-4.6217~$\pmtt$~0.0231}}
& {\ttlf\texttt{0.500}}
\\
\cline{4-4}\cline{5-5}\cline{6-6}\cline{7-7}\cline{8-8}& & & {\ttlf5}
& {\ttlf\ttlf{$\texttt{3.1}$}}
& {\ttlf\ttlf$\texttt{10}^{\texttt{-2}}$}
& {\ttlf\texttt{-5.0834~$\pmtt$~0.0253}}
& {\ttlf\texttt{0.497}}
\\
\cline{4-4}\cline{5-5}\cline{6-6}\cline{7-7}\cline{8-8}& & & {\ttlf6}
& {\ttlf\ttlf{$\texttt{3.1}$}}
& {\ttlf\ttlf$\texttt{10}^{\texttt{-2}}$}
& {\ttlf\texttt{-7.4490~$\pmtt$~0.0361}}
& {\ttlf\texttt{0.484}}
\\
\cline{3-3}\cline{4-4}\cline{5-5}\cline{6-6}\cline{7-7}\cline{8-8}& & \multirow{3}{*}{${\twoloopfloopsnonplanar}$}
& {\ttlf1}
& {\ttlf\ttlf{$\texttt{4.8}$}}
& {\ttlf\ttlf$\texttt{10}^{\texttt{-2}}$}
& {\ttlf\texttt{~3.8228~$\pmtt$~0.0190}}
& {\ttlf\texttt{0.496}}
\\
\cline{4-4}\cline{5-5}\cline{6-6}\cline{7-7}\cline{8-8}& & & {\ttlf2}
& {\ttlf\ttlf{$\texttt{1.1}$}}
& {\ttlf\ttlf$\texttt{10}^{\texttt{-2}}$}
& {\ttlf\texttt{~4.8366~$\pmtt$~0.0238}}
& {\ttlf\texttt{0.492}}
\\
\cline{4-4}\cline{5-5}\cline{6-6}\cline{7-7}\cline{8-8}& & & {\ttlf3}
& {\ttlf\ttlf{$\texttt{1.3}$}}
& {\ttlf\ttlf$\texttt{10}^{\texttt{-2}}$}
& {\ttlf\texttt{~7.3591~$\pmtt$~0.0338}}
& {\ttlf\texttt{0.459}}
\\
\cline{3-3}\cline{4-4}\cline{5-5}\cline{6-6}\cline{7-7}\cline{8-8}& & \multirow{3}{*}{$\twoloopfloopsuv$}
& {\ttlf1}
& {\ttlf\ttlf{$\texttt{3.6}$}}
& {\ttlf\ttlf$\texttt{10}^{\texttt{-4}}$}
& {\ttlf\texttt{~3.7714~$\pmtt$~0.0188}}
& {\ttlf\texttt{0.498}}
\\
\cline{4-4}\cline{5-5}\cline{6-6}\cline{7-7}\cline{8-8}& & & {\ttlf2}
& {\ttlf\ttlf{$\texttt{0.2}$}}
& {\ttlf\ttlf$\texttt{10}^{\texttt{-3}}$}
& {\ttlf\texttt{-7.9544~$\pmtt$~0.0062}}
& {\ttlf\texttt{0.079}}
\\
\cline{4-4}\cline{5-5}\cline{6-6}\cline{7-7}\cline{8-8}& & & {\ttlf3}
& {\ttlf\ttlf{$\texttt{0.2}$}}
& {\ttlf\ttlf$\texttt{10}^{\texttt{-2}}$}
& {\ttlf\texttt{-1.6696~$\pmtt$~0.0011}}
& {\ttlf\texttt{0.067}}
\\
[\doublerulesep]\ccline{2-2}\ccline{3-3}\ccline{4-4}\ccline{5-5}\ccline{6-6}\ccline{7-7}\ccline{8-8}& \multirow{12}{*}{2}
& \multirow{6}{*}{${\twoloopfloopsplanar}$}
& {\ttlf1}
& {\ttlf\ttlf{$\texttt{46.1}$}}
& {\ttlf\ttlf$\texttt{10}^{\texttt{-1}}$}
& {\ttlf\texttt{-5.6166~$\pmtt$~0.0281}}
& {\ttlf\texttt{0.500}}
\\
\cline{4-4}\cline{5-5}\cline{6-6}\cline{7-7}\cline{8-8}& & & {\ttlf2}
& {\ttlf\ttlf{$\texttt{11.8}$}}
& {\ttlf\ttlf$\texttt{10}^{\texttt{-1}}$}
& {\ttlf\texttt{-1.8812~$\pmtt$~0.0094}}
& {\ttlf\texttt{0.500}}
\\
\cline{4-4}\cline{5-5}\cline{6-6}\cline{7-7}\cline{8-8}& & & {\ttlf3}
& {\ttlf\ttlf{$\texttt{14.4}$}}
& {\ttlf\ttlf$\texttt{10}^{\texttt{-1}}$}
& {\ttlf\texttt{-1.3839~$\pmtt$~0.0068}}
& {\ttlf\texttt{0.495}}
\\
\cline{4-4}\cline{5-5}\cline{6-6}\cline{7-7}\cline{8-8}& & & {\ttlf4}
& {\ttlf\ttlf{$\texttt{40.5}$}}
& {\ttlf\ttlf$\texttt{10}^{\texttt{-1}}$}
& {\ttlf\texttt{~6.0761~$\pmtt$~0.0299}}
& {\ttlf\texttt{0.492}}
\\
\cline{4-4}\cline{5-5}\cline{6-6}\cline{7-7}\cline{8-8}& & & {\ttlf5}
& {\ttlf\ttlf{$\texttt{7.8}$}}
& {\ttlf\ttlf$\texttt{10}^{\texttt{-1}}$}
& {\ttlf\texttt{-2.3779~$\pmtt$~0.0115}}
& {\ttlf\texttt{0.483}}
\\
\cline{4-4}\cline{5-5}\cline{6-6}\cline{7-7}\cline{8-8}& & & {\ttlf6}
& {\ttlf\ttlf{$\texttt{3.6}$}}
& {\ttlf\ttlf$\texttt{10}^{\texttt{-1}}$}
& {\ttlf\texttt{-2.2210~$\pmtt$~0.0108}}
& {\ttlf\texttt{0.486}}
\\
\cline{3-3}\cline{4-4}\cline{5-5}\cline{6-6}\cline{7-7}\cline{8-8}& & \multirow{3}{*}{${\twoloopfloopsnonplanar}$}
& {\ttlf1}
& {\ttlf\ttlf{$\texttt{37.8}$}}
& {\ttlf\ttlf$\texttt{10}^{\texttt{-1}}$}
& {\ttlf\texttt{-3.6330~$\pmtt$~0.0181}}
& {\ttlf\texttt{0.497}}
\\
\cline{4-4}\cline{5-5}\cline{6-6}\cline{7-7}\cline{8-8}& & & {\ttlf2}
& {\ttlf\ttlf{$\texttt{3.3}$}}
& {\ttlf\ttlf$\texttt{10}^{\texttt{-1}}$}
& {\ttlf\texttt{~1.8422~$\pmtt$~0.0090}}
& {\ttlf\texttt{0.487}}
\\
\cline{4-4}\cline{5-5}\cline{6-6}\cline{7-7}\cline{8-8}& & & {\ttlf3}
& {\ttlf\ttlf{$\texttt{1.6}$}}
& {\ttlf\ttlf$\texttt{10}^{\texttt{-1}}$}
& {\ttlf\texttt{~1.9205~$\pmtt$~0.0096}}
& {\ttlf\texttt{0.499}}
\\
\cline{3-3}\cline{4-4}\cline{5-5}\cline{6-6}\cline{7-7}\cline{8-8}& & \multirow{3}{*}{$\twoloopfloopsuv$}
& {\ttlf1}
& {\ttlf\ttlf{$\texttt{0.2}$}}
& {\ttlf\ttlf$\texttt{10}^{\texttt{-1}}$}
& {\ttlf\texttt{~1.1028~$\pmtt$~0.0014}}
& {\ttlf\texttt{0.131}}
\\
\cline{4-4}\cline{5-5}\cline{6-6}\cline{7-7}\cline{8-8}& & & {\ttlf2}
& {\ttlf\ttlf{$\texttt{0.2}$}}
& {\ttlf\ttlf$\texttt{10}^{\texttt{-2}}$}
& {\ttlf\texttt{-6.5188~$\pmtt$~0.0061}}
& {\ttlf\texttt{0.093}}
\\
\cline{4-4}\cline{5-5}\cline{6-6}\cline{7-7}\cline{8-8}& & & {\ttlf3}
& {\ttlf\ttlf{$\texttt{0.2}$}}
& {\ttlf\ttlf$\texttt{10}^{\texttt{-2}}$}
& {\ttlf\texttt{-2.3606~$\pmtt$~0.0026}}
& {\ttlf\texttt{0.110}}
\\
[\doublerulesep]\ccline{2-2}\ccline{3-3}\ccline{4-4}\ccline{5-5}\ccline{6-6}\ccline{7-7}\ccline{8-8}& \multirow{12}{*}{3}
& \multirow{6}{*}{${\twoloopfloopsplanar}$}
& {\ttlf1}
& {\ttlf\ttlf{$\texttt{2.5}$}}
& {\ttlf\ttlf$\texttt{10}^{\texttt{-2}}$}
& {\ttlf\texttt{-6.3624~$\pmtt$~0.0311}}
& {\ttlf\texttt{0.489}}
\\
\cline{4-4}\cline{5-5}\cline{6-6}\cline{7-7}\cline{8-8}& & & {\ttlf2}
& {\ttlf\ttlf{$\texttt{4.2}$}}
& {\ttlf\ttlf$\texttt{10}^{\texttt{-2}}$}
& {\ttlf\texttt{-6.9106~$\pmtt$~0.0344}}
& {\ttlf\texttt{0.498}}
\\
\cline{4-4}\cline{5-5}\cline{6-6}\cline{7-7}\cline{8-8}& & & {\ttlf3}
& {\ttlf\ttlf{$\texttt{9.9}$}}
& {\ttlf\ttlf$\texttt{10}^{\texttt{-2}}$}
& {\ttlf\texttt{-3.5816~$\pmtt$~0.0174}}
& {\ttlf\texttt{0.487}}
\\
\cline{4-4}\cline{5-5}\cline{6-6}\cline{7-7}\cline{8-8}& & & {\ttlf4}
& {\ttlf\ttlf{$\texttt{5.2}$}}
& {\ttlf\ttlf$\texttt{10}^{\texttt{-2}}$}
& {\ttlf\texttt{-5.7642~$\pmtt$~0.0280}}
& {\ttlf\texttt{0.485}}
\\
\cline{4-4}\cline{5-5}\cline{6-6}\cline{7-7}\cline{8-8}& & & {\ttlf5}
& {\ttlf\ttlf{$\texttt{16.7}$}}
& {\ttlf\ttlf$\texttt{10}^{\texttt{-2}}$}
& {\ttlf\texttt{-3.3644~$\pmtt$~0.0164}}
& {\ttlf\texttt{0.487}}
\\
\cline{4-4}\cline{5-5}\cline{6-6}\cline{7-7}\cline{8-8}& & & {\ttlf6}
& {\ttlf\ttlf{$\texttt{2.3}$}}
& {\ttlf\ttlf$\texttt{10}^{\texttt{-2}}$}
& {\ttlf\texttt{-7.6106~$\pmtt$~0.0351}}
& {\ttlf\texttt{0.461}}
\\
\cline{3-3}\cline{4-4}\cline{5-5}\cline{6-6}\cline{7-7}\cline{8-8}& & \multirow{3}{*}{${\twoloopfloopsnonplanar}$}
& {\ttlf1}
& {\ttlf\ttlf{$\texttt{1.3}$}}
& {\ttlf\ttlf$\texttt{10}^{\texttt{-2}}$}
& {\ttlf\texttt{~6.6713~$\pmtt$~0.0292}}
& {\ttlf\texttt{0.438}}
\\
\cline{4-4}\cline{5-5}\cline{6-6}\cline{7-7}\cline{8-8}& & & {\ttlf2}
& {\ttlf\ttlf{$\texttt{4.2}$}}
& {\ttlf\ttlf$\texttt{10}^{\texttt{-2}}$}
& {\ttlf\texttt{~4.8345~$\pmtt$~0.0237}}
& {\ttlf\texttt{0.491}}
\\
\cline{4-4}\cline{5-5}\cline{6-6}\cline{7-7}\cline{8-8}& & & {\ttlf3}
& {\ttlf\ttlf{$\texttt{1.1}$}}
& {\ttlf\ttlf$\texttt{10}^{\texttt{-2}}$}
& {\ttlf\texttt{~6.3474~$\pmtt$~0.0287}}
& {\ttlf\texttt{0.452}}
\\
\cline{3-3}\cline{4-4}\cline{5-5}\cline{6-6}\cline{7-7}\cline{8-8}& & \multirow{3}{*}{$\twoloopfloopsuv$}
& {\ttlf1}
& {\ttlf\ttlf{$\texttt{0.2}$}}
& {\ttlf\ttlf$\texttt{10}^{\texttt{-2}}$}
& {\ttlf\texttt{-1.4236~$\pmtt$~0.0010}}
& {\ttlf\texttt{0.068}}
\\
\cline{4-4}\cline{5-5}\cline{6-6}\cline{7-7}\cline{8-8}& & & {\ttlf2}
& {\ttlf\ttlf{$\texttt{1.1}$}}
& {\ttlf\ttlf$\texttt{10}^{\texttt{-4}}$}
& {\ttlf\texttt{-7.8267~$\pmtt$~0.0389}}
& {\ttlf\texttt{0.497}}
\\
\cline{4-4}\cline{5-5}\cline{6-6}\cline{7-7}\cline{8-8}& & & {\ttlf3}
& {\ttlf\ttlf{$\texttt{0.2}$}}
& {\ttlf\ttlf$\texttt{10}^{\texttt{-3}}$}
& {\ttlf\texttt{-9.4734~$\pmtt$~0.0073}}
& {\ttlf\texttt{0.077}}
\\
\cline{4-4}\cline{5-5}\cline{6-6}\cline{7-7}\cline{8-8}\hline\hline
\end{tabular}%
}%
\caption{\label{tab:me_single1}Contributions to the two-loop squared matrix element of $q\bar{q}\to \gamma\gamma\gamma$ via a light-quark $(m=0)$ loop, as defined in section~\ref{sec:individual_me}.}%

\end{table}

\begin{table}[H]
\centering
\renewcommand{\arraystretch}{1.1}
\resizebox{\columnwidth}
{!}{%
\begin{tabular}{llllrrlr}
\hline\hline
Process & PSP & Part & Perm. & $N_p \ [10^8]$ & Exp. & \ \ \ \ \  \ \ \ \ Result & $\Delta \ [\%]$ \\
\hline\hline\multirow{36}{*}{\makecell[l]{$q\bar{q}\to\gamma^*\gamma^*\gamma^*$\\$m=0$}}
& \multirow{12}{*}{1}
& \multirow{6}{*}{${\twoloopfloopsplanar}$}
& {\ttlf1}
& {\ttlf\ttlf{$\texttt{1.2}$}}
& {\ttlf\ttlf$\texttt{10}^{\texttt{-2}}$}
& {\ttlf\texttt{-7.7436~$\pmtt$~0.0165}}
& {\ttlf\texttt{0.213}}
\\
\cline{4-4}\cline{5-5}\cline{6-6}\cline{7-7}\cline{8-8}& & & {\ttlf2}
& {\ttlf\ttlf{$\texttt{6.4}$}}
& {\ttlf\ttlf$\texttt{10}^{\texttt{-4}}$}
& {\ttlf\texttt{~3.9565~$\pmtt$~1.4241}}
& {\ttlf\texttt{{35.994}}}
\\
\cline{4-4}\cline{5-5}\cline{6-6}\cline{7-7}\cline{8-8}& & & {\ttlf3}
& {\ttlf\ttlf{$\texttt{1.7}$}}
& {\ttlf\ttlf$\texttt{10}^{\texttt{-1}}$}
& {\ttlf\texttt{-1.0860~$\pmtt$~0.0043}}
& {\ttlf\texttt{0.398}}
\\
\cline{4-4}\cline{5-5}\cline{6-6}\cline{7-7}\cline{8-8}& & & {\ttlf4}
& {\ttlf\ttlf{$\texttt{1.2}$}}
& {\ttlf\ttlf$\texttt{10}^{\texttt{-2}}$}
& {\ttlf\texttt{-6.4157~$\pmtt$~0.0186}}
& {\ttlf\texttt{0.289}}
\\
\cline{4-4}\cline{5-5}\cline{6-6}\cline{7-7}\cline{8-8}& & & {\ttlf5}
& {\ttlf\ttlf{$\texttt{1.2}$}}
& {\ttlf\ttlf$\texttt{10}^{\texttt{-1}}$}
& {\ttlf\texttt{-1.0801~$\pmtt$~0.0041}}
& {\ttlf\texttt{0.378}}
\\
\cline{4-4}\cline{5-5}\cline{6-6}\cline{7-7}\cline{8-8}& & & {\ttlf6}
& {\ttlf\ttlf{$\texttt{1.2}$}}
& {\ttlf\ttlf$\texttt{10}^{\texttt{-2}}$}
& {\ttlf\texttt{-8.8469~$\pmtt$~0.0351}}
& {\ttlf\texttt{0.396}}
\\
\cline{3-3}\cline{4-4}\cline{5-5}\cline{6-6}\cline{7-7}\cline{8-8}& & \multirow{3}{*}{${\twoloopfloopsnonplanar}$}
& {\ttlf1}
& {\ttlf\ttlf{$\texttt{1.0}$}}
& {\ttlf\ttlf$\texttt{10}^{\texttt{-1}}$}
& {\ttlf\texttt{~1.1219~$\pmtt$~0.0027}}
& {\ttlf\texttt{0.237}}
\\
\cline{4-4}\cline{5-5}\cline{6-6}\cline{7-7}\cline{8-8}& & & {\ttlf2}
& {\ttlf\ttlf{$\texttt{1.0}$}}
& {\ttlf\ttlf$\texttt{10}^{\texttt{-2}}$}
& {\ttlf\texttt{~8.1511~$\pmtt$~0.0336}}
& {\ttlf\texttt{0.412}}
\\
\cline{4-4}\cline{5-5}\cline{6-6}\cline{7-7}\cline{8-8}& & & {\ttlf3}
& {\ttlf\ttlf{$\texttt{1.0}$}}
& {\ttlf\ttlf$\texttt{10}^{\texttt{-1}}$}
& {\ttlf\texttt{~1.6781~$\pmtt$~0.0059}}
& {\ttlf\texttt{0.353}}
\\
\cline{3-3}\cline{4-4}\cline{5-5}\cline{6-6}\cline{7-7}\cline{8-8}& & \multirow{3}{*}{$\twoloopfloopsuv$}
& {\ttlf1}
& {\ttlf\ttlf{$\texttt{1.0}$}}
& {\ttlf\ttlf$\texttt{10}^{\texttt{-3}}$}
& {\ttlf\texttt{-7.1181~$\pmtt$~0.0035}}
& {\ttlf\texttt{0.049}}
\\
\cline{4-4}\cline{5-5}\cline{6-6}\cline{7-7}\cline{8-8}& & & {\ttlf2}
& {\ttlf\ttlf{$\texttt{1.0}$}}
& {\ttlf\ttlf$\texttt{10}^{\texttt{-2}}$}
& {\ttlf\texttt{-1.2854~$\pmtt$~0.0003}}
& {\ttlf\texttt{0.024}}
\\
\cline{4-4}\cline{5-5}\cline{6-6}\cline{7-7}\cline{8-8}& & & {\ttlf3}
& {\ttlf\ttlf{$\texttt{1.0}$}}
& {\ttlf\ttlf$\texttt{10}^{\texttt{-2}}$}
& {\ttlf\texttt{-2.9185~$\pmtt$~0.0005}}
& {\ttlf\texttt{0.016}}
\\
[\doublerulesep]\ccline{2-2}\ccline{3-3}\ccline{4-4}\ccline{5-5}\ccline{6-6}\ccline{7-7}\ccline{8-8}& \multirow{12}{*}{2}
& \multirow{6}{*}{${\twoloopfloopsplanar}$}
& {\ttlf1}
& {\ttlf\ttlf{$\texttt{7.2}$}}
& {\ttlf\ttlf$\texttt{10}^{\texttt{-2}}$}
& {\ttlf\texttt{~4.9402~$\pmtt$~0.0203}}
& {\ttlf\texttt{0.411}}
\\
\cline{4-4}\cline{5-5}\cline{6-6}\cline{7-7}\cline{8-8}& & & {\ttlf2}
& {\ttlf\ttlf{$\texttt{12.0}$}}
& {\ttlf\ttlf$\texttt{10}^{\texttt{-1}}$}
& {\ttlf\texttt{-2.1192~$\pmtt$~0.0151}}
& {\ttlf\texttt{0.713}}
\\
\cline{4-4}\cline{5-5}\cline{6-6}\cline{7-7}\cline{8-8}& & & {\ttlf3}
& {\ttlf\ttlf{$\texttt{7.2}$}}
& {\ttlf\ttlf$\texttt{10}^{\texttt{-1}}$}
& {\ttlf\texttt{~1.5533~$\pmtt$~0.0061}}
& {\ttlf\texttt{0.394}}
\\
\cline{4-4}\cline{5-5}\cline{6-6}\cline{7-7}\cline{8-8}& & & {\ttlf4}
& {\ttlf\ttlf{$\texttt{3.2}$}}
& {\ttlf\ttlf$\texttt{10}^{\texttt{-1}}$}
& {\ttlf\texttt{-3.9710~$\pmtt$~0.0031}}
& {\ttlf\texttt{0.079}}
\\
\cline{4-4}\cline{5-5}\cline{6-6}\cline{7-7}\cline{8-8}& & & {\ttlf5}
& {\ttlf\ttlf{$\texttt{4.5}$}}
& {\ttlf\ttlf$\texttt{10}^{\texttt{-1}}$}
& {\ttlf\texttt{-1.7658~$\pmtt$~0.0057}}
& {\ttlf\texttt{0.325}}
\\
\cline{4-4}\cline{5-5}\cline{6-6}\cline{7-7}\cline{8-8}& & & {\ttlf6}
& {\ttlf\ttlf{$\texttt{7.5}$}}
& {\ttlf\ttlf$\texttt{10}^{\texttt{-1}}$}
& {\ttlf\texttt{-3.9478~$\pmtt$~0.0075}}
& {\ttlf\texttt{0.191}}
\\
\cline{3-3}\cline{4-4}\cline{5-5}\cline{6-6}\cline{7-7}\cline{8-8}& & \multirow{3}{*}{${\twoloopfloopsnonplanar}$}
& {\ttlf1}
& {\ttlf\ttlf{$\texttt{2.5}$}}
& {\ttlf\ttlf$\texttt{10}^{\texttt{-1}}$}
& {\ttlf\texttt{~2.5209~$\pmtt$~0.0058}}
& {\ttlf\texttt{0.232}}
\\
\cline{4-4}\cline{5-5}\cline{6-6}\cline{7-7}\cline{8-8}& & & {\ttlf2}
& {\ttlf\ttlf{$\texttt{6.9}$}}
& {\ttlf\ttlf$\texttt{10}^{\texttt{-1}}$}
& {\ttlf\texttt{~3.6117~$\pmtt$~0.0102}}
& {\ttlf\texttt{0.283}}
\\
\cline{4-4}\cline{5-5}\cline{6-6}\cline{7-7}\cline{8-8}& & & {\ttlf3}
& {\ttlf\ttlf{$\texttt{3.6}$}}
& {\ttlf\ttlf$\texttt{10}^{\texttt{-1}}$}
& {\ttlf\texttt{~1.7200~$\pmtt$~0.0084}}
& {\ttlf\texttt{0.488}}
\\
\cline{3-3}\cline{4-4}\cline{5-5}\cline{6-6}\cline{7-7}\cline{8-8}& & \multirow{3}{*}{$\twoloopfloopsuv$}
& {\ttlf1}
& {\ttlf\ttlf{$\texttt{1.0}$}}
& {\ttlf\ttlf$\texttt{10}^{\texttt{-3}}$}
& {\ttlf\texttt{-5.1845~$\pmtt$~0.0110}}
& {\ttlf\texttt{0.213}}
\\
\cline{4-4}\cline{5-5}\cline{6-6}\cline{7-7}\cline{8-8}& & & {\ttlf2}
& {\ttlf\ttlf{$\texttt{1.0}$}}
& {\ttlf\ttlf$\texttt{10}^{\texttt{-2}}$}
& {\ttlf\texttt{-6.6730~$\pmtt$~0.0014}}
& {\ttlf\texttt{0.021}}
\\
\cline{4-4}\cline{5-5}\cline{6-6}\cline{7-7}\cline{8-8}& & & {\ttlf3}
& {\ttlf\ttlf{$\texttt{1.0}$}}
& {\ttlf\ttlf$\texttt{10}^{\texttt{-2}}$}
& {\ttlf\texttt{-3.0323~$\pmtt$~0.0008}}
& {\ttlf\texttt{0.026}}
\\
[\doublerulesep]\ccline{2-2}\ccline{3-3}\ccline{4-4}\ccline{5-5}\ccline{6-6}\ccline{7-7}\ccline{8-8}& \multirow{12}{*}{3}
& \multirow{6}{*}{${\twoloopfloopsplanar}$}
& {\ttlf1}
& {\ttlf\ttlf{$\texttt{1.1}$}}
& {\ttlf\ttlf$\texttt{10}^{\texttt{-2}}$}
& {\ttlf\texttt{-7.4930~$\pmtt$~0.0264}}
& {\ttlf\texttt{0.352}}
\\
\cline{4-4}\cline{5-5}\cline{6-6}\cline{7-7}\cline{8-8}& & & {\ttlf2}
& {\ttlf\ttlf{$\texttt{3.5}$}}
& {\ttlf\ttlf$\texttt{10}^{\texttt{-1}}$}
& {\ttlf\texttt{-1.1061~$\pmtt$~0.0046}}
& {\ttlf\texttt{0.414}}
\\
\cline{4-4}\cline{5-5}\cline{6-6}\cline{7-7}\cline{8-8}& & & {\ttlf3}
& {\ttlf\ttlf{$\texttt{4.5}$}}
& {\ttlf\ttlf$\texttt{10}^{\texttt{-2}}$}
& {\ttlf\texttt{-1.5962~$\pmtt$~0.0101}}
& {\ttlf\texttt{0.632}}
\\
\cline{4-4}\cline{5-5}\cline{6-6}\cline{7-7}\cline{8-8}& & & {\ttlf4}
& {\ttlf\ttlf{$\texttt{1.6}$}}
& {\ttlf\ttlf$\texttt{10}^{\texttt{-2}}$}
& {\ttlf\texttt{-8.1118~$\pmtt$~0.0321}}
& {\ttlf\texttt{0.395}}
\\
\cline{4-4}\cline{5-5}\cline{6-6}\cline{7-7}\cline{8-8}& & & {\ttlf5}
& {\ttlf\ttlf{$\texttt{4.5}$}}
& {\ttlf\ttlf$\texttt{10}^{\texttt{-2}}$}
& {\ttlf\texttt{-4.5936~$\pmtt$~0.0340}}
& {\ttlf\texttt{0.739}}
\\
\cline{4-4}\cline{5-5}\cline{6-6}\cline{7-7}\cline{8-8}& & & {\ttlf6}
& {\ttlf\ttlf{$\texttt{1.1}$}}
& {\ttlf\ttlf$\texttt{10}^{\texttt{-1}}$}
& {\ttlf\texttt{-1.0366~$\pmtt$~0.0021}}
& {\ttlf\texttt{0.204}}
\\
\cline{3-3}\cline{4-4}\cline{5-5}\cline{6-6}\cline{7-7}\cline{8-8}& & \multirow{3}{*}{${\twoloopfloopsnonplanar}$}
& {\ttlf1}
& {\ttlf\ttlf{$\texttt{1.0}$}}
& {\ttlf\ttlf$\texttt{10}^{\texttt{-1}}$}
& {\ttlf\texttt{~1.2999~$\pmtt$~0.0043}}
& {\ttlf\texttt{0.333}}
\\
\cline{4-4}\cline{5-5}\cline{6-6}\cline{7-7}\cline{8-8}& & & {\ttlf2}
& {\ttlf\ttlf{$\texttt{1.0}$}}
& {\ttlf\ttlf$\texttt{10}^{\texttt{-1}}$}
& {\ttlf\texttt{~1.2591~$\pmtt$~0.0046}}
& {\ttlf\texttt{0.363}}
\\
\cline{4-4}\cline{5-5}\cline{6-6}\cline{7-7}\cline{8-8}& & & {\ttlf3}
& {\ttlf\ttlf{$\texttt{1.0}$}}
& {\ttlf\ttlf$\texttt{10}^{\texttt{-2}}$}
& {\ttlf\texttt{~9.1649~$\pmtt$~0.0357}}
& {\ttlf\texttt{0.390}}
\\
\cline{3-3}\cline{4-4}\cline{5-5}\cline{6-6}\cline{7-7}\cline{8-8}& & \multirow{3}{*}{$\twoloopfloopsuv$}
& {\ttlf1}
& {\ttlf\ttlf{$\texttt{1.0}$}}
& {\ttlf\ttlf$\texttt{10}^{\texttt{-2}}$}
& {\ttlf\texttt{-2.3057~$\pmtt$~0.0004}}
& {\ttlf\texttt{0.017}}
\\
\cline{4-4}\cline{5-5}\cline{6-6}\cline{7-7}\cline{8-8}& & & {\ttlf2}
& {\ttlf\ttlf{$\texttt{1.0}$}}
& {\ttlf\ttlf$\texttt{10}^{\texttt{-2}}$}
& {\ttlf\texttt{-1.0042~$\pmtt$~0.0004}}
& {\ttlf\texttt{0.044}}
\\
\cline{4-4}\cline{5-5}\cline{6-6}\cline{7-7}\cline{8-8}& & & {\ttlf3}
& {\ttlf\ttlf{$\texttt{1.0}$}}
& {\ttlf\ttlf$\texttt{10}^{\texttt{-2}}$}
& {\ttlf\texttt{-1.5832~$\pmtt$~0.0003}}
& {\ttlf\texttt{0.018}}
\\
\cline{4-4}\cline{5-5}\cline{6-6}\cline{7-7}\cline{8-8}\hline\hline
\end{tabular}%
}%
\caption{\label{tab:me_single2}Contributions to the two-loop squared matrix element of $q\bar{q}\to \gamma^*\gamma^*\gamma^*$ via a light-quark $(m=0)$ loop, as defined in section~\ref{sec:individual_me}.}%

\end{table}

\begin{table}[H]
\centering
\renewcommand{\arraystretch}{1.1}
\resizebox{\columnwidth}
{!}{%
\begin{tabular}{llllrrlr}
\hline\hline
Process & PSP & Part & Perm. & $N_p \ [10^8]$ & Exp. & \ \ \ \ \  \ \ \ \ Result & $\Delta \ [\%]$ \\
\hline\hline\multirow{12}{*}{\makecell[l]{$q\bar{q}\to\gamma\gamma$\\$m>0$}}
& \multirow{4}{*}{1}
& \multirow{2}{*}{${\twoloopfloopsplanar}$}
& {\ttlf1}
& {\ttlf\ttlf{$\texttt{158.9}$}}
& {\ttlf\ttlf$\texttt{10}^{\texttt{+1}}$}
& {\ttlf\texttt{-6.4795~$\pmtt$~0.0006}}
& {\ttlf\texttt{0.010}}
\\
\cline{4-4}\cline{5-5}\cline{6-6}\cline{7-7}\cline{8-8}& & & {\ttlf2}
& {\ttlf\ttlf{$\texttt{160.7}$}}
& {\ttlf\ttlf$\texttt{10}^{\texttt{+1}}$}
& {\ttlf\texttt{-6.4866~$\pmtt$~0.0006}}
& {\ttlf\texttt{0.010}}
\\
\cline{3-3}\cline{4-4}\cline{5-5}\cline{6-6}\cline{7-7}\cline{8-8}& & \multirow{1}{*}{${\twoloopfloopsnonplanar}$}
& {\ttlf1}
& {\ttlf\ttlf{$\texttt{102.5}$}}
& {\ttlf\ttlf$\texttt{10}^{\texttt{+2}}$}
& {\ttlf\texttt{~1.5316~$\pmtt$~0.0002}}
& {\ttlf\texttt{0.010}}
\\
\cline{3-3}\cline{4-4}\cline{5-5}\cline{6-6}\cline{7-7}\cline{8-8}& & \multirow{1}{*}{$\twoloopfloopsuv$}
& {\ttlf1}
& {\ttlf\ttlf{$\texttt{4.5}$}}
& {\ttlf\ttlf$\texttt{10}^{\texttt{+1}}$}
& {\ttlf\texttt{-4.2919~$\pmtt$~0.0004}}
& {\ttlf\texttt{0.010}}
\\
[\doublerulesep]\ccline{2-2}\ccline{3-3}\ccline{4-4}\ccline{5-5}\ccline{6-6}\ccline{7-7}\ccline{8-8}& \multirow{4}{*}{2}
& \multirow{2}{*}{${\twoloopfloopsplanar}$}
& {\ttlf1}
& {\ttlf\ttlf{$\texttt{74.6}$}}
& {\ttlf\ttlf$\texttt{10}^{\texttt{+2}}$}
& {\ttlf\texttt{-2.0226~$\pmtt$~0.0002}}
& {\ttlf\texttt{0.010}}
\\
\cline{4-4}\cline{5-5}\cline{6-6}\cline{7-7}\cline{8-8}& & & {\ttlf2}
& {\ttlf\ttlf{$\texttt{72.2}$}}
& {\ttlf\ttlf$\texttt{10}^{\texttt{+2}}$}
& {\ttlf\texttt{-1.9763~$\pmtt$~0.0002}}
& {\ttlf\texttt{0.010}}
\\
\cline{3-3}\cline{4-4}\cline{5-5}\cline{6-6}\cline{7-7}\cline{8-8}& & \multirow{1}{*}{${\twoloopfloopsnonplanar}$}
& {\ttlf1}
& {\ttlf\ttlf{$\texttt{57.3}$}}
& {\ttlf\ttlf$\texttt{10}^{\texttt{+2}}$}
& {\ttlf\texttt{~3.9912~$\pmtt$~0.0004}}
& {\ttlf\texttt{0.010}}
\\
\cline{3-3}\cline{4-4}\cline{5-5}\cline{6-6}\cline{7-7}\cline{8-8}& & \multirow{1}{*}{$\twoloopfloopsuv$}
& {\ttlf1}
& {\ttlf\ttlf{$\texttt{8.6}$}}
& {\ttlf\ttlf$\texttt{10}^{\texttt{+1}}$}
& {\ttlf\texttt{-4.8301~$\pmtt$~0.0005}}
& {\ttlf\texttt{0.010}}
\\
[\doublerulesep]\ccline{2-2}\ccline{3-3}\ccline{4-4}\ccline{5-5}\ccline{6-6}\ccline{7-7}\ccline{8-8}& \multirow{4}{*}{3}
& \multirow{2}{*}{${\twoloopfloopsplanar}$}
& {\ttlf1}
& {\ttlf\ttlf{$\texttt{2.8}$}}
& {\ttlf\ttlf$\texttt{10}^{\texttt{+2}}$}
& {\ttlf\texttt{-3.7844~$\pmtt$~0.0018}}
& {\ttlf\texttt{0.047}}
\\
\cline{4-4}\cline{5-5}\cline{6-6}\cline{7-7}\cline{8-8}& & & {\ttlf2}
& {\ttlf\ttlf{$\texttt{2.5}$}}
& {\ttlf\ttlf$\texttt{10}^{\texttt{+2}}$}
& {\ttlf\texttt{-4.0130~$\pmtt$~0.0019}}
& {\ttlf\texttt{0.048}}
\\
\cline{3-3}\cline{4-4}\cline{5-5}\cline{6-6}\cline{7-7}\cline{8-8}& & \multirow{1}{*}{${\twoloopfloopsnonplanar}$}
& {\ttlf1}
& {\ttlf\ttlf{$\texttt{1.4}$}}
& {\ttlf\ttlf$\texttt{10}^{\texttt{+2}}$}
& {\ttlf\texttt{~6.6582~$\pmtt$~0.0027}}
& {\ttlf\texttt{0.040}}
\\
\cline{3-3}\cline{4-4}\cline{5-5}\cline{6-6}\cline{7-7}\cline{8-8}& & \multirow{1}{*}{$\twoloopfloopsuv$}
& {\ttlf1}
& {\ttlf\ttlf{$\texttt{1.4}$}}
& {\ttlf\ttlf$\texttt{10}^{\texttt{+1}}$}
& {\ttlf\texttt{-5.6474~$\pmtt$~0.0019}}
& {\ttlf\texttt{0.034}}
\\
[\doublerulesep]\ccline{1-1}\ccline{2-2}\ccline{3-3}\ccline{4-4}\ccline{5-5}\ccline{6-6}\ccline{7-7}\ccline{8-8}\multirow{12}{*}{\makecell[l]{$q\bar{q}\to\gamma^*\gamma^*$\\$m>0$}}
& \multirow{4}{*}{1}
& \multirow{2}{*}{${\twoloopfloopsplanar}$}
& {\ttlf1}
& {\ttlf\ttlf{$\texttt{509.9}$}}
& {\ttlf\ttlf$\texttt{10}^{\texttt{+2}}$}
& {\ttlf\texttt{-1.4912~$\pmtt$~0.0001}}
& {\ttlf\texttt{0.005}}
\\
\cline{4-4}\cline{5-5}\cline{6-6}\cline{7-7}\cline{8-8}& & & {\ttlf2}
& {\ttlf\ttlf{$\texttt{494.1}$}}
& {\ttlf\ttlf$\texttt{10}^{\texttt{+2}}$}
& {\ttlf\texttt{-1.5267~$\pmtt$~0.0001}}
& {\ttlf\texttt{0.005}}
\\
\cline{3-3}\cline{4-4}\cline{5-5}\cline{6-6}\cline{7-7}\cline{8-8}& & \multirow{1}{*}{${\twoloopfloopsnonplanar}$}
& {\ttlf1}
& {\ttlf\ttlf{$\texttt{238.3}$}}
& {\ttlf\ttlf$\texttt{10}^{\texttt{+2}}$}
& {\ttlf\texttt{~3.4722~$\pmtt$~0.0002}}
& {\ttlf\texttt{0.005}}
\\
\cline{3-3}\cline{4-4}\cline{5-5}\cline{6-6}\cline{7-7}\cline{8-8}& & \multirow{1}{*}{$\twoloopfloopsuv$}
& {\ttlf1}
& {\ttlf\ttlf{$\texttt{407.1}$}}
& {\ttlf\ttlf$\texttt{10}^{\texttt{+1}}$}
& {\ttlf\texttt{-8.6782~$\pmtt$~0.0001}}
& {\ttlf\texttt{0.001}}
\\
[\doublerulesep]\ccline{2-2}\ccline{3-3}\ccline{4-4}\ccline{5-5}\ccline{6-6}\ccline{7-7}\ccline{8-8}& \multirow{4}{*}{2}
& \multirow{2}{*}{${\twoloopfloopsplanar}$}
& {\ttlf1}
& {\ttlf\ttlf{$\texttt{98.2}$}}
& {\ttlf\ttlf$\texttt{10}^{\texttt{+2}}$}
& {\ttlf\texttt{-4.0071~$\pmtt$~0.0004}}
& {\ttlf\texttt{0.010}}
\\
\cline{4-4}\cline{5-5}\cline{6-6}\cline{7-7}\cline{8-8}& & & {\ttlf2}
& {\ttlf\ttlf{$\texttt{99.7}$}}
& {\ttlf\ttlf$\texttt{10}^{\texttt{+2}}$}
& {\ttlf\texttt{-3.4199~$\pmtt$~0.0003}}
& {\ttlf\texttt{0.010}}
\\
\cline{3-3}\cline{4-4}\cline{5-5}\cline{6-6}\cline{7-7}\cline{8-8}& & \multirow{1}{*}{${\twoloopfloopsnonplanar}$}
& {\ttlf1}
& {\ttlf\ttlf{$\texttt{40.5}$}}
& {\ttlf\ttlf$\texttt{10}^{\texttt{+2}}$}
& {\ttlf\texttt{~7.5923~$\pmtt$~0.0008}}
& {\ttlf\texttt{0.010}}
\\
\cline{3-3}\cline{4-4}\cline{5-5}\cline{6-6}\cline{7-7}\cline{8-8}& & \multirow{1}{*}{$\twoloopfloopsuv$}
& {\ttlf1}
& {\ttlf\ttlf{$\texttt{10.8}$}}
& {\ttlf\ttlf$\texttt{10}^{\texttt{+1}}$}
& {\ttlf\texttt{-8.7016~$\pmtt$~0.0009}}
& {\ttlf\texttt{0.010}}
\\
[\doublerulesep]\ccline{2-2}\ccline{3-3}\ccline{4-4}\ccline{5-5}\ccline{6-6}\ccline{7-7}\ccline{8-8}& \multirow{4}{*}{3}
& \multirow{2}{*}{${\twoloopfloopsplanar}$}
& {\ttlf1}
& {\ttlf\ttlf{$\texttt{5.8}$}}
& {\ttlf\ttlf$\texttt{10}^{\texttt{+2}}$}
& {\ttlf\texttt{-4.9630~$\pmtt$~0.0025}}
& {\ttlf\texttt{0.050}}
\\
\cline{4-4}\cline{5-5}\cline{6-6}\cline{7-7}\cline{8-8}& & & {\ttlf2}
& {\ttlf\ttlf{$\texttt{4.8}$}}
& {\ttlf\ttlf$\texttt{10}^{\texttt{+2}}$}
& {\ttlf\texttt{-6.9926~$\pmtt$~0.0034}}
& {\ttlf\texttt{0.049}}
\\
\cline{3-3}\cline{4-4}\cline{5-5}\cline{6-6}\cline{7-7}\cline{8-8}& & \multirow{1}{*}{${\twoloopfloopsnonplanar}$}
& {\ttlf1}
& {\ttlf\ttlf{$\texttt{1.4}$}}
& {\ttlf\ttlf$\texttt{10}^{\texttt{+3}}$}
& {\ttlf\texttt{~1.0919~$\pmtt$~0.0004}}
& {\ttlf\texttt{0.038}}
\\
\cline{3-3}\cline{4-4}\cline{5-5}\cline{6-6}\cline{7-7}\cline{8-8}& & \multirow{1}{*}{$\twoloopfloopsuv$}
& {\ttlf1}
& {\ttlf\ttlf{$\texttt{1.4}$}}
& {\ttlf\ttlf$\texttt{10}^{\texttt{+1}}$}
& {\ttlf\texttt{-8.7440~$\pmtt$~0.0033}}
& {\ttlf\texttt{0.038}}
\\
\cline{4-4}\cline{5-5}\cline{6-6}\cline{7-7}\cline{8-8}\hline\hline
\end{tabular}%
}%
\caption{\label{tab:me_single3}Contributions to the two-loop squared matrix element of $q\bar{q}\to \gamma\gamma$ and $q\bar{q}\to \gamma^*\gamma^*$ via a heavy-quark $(m>0)$ loop, as defined in section~\ref{sec:individual_me}.}%

\end{table}

\begin{table}[H]
\centering
\renewcommand{\arraystretch}{1.1}
\resizebox{\columnwidth}
{!}{%
\begin{tabular}{llllrrlr}
\hline\hline
Process & PSP & Part & Perm. & $N_p \ [10^8]$ & Exp. & \ \ \ \ \  \ \ \ \ Result & $\Delta \ [\%]$ \\
\hline\hline\multirow{36}{*}{\makecell[l]{$q\bar{q}\to\gamma\gamma\gamma$\\$m>0$}}
& \multirow{12}{*}{1}
& \multirow{6}{*}{${\twoloopfloopsplanar}$}
& {\ttlf1}
& {\ttlf\ttlf{$\texttt{25.8}$}}
& {\ttlf\ttlf$\texttt{10}^{\texttt{-3}}$}
& {\ttlf\texttt{-5.3742~$\pmtt$~0.0159}}
& {\ttlf\texttt{0.297}}
\\
\cline{4-4}\cline{5-5}\cline{6-6}\cline{7-7}\cline{8-8}& & & {\ttlf2}
& {\ttlf\ttlf{$\texttt{1.3}$}}
& {\ttlf\ttlf$\texttt{10}^{\texttt{-2}}$}
& {\ttlf\texttt{-4.9746~$\pmtt$~0.0055}}
& {\ttlf\texttt{0.110}}
\\
\cline{4-4}\cline{5-5}\cline{6-6}\cline{7-7}\cline{8-8}& & & {\ttlf3}
& {\ttlf\ttlf{$\texttt{1.3}$}}
& {\ttlf\ttlf$\texttt{10}^{\texttt{-2}}$}
& {\ttlf\texttt{-9.2242~$\pmtt$~0.0135}}
& {\ttlf\texttt{0.147}}
\\
\cline{4-4}\cline{5-5}\cline{6-6}\cline{7-7}\cline{8-8}& & & {\ttlf4}
& {\ttlf\ttlf{$\texttt{1.9}$}}
& {\ttlf\ttlf$\texttt{10}^{\texttt{-2}}$}
& {\ttlf\texttt{-2.1012~$\pmtt$~0.0062}}
& {\ttlf\texttt{0.296}}
\\
\cline{4-4}\cline{5-5}\cline{6-6}\cline{7-7}\cline{8-8}& & & {\ttlf5}
& {\ttlf\ttlf{$\texttt{1.3}$}}
& {\ttlf\ttlf$\texttt{10}^{\texttt{-2}}$}
& {\ttlf\texttt{-5.4950~$\pmtt$~0.0055}}
& {\ttlf\texttt{0.101}}
\\
\cline{4-4}\cline{5-5}\cline{6-6}\cline{7-7}\cline{8-8}& & & {\ttlf6}
& {\ttlf\ttlf{$\texttt{1.3}$}}
& {\ttlf\ttlf$\texttt{10}^{\texttt{-2}}$}
& {\ttlf\texttt{-9.2726~$\pmtt$~0.0124}}
& {\ttlf\texttt{0.133}}
\\
\cline{3-3}\cline{4-4}\cline{5-5}\cline{6-6}\cline{7-7}\cline{8-8}& & \multirow{3}{*}{${\twoloopfloopsnonplanar}$}
& {\ttlf1}
& {\ttlf\ttlf{$\texttt{1.4}$}}
& {\ttlf\ttlf$\texttt{10}^{\texttt{-2}}$}
& {\ttlf\texttt{~3.3946~$\pmtt$~0.0094}}
& {\ttlf\texttt{0.277}}
\\
\cline{4-4}\cline{5-5}\cline{6-6}\cline{7-7}\cline{8-8}& & & {\ttlf2}
& {\ttlf\ttlf{$\texttt{1.3}$}}
& {\ttlf\ttlf$\texttt{10}^{\texttt{-2}}$}
& {\ttlf\texttt{~8.8332~$\pmtt$~0.0074}}
& {\ttlf\texttt{0.083}}
\\
\cline{4-4}\cline{5-5}\cline{6-6}\cline{7-7}\cline{8-8}& & & {\ttlf3}
& {\ttlf\ttlf{$\texttt{1.3}$}}
& {\ttlf\ttlf$\texttt{10}^{\texttt{-1}}$}
& {\ttlf\texttt{~1.7806~$\pmtt$~0.0012}}
& {\ttlf\texttt{0.067}}
\\
\cline{3-3}\cline{4-4}\cline{5-5}\cline{6-6}\cline{7-7}\cline{8-8}& & \multirow{3}{*}{$\twoloopfloopsuv$}
& {\ttlf1}
& {\ttlf\ttlf{$\texttt{10.8}$}}
& {\ttlf\ttlf$\texttt{10}^{\texttt{-4}}$}
& {\ttlf\texttt{~3.7726~$\pmtt$~0.0112}}
& {\ttlf\texttt{0.297}}
\\
\cline{4-4}\cline{5-5}\cline{6-6}\cline{7-7}\cline{8-8}& & & {\ttlf2}
& {\ttlf\ttlf{$\texttt{1.3}$}}
& {\ttlf\ttlf$\texttt{10}^{\texttt{-3}}$}
& {\ttlf\texttt{-7.9525~$\pmtt$~0.0023}}
& {\ttlf\texttt{0.029}}
\\
\cline{4-4}\cline{5-5}\cline{6-6}\cline{7-7}\cline{8-8}& & & {\ttlf3}
& {\ttlf\ttlf{$\texttt{1.3}$}}
& {\ttlf\ttlf$\texttt{10}^{\texttt{-2}}$}
& {\ttlf\texttt{-1.6705~$\pmtt$~0.0004}}
& {\ttlf\texttt{0.025}}
\\
[\doublerulesep]\ccline{2-2}\ccline{3-3}\ccline{4-4}\ccline{5-5}\ccline{6-6}\ccline{7-7}\ccline{8-8}& \multirow{12}{*}{2}
& \multirow{6}{*}{${\twoloopfloopsplanar}$}
& {\ttlf1}
& {\ttlf\ttlf{$\texttt{45.1}$}}
& {\ttlf\ttlf$\texttt{10}^{\texttt{-1}}$}
& {\ttlf\texttt{-1.5662~$\pmtt$~0.0031}}
& {\ttlf\texttt{0.199}}
\\
\cline{4-4}\cline{5-5}\cline{6-6}\cline{7-7}\cline{8-8}& & & {\ttlf2}
& {\ttlf\ttlf{$\texttt{3.3}$}}
& {\ttlf\ttlf$\texttt{10}^{\texttt{-1}}$}
& {\ttlf\texttt{-3.0457~$\pmtt$~0.0060}}
& {\ttlf\texttt{0.195}}
\\
\cline{4-4}\cline{5-5}\cline{6-6}\cline{7-7}\cline{8-8}& & & {\ttlf3}
& {\ttlf\ttlf{$\texttt{1.4}$}}
& {\ttlf\ttlf$\texttt{10}^{\texttt{-1}}$}
& {\ttlf\texttt{-1.7626~$\pmtt$~0.0031}}
& {\ttlf\texttt{0.178}}
\\
\cline{4-4}\cline{5-5}\cline{6-6}\cline{7-7}\cline{8-8}& & & {\ttlf4}
& {\ttlf\ttlf{$\texttt{8.6}$}}
& {\ttlf\ttlf$\texttt{10}^{\texttt{-1}}$}
& {\ttlf\texttt{~3.7794~$\pmtt$~0.0075}}
& {\ttlf\texttt{0.199}}
\\
\cline{4-4}\cline{5-5}\cline{6-6}\cline{7-7}\cline{8-8}& & & {\ttlf5}
& {\ttlf\ttlf{$\texttt{3.3}$}}
& {\ttlf\ttlf$\texttt{10}^{\texttt{-1}}$}
& {\ttlf\texttt{-2.7911~$\pmtt$~0.0055}}
& {\ttlf\texttt{0.195}}
\\
\cline{4-4}\cline{5-5}\cline{6-6}\cline{7-7}\cline{8-8}& & & {\ttlf6}
& {\ttlf\ttlf{$\texttt{1.4}$}}
& {\ttlf\ttlf$\texttt{10}^{\texttt{-1}}$}
& {\ttlf\texttt{-2.4841~$\pmtt$~0.0042}}
& {\ttlf\texttt{0.169}}
\\
\cline{3-3}\cline{4-4}\cline{5-5}\cline{6-6}\cline{7-7}\cline{8-8}& & \multirow{3}{*}{${\twoloopfloopsnonplanar}$}
& {\ttlf1}
& {\ttlf\ttlf{$\texttt{28.1}$}}
& {\ttlf\ttlf$\texttt{10}^{\texttt{-1}}$}
& {\ttlf\texttt{-2.6376~$\pmtt$~0.0052}}
& {\ttlf\texttt{0.199}}
\\
\cline{4-4}\cline{5-5}\cline{6-6}\cline{7-7}\cline{8-8}& & & {\ttlf2}
& {\ttlf\ttlf{$\texttt{1.4}$}}
& {\ttlf\ttlf$\texttt{10}^{\texttt{-1}}$}
& {\ttlf\texttt{~4.8343~$\pmtt$~0.0068}}
& {\ttlf\texttt{0.140}}
\\
\cline{4-4}\cline{5-5}\cline{6-6}\cline{7-7}\cline{8-8}& & & {\ttlf3}
& {\ttlf\ttlf{$\texttt{1.4}$}}
& {\ttlf\ttlf$\texttt{10}^{\texttt{-1}}$}
& {\ttlf\texttt{~3.7390~$\pmtt$~0.0034}}
& {\ttlf\texttt{0.090}}
\\
\cline{3-3}\cline{4-4}\cline{5-5}\cline{6-6}\cline{7-7}\cline{8-8}& & \multirow{3}{*}{$\twoloopfloopsuv$}
& {\ttlf1}
& {\ttlf\ttlf{$\texttt{1.4}$}}
& {\ttlf\ttlf$\texttt{10}^{\texttt{-1}}$}
& {\ttlf\texttt{~1.1025~$\pmtt$~0.0004}}
& {\ttlf\texttt{0.033}}
\\
\cline{4-4}\cline{5-5}\cline{6-6}\cline{7-7}\cline{8-8}& & & {\ttlf2}
& {\ttlf\ttlf{$\texttt{1.4}$}}
& {\ttlf\ttlf$\texttt{10}^{\texttt{-2}}$}
& {\ttlf\texttt{-6.5237~$\pmtt$~0.0019}}
& {\ttlf\texttt{0.030}}
\\
\cline{4-4}\cline{5-5}\cline{6-6}\cline{7-7}\cline{8-8}& & & {\ttlf3}
& {\ttlf\ttlf{$\texttt{1.4}$}}
& {\ttlf\ttlf$\texttt{10}^{\texttt{-2}}$}
& {\ttlf\texttt{-2.3608~$\pmtt$~0.0009}}
& {\ttlf\texttt{0.037}}
\\
[\doublerulesep]\ccline{2-2}\ccline{3-3}\ccline{4-4}\ccline{5-5}\ccline{6-6}\ccline{7-7}\ccline{8-8}& \multirow{12}{*}{3}
& \multirow{6}{*}{${\twoloopfloopsplanar}$}
& {\ttlf1}
& {\ttlf\ttlf{$\texttt{1.3}$}}
& {\ttlf\ttlf$\texttt{10}^{\texttt{-2}}$}
& {\ttlf\texttt{-6.2694~$\pmtt$~0.0054}}
& {\ttlf\texttt{0.086}}
\\
\cline{4-4}\cline{5-5}\cline{6-6}\cline{7-7}\cline{8-8}& & & {\ttlf2}
& {\ttlf\ttlf{$\texttt{21.0}$}}
& {\ttlf\ttlf$\texttt{10}^{\texttt{-3}}$}
& {\ttlf\texttt{-1.4103~$\pmtt$~0.0126}}
& {\ttlf\texttt{0.891}}
\\
\cline{4-4}\cline{5-5}\cline{6-6}\cline{7-7}\cline{8-8}& & & {\ttlf3}
& {\ttlf\ttlf{$\texttt{1.3}$}}
& {\ttlf\ttlf$\texttt{10}^{\texttt{-2}}$}
& {\ttlf\texttt{-3.2543~$\pmtt$~0.0032}}
& {\ttlf\texttt{0.099}}
\\
\cline{4-4}\cline{5-5}\cline{6-6}\cline{7-7}\cline{8-8}& & & {\ttlf4}
& {\ttlf\ttlf{$\texttt{1.3}$}}
& {\ttlf\ttlf$\texttt{10}^{\texttt{-2}}$}
& {\ttlf\texttt{-6.8610~$\pmtt$~0.0055}}
& {\ttlf\texttt{0.080}}
\\
\cline{4-4}\cline{5-5}\cline{6-6}\cline{7-7}\cline{8-8}& & & {\ttlf5}
& {\ttlf\ttlf{$\texttt{1.3}$}}
& {\ttlf\ttlf$\texttt{10}^{\texttt{-2}}$}
& {\ttlf\texttt{-1.7552~$\pmtt$~0.0053}}
& {\ttlf\texttt{0.300}}
\\
\cline{4-4}\cline{5-5}\cline{6-6}\cline{7-7}\cline{8-8}& & & {\ttlf6}
& {\ttlf\ttlf{$\texttt{1.3}$}}
& {\ttlf\ttlf$\texttt{10}^{\texttt{-2}}$}
& {\ttlf\texttt{-2.3023~$\pmtt$~0.0032}}
& {\ttlf\texttt{0.138}}
\\
\cline{3-3}\cline{4-4}\cline{5-5}\cline{6-6}\cline{7-7}\cline{8-8}& & \multirow{3}{*}{${\twoloopfloopsnonplanar}$}
& {\ttlf1}
& {\ttlf\ttlf{$\texttt{0.3}$}}
& {\ttlf\ttlf$\texttt{10}^{\texttt{-1}}$}
& {\ttlf\texttt{~1.1829~$\pmtt$~0.0010}}
& {\ttlf\texttt{0.088}}
\\
\cline{4-4}\cline{5-5}\cline{6-6}\cline{7-7}\cline{8-8}& & & {\ttlf2}
& {\ttlf\ttlf{$\texttt{3.1}$}}
& {\ttlf\ttlf$\texttt{10}^{\texttt{-2}}$}
& {\ttlf\texttt{~2.0052~$\pmtt$~0.0057}}
& {\ttlf\texttt{0.285}}
\\
\cline{4-4}\cline{5-5}\cline{6-6}\cline{7-7}\cline{8-8}& & & {\ttlf3}
& {\ttlf\ttlf{$\texttt{1.3}$}}
& {\ttlf\ttlf$\texttt{10}^{\texttt{-2}}$}
& {\ttlf\texttt{~6.1770~$\pmtt$~0.0055}}
& {\ttlf\texttt{0.089}}
\\
\cline{3-3}\cline{4-4}\cline{5-5}\cline{6-6}\cline{7-7}\cline{8-8}& & \multirow{3}{*}{$\twoloopfloopsuv$}
& {\ttlf1}
& {\ttlf\ttlf{$\texttt{1.3}$}}
& {\ttlf\ttlf$\texttt{10}^{\texttt{-2}}$}
& {\ttlf\texttt{-1.4232~$\pmtt$~0.0004}}
& {\ttlf\texttt{0.025}}
\\
\cline{4-4}\cline{5-5}\cline{6-6}\cline{7-7}\cline{8-8}& & & {\ttlf2}
& {\ttlf\ttlf{$\texttt{3.1}$}}
& {\ttlf\ttlf$\texttt{10}^{\texttt{-4}}$}
& {\ttlf\texttt{-7.8155~$\pmtt$~0.0226}}
& {\ttlf\texttt{0.289}}
\\
\cline{4-4}\cline{5-5}\cline{6-6}\cline{7-7}\cline{8-8}& & & {\ttlf3}
& {\ttlf\ttlf{$\texttt{1.3}$}}
& {\ttlf\ttlf$\texttt{10}^{\texttt{-3}}$}
& {\ttlf\texttt{-9.4731~$\pmtt$~0.0027}}
& {\ttlf\texttt{0.029}}
\\
\cline{4-4}\cline{5-5}\cline{6-6}\cline{7-7}\cline{8-8}\hline\hline
\end{tabular}%
}%
\caption{\label{tab:me_single4}Contributions to the two-loop squared matrix element of $q\bar{q}\to \gamma\gamma\gamma$ via a heavy-quark $(m>0)$ loop, as defined in section~\ref{sec:individual_me}.}%

\end{table}

\section{Phase-space points}
\label{sec:ps_points}
\begin{lstlisting}[language=Python,frame=single,caption={Phase-space points for $q(p_1)\bar{q}(p_2)\to\gamma(q_1)\gamma(q_2)$.},commentstyle=\color{teal},captionpos=b,label={lst:1}]
#PSP1
0.5000000000000000E+03  0.0000000000000000E+00  0.0000000000000000E+00  0.5000000000000000E+03
0.5000000000000000E+03  0.0000000000000000E+00  0.0000000000000000E+00 -0.5000000000000000E+03
0.4999999999999999E+03  0.1109242844438328E+03  0.4448307894881214E+03 -0.1995529299308788E+03
0.4999999999999999E+03 -0.1109242844438328E+03 -0.4448307894881214E+03  0.1995529299308787E+03
#PSP2
0.5000000000000000E+03  0.0000000000000000E+00  0.0000000000000000E+00  0.5000000000000000E+03
0.5000000000000000E+03  0.0000000000000000E+00  0.0000000000000000E+00 -0.5000000000000000E+03
0.5000000000000000E+03 -0.2819155058093908E+03 -0.3907909666011574E+03  0.1334393795218208E+03
0.5000000000000002E+03  0.2819155058093908E+03  0.3907909666011575E+03 -0.1334393795218208E+03
#PSP3
0.5000000000000000E+03  0.0000000000000000E+00  0.0000000000000000E+00  0.5000000000000000E+03
0.5000000000000000E+03  0.0000000000000000E+00  0.0000000000000000E+00 -0.5000000000000000E+03
0.4999999999999990E+03 -0.2406210290937675E+03  0.2804720059911282E+03 -0.3368040590805997E+03
0.5000000000000012E+03  0.2406210290937674E+03 -0.2804720059911282E+03  0.3368040590805997E+03
\end{lstlisting}

\begin{lstlisting}[language=Python,frame=single,caption={Phase-space points for $q(p_1)\bar{q}(p_2)\to\gamma^*(q_1)\gamma^*(q_2)$.
},commentstyle=\color{teal},captionpos=b,label={lst:2}]
#PSP1
0.5000000000000000E+03  0.0000000000000000E+00  0.0000000000000000E+00  0.5000000000000000E+03
0.5000000000000000E+03  0.0000000000000000E+00  0.0000000000000000E+00 -0.5000000000000000E+03
0.4999999999999998E+03  0.1090639744303390E+03  0.4373705369731355E+03 -0.1962062298314933E+03
0.4999999999999998E+03 -0.1090639744303390E+03 -0.4373705369731355E+03  0.1962062298314931E+03
#PSP2
0.5000000000000000E+03  0.0000000000000000E+00  0.0000000000000000E+00  0.5000000000000000E+03
0.5000000000000000E+03  0.0000000000000000E+00  0.0000000000000000E+00 -0.5000000000000000E+03
0.4999999999999999E+03 -0.2771875038119387E+03 -0.3842370153902377E+03  0.1312014690844130E+03
0.5000000000000002E+03  0.2771875038119388E+03  0.3842370153902378E+03 -0.1312014690844130E+03
#PSP3
0.5000000000000000E+03  0.0000000000000000E+00  0.0000000000000000E+00  0.5000000000000000E+03
0.5000000000000000E+03  0.0000000000000000E+00  0.0000000000000000E+00 -0.5000000000000000E+03
0.4999999999999990E+03 -0.2365855763331326E+03  0.2757682129140208E+03 -0.3311555217306899E+03
0.5000000000000011E+03  0.2365855763331325E+03 -0.2757682129140208E+03  0.3311555217306899E+03
\end{lstlisting}
\begin{lstlisting}[language=Python,frame=single,caption={Phase-space points for $q(p_1)\bar{q}(p_2)\to\gamma^*(q_1)\gamma^*_2(q_2)$ with $q_1^2 = M_Z^2$ and $q_2^2 = (50\, \text{GeV})^2$.
},commentstyle=\color{teal},captionpos=b,label={lst:2}]
#PSP1
0.5000000000000000E+03  0.0000000000000000E+00  0.0000000000000000E+00  0.5000000000000000E+03
0.5000000000000000E+03  0.0000000000000000E+00  0.0000000000000000E+00 -0.5000000000000000E+03
0.5029075891223002E+03  0.1097199562383736E+03  0.4400011683720305E+03 -0.1973863419451843E+03
0.4970924108776998E+03 -0.1097199562383737E+03 -0.4400011683720306E+03  0.1973863419451842E+03
#PSP2
0.5000000000000000E+03  0.0000000000000000E+00  0.0000000000000000E+00  0.5000000000000000E+03
0.5000000000000000E+03  0.0000000000000000E+00  0.0000000000000000E+00 -0.5000000000000000E+03
0.5029075891223002E+03 -0.2788546900745418E+03 -0.3865480671686644E+03  0.1319906002100329E+03
0.4970924108776998E+03  0.2788546900745419E+03  0.3865480671686645E+03 -0.1319906002100329E+03
#PSP3
0.5000000000000000E+03  0.0000000000000000E+00  0.0000000000000000E+00  0.5000000000000000E+03
0.5000000000000000E+03  0.0000000000000000E+00  0.0000000000000000E+00 -0.5000000000000000E+03
0.5029075891222992E+03 -0.2380085561477647E+03  0.2774268626363575E+03 -0.3331473068184808E+03
0.4970924108777007E+03  0.2380085561477645E+03 -0.2774268626363575E+03  0.3331473068184808E+03
\end{lstlisting}
\newpage
\begin{lstlisting}[language=Python,frame=single,caption={Phase-space points for $q(p_1)\bar{q}(p_2)\to\gamma(q_1)\gamma(q_2)\gamma(q_3)$.
The heavy-quark loop contribution is evaluated with quark masses $m=m_t$, $m_t/2$, $2m_t$, where $m_t=172$ is the top-quark mass in GeV.},commentstyle=\color{teal},captionpos=b,label={lst:3}]
#PSP1
0.5000000000000000E+03  0.0000000000000000E+00  0.0000000000000000E+00  0.5000000000000000E+03
0.5000000000000000E+03  0.0000000000000000E+00  0.0000000000000000E+00 -0.5000000000000000E+03
0.4585787878854403E+03  0.1694532203096799E+03  0.3796536620781987E+03 -0.1935024746502525E+03
0.3640666207368177E+03 -0.1832986929319185E+02 -0.3477043013193672E+03  0.1063496077587081E+03
0.1773545913777421E+03 -0.1511233510164880E+03 -0.3194936075883155E+02  0.8715286689154436E+02
#PSP2
0.5000000000000000E+03  0.0000000000000000E+00  0.0000000000000000E+00  0.5000000000000000E+03
0.5000000000000000E+03  0.0000000000000000E+00  0.0000000000000000E+00 -0.5000000000000000E+03
0.4951533920773834E+03  0.1867229157692120E+03  0.3196835780850242E+03 -0.3288066974913060E+03
0.1026779138750091E+03 -0.7062042730772224E+02 -0.4345595295696615E+02  0.6055649756384902E+02
0.4021686940476072E+03 -0.1161024884614897E+03 -0.2762276251280580E+03  0.2682501999274569E+03
#PSP3
0.5000000000000000E+03  0.0000000000000000E+00  0.0000000000000000E+00  0.5000000000000000E+03
0.5000000000000000E+03  0.0000000000000000E+00  0.0000000000000000E+00 -0.5000000000000000E+03
0.2014347360150499E+03  0.1460392683472161E+03 -0.1127399318336771E+03 -0.8085909190808647E+02
0.4577010828828537E+03 -0.3806771896263719E+03 -0.7377601450610842E+02  0.2431712529348376E+03
0.3408641811020967E+03  0.2346379212791557E+03  0.1865159463397859E+03 -0.1623121610267509E+03
\end{lstlisting}
\begin{lstlisting}[language=Python,frame=single,caption={Phase-space points for $q(p_1)\bar{q}(p_2)\to\gamma^*(q_1)\gamma^*(q_2)\gamma^*(q_3)$.},commentstyle=\color{teal},captionpos=b,label={lst:4}]
#PSP1
0.5000000000000000E+03  0.0000000000000000E+00  0.0000000000000000E+00  0.5000000000000000E+03
0.5000000000000000E+03  0.0000000000000000E+00  0.0000000000000000E+00 -0.5000000000000000E+03
0.4477393491286647E+03  0.1619802634433327E+03  0.3629107790826664E+03 -0.1849689357540892E+03
0.3597595401382783E+03 -0.1752151450156574E+02 -0.3323703983032166E+03  0.1016595462179398E+03
0.1925011107330571E+03 -0.1444587489417669E+03 -0.3054038077944974E+02  0.8330938953614938E+02
#PSP2
0.5000000000000000E+03  0.0000000000000000E+00  0.0000000000000000E+00  0.5000000000000000E+03
0.5000000000000000E+03  0.0000000000000000E+00  0.0000000000000000E+00 -0.5000000000000000E+03
0.4763584456437345E+03  0.1763133172889205E+03  0.3018615679964441E+03 -0.3104760834666963E+03
0.1330984765879482E+03 -0.6668341566803356E+02 -0.4103333107364272E+02  0.5718053900827706E+02
0.3905430777683173E+03 -0.1096299016208870E+03 -0.2608282369228013E+03  0.2532955444584192E+03
#PSP3
0.5000000000000000E+03  0.0000000000000000E+00  0.0000000000000000E+00  0.5000000000000000E+03
0.5000000000000000E+03  0.0000000000000000E+00  0.0000000000000000E+00 -0.5000000000000000E+03
0.2133710236616980E+03  0.1398546405276659E+03 -0.1079655000888193E+03 -0.7743478422058600E+02
0.4477027246606839E+03 -0.3645558630552619E+03 -0.7065166858946638E+02  0.2328731754873389E+03
0.3389262516776180E+03  0.2247012225275959E+03  0.1786171686782860E+03 -0.1554383912667527E+03
\end{lstlisting}
\begin{lstlisting}[language=Python,frame=single,caption={Phase-space points for $q(p_1)\bar{q}(p_2)\to\gamma(q_1)\gamma(q_2)$ via a heavy-quark loop with quark masses $m^2 = 10/13$, $3/11$, $10/51$ in $\text{GeV}^2$ corresponding to the invariants $(s/m^2,t/m^2)=$ $(13/10,-3/5)$, $(11/3,-5/2)$, $(51/10,-11/10)$ of refs.~\cite{Ahmed:2025osb,Becchetti:2025rrz}.},commentstyle=\color{teal},captionpos=b,label={lst:psp}]
#PSP1
1/2               0  0     1/2
1/2               0  0    -1/2
1/2     sqrt(42)/13  0    1/26
1/2    -sqrt(42)/13  0   -1/26
#PSP2
1/2               0  0     1/2
1/2               0  0    -1/2
1/2    sqrt(105)/22  0   -2/11
1/2   -sqrt(105)/22  0    2/11
#PSP3
1/2               0  0     1/2
1/2               0  0    -1/2
1/2  2*sqrt(110)/51  0  29/102
1/2 -2*sqrt(110)/51  0 -29/102
\end{lstlisting}
\newpage
\begin{lstlisting}[language=Python,frame=single,caption={Phase-space points for $q(p_1)\bar{q}(p_2)\to\gamma^*(q_1)\gamma^*(q_2)$ via a heavy-quark loop with quark masses $m^2 = 10/13$, $3/11$, $10/51$ in $\text{GeV}^2$ corresponding to the invariants $(s/m^2,t/m^2)=$ $(13/10,-3/5)$, $(11/3,-5/2)$, $(51/10,-11/10)$.},commentstyle=\color{teal},captionpos=b,label={lst:psp_massive_full}]
#PSP1
0.5000000000000000E+00  0.0000000000000000E+00  0.0000000000000000E+00  0.5000000000000000E+00
0.5000000000000000E+00  0.0000000000000000E+00  0.0000000000000000E+00 -0.5000000000000000E+00
0.5000000000000000E+00  0.4906893300051509E+00  0.0000000000000000E+00  0.0301463600677784E+00
0.5000000000000000E+00 -0.4906893300051509E+00  0.0000000000000000E+00 -0.0301463600677784E+00
#PSP2
0.5000000000000000E+00  0.0000000000000000E+00  0.0000000000000000E+00  0.5000000000000000E+00
0.5000000000000000E+00  0.0000000000000000E+00  0.0000000000000000E+00 -0.5000000000000000E+00
0.5000000000000000E+00  0.4533587177288598E+00  0.0000000000000000E+00 -0.1901333602119417E+00
0.5000000000000000E+00 -0.4533587177288598E+00  0.0000000000000000E+00  0.1901333602119417E+00
#PSP3
0.5000000000000000E+00  0.0000000000000000E+00  0.0000000000000000E+00  0.5000000000000000E+00
0.5000000000000000E+00  0.0000000000000000E+00  0.0000000000000000E+00 -0.5000000000000000E+00
0.5000000000000000E+00  0.4068287398978793E+00  0.0000000000000000E+00  0.2759985470964361E+00
0.5000000000000000E+00 -0.4068287398978793E+00  0.0000000000000000E+00 -0.2759985470964361E+00
\end{lstlisting}

\newpage
\bibliographystyle{JHEP}
\bibliography{biblio}
\end{document}